%% file: matComp_HSBM_main.tex
\documentclass[journal,final,onecolumn,10pt,twoside]{IEEEtran}

\usepackage{amsmath}
\usepackage{amsfonts}
\usepackage{amssymb}
\usepackage{amsthm}
\usepackage{amstext}
\usepackage{mathrsfs}
\usepackage{mathtools}
\usepackage{dsfont}
\usepackage{cite}
\usepackage{times}
\usepackage{ifthen}
\usepackage{latexsym}
\usepackage{verbatim}
\usepackage{psfrag}
\usepackage{bbm}
\usepackage{float}
\usepackage{enumitem}
\usepackage{graphics}
\usepackage{graphicx}
\usepackage{epsfig}
\usepackage{epsf}
\usepackage{algorithm}
\usepackage{algpseudocode}
\usepackage[dvipsnames]{xcolor} 
\usepackage{hyperref}

\ifCLASSOPTIONcompsoc
  \usepackage[caption=false,font=normalsize,labelfont=sf,textfont=sf]{subfig}
\else
  \usepackage[caption=false,font=footnotesize]{subfig}
\fi

\input{mySymb.tex}

\newcommand{\red}{\textcolor[rgb]{0,0,0}}

\newenvironment{JA_AE}{\par\color{black}}{\par}

\newcommand{\given}{\;|\;}
\newcommand{\indicatorFn}[1]{\mathbbm{1} \left[#1\right]}

\newcommand{\gtMat}[0]{X_0}

\newcommand{\matSet}[0]{\mathcal{M}^{(\diff)}}
\newcommand{\partitionSet}[0]{\mathcal{Z}}
\newcommand{\vecU}[2]{u_{#1}^{(#2)}}
\newcommand{\vecV}[2]{v_{#1}^{(#2)}}

\newcommand{\grpG}[2]{Z_0(#2,#1)}

\newcommand{\kUsrs}[4]{n_{#1, #2}^{(#3, #4)}}

\newcommand{\estML}[0]{\psi_{\text{ML}}}

\newcommand{\numEdge}[3]{e_{#1} \left(#2,#3\right)}
\newcommand{\userPairSet}[2]{\mathcal{P}_{#1}\left(#2\right)}
\newcommand{\numDiffElmnt}[2]{\Lambda \left(#1,#2\right)}

\newcommand{\hamDist}[2]{d_{\text{H}} \left(#1,#2\right)}
\newcommand{\diff}{\delta}
\newcommand{\intraTau}[0]{\delta_g}
\newcommand{\interTau}[0]{\delta_c}

\newcommand{\Ir}{I_r}

\newcommand{\Ig}[0]{I_{\alpha,\beta}}
\newcommand{\IcOne}[0]{I_{\alpha,\gamma}}
\newcommand{\IcTwo}[0]{I_{\beta,\gamma}}
\newcommand{\tG}[2]{\widetilde{Z}_0(#2,#1)}
\newcommand{\uG}[2]{\overline{Z}_0(#2,#1)}

\newcommand{\colBlock}[1]{\mathcal{S}_{#1}}
\newcommand{\colBlockSize}[1]{s_{#1}}

\newcommand{\tupleSetDelta}[0]{\mathcal{T}^{(\delta)}}

\newcommand{\misClassfSet}[2]{\mathcal{P}_{#1 \rightarrow #2}}
\newcommand{\diffEntriesSet}[0]{\mathcal{P}_{\text{d}}}
\newcommand{\Pleftrightarrow}[2]{P_{#1 \leftrightarrow #2}}

\newcommand{\E}{\mathbb{E}}
\newcommand{\M}[1]{X_{\langle #1\rangle}}
\newcommand{\Z}[1]{Z_{\langle #1\rangle}}

\newcommand{\alphaEdge}[0]{\widetilde{\alpha}}
\newcommand{\betaEdge}[0]{\widetilde{\beta}}
\newcommand{\gammaEdge}[0]{\widetilde{\gamma}}
\newcommand{\alphaConst}[0]{\alpha}
\newcommand{\betaConst}[0]{\beta}
\newcommand{\gammaConst}[0]{\gamma}

\newcommand{\ratingCluster}[1]{R^{(#1)}}
\newcommand{\gtRatingCluster}[1]{R_{0}^{(#1)}}
\newcommand{\gtRatingClusterHat}[1]{\widehat{R}^{(#1)}}
\newcommand{\iGenerator}[1]{\Phi^{(#1)}}
\newcommand{\iBasis}[1]{W^{(#1)}}

\newcommand{\frakz}{Z}
\newcommand{\cV}{\mathcal{V}}
\newcommand{\cZ}{\mathcal{Z}}

\newcommand{\dElmntsGen}[4]{d_{#1, #2}^{(#3, #4)}}

\newcommand{\XRatingCluster}[1]{R^{(#1)}}

\newcommand{\TlargeOne}[0]{\mathcal{R}_{1}}
\newcommand{\TlargeTwo}[0]{\mathcal{R}_{2}}

\newcommand{\TlargeOneOne}[0]{\mathcal{R}_{1,1}}
\newcommand{\TlargeOneTwo}[0]{\mathcal{R}_{1,2}}
\newcommand{\TlargeOneThree}[0]{\mathcal{R}_{1,3}}
\newcommand{\TsmallErr}[0]{\mathcal{T}_{\textrm{small}}^{(\delta)}}
\newcommand{\TlargeErr}[0]{\mathcal{T}_{\textrm{large}}^{(\delta)}}

\newcommand{\relabelConst}[0]{\tau}
\newcommand{\numErrColConfig}[0]{\kappa}
\newcommand{\totNumErrCol}[1]{f^{(#1)}}
\newcommand{\numErrColElemnt}[2]{f_{#1}^{(#2)}}
\newcommand{\colVecRhat}[1]{w_{#1}}
\newcommand{\ratingVecSet}[1]{\mathcal{R}^{(#1)}}
\newcommand{\kUsrsHat}[4]{\widehat{n}_{#1, #2}^{(#3, #4)}}
\newcommand{\dElmntsGenHat}[4]{\widehat{d}_{#1, #2}^{\:(#3, #4)}}

\allowdisplaybreaks

\begin{document}

\title{On the Fundamental Limits of Matrix Completion: Leveraging Hierarchical Similarity Graphs}

\author{
Junhyung Ahn$^*$, Adel Elmahdy$^*$, Soheil Mohajer, and Changho Suh
\thanks{
$^*$The first two authors contributed equally to this work.
}
\thanks{
The work of Junhyung Ahn and Changho Suh was supported by the National Research Foundation of Korea (NRF) grant funded by the Korea government (MSIP) (No.2018R1A1A1A05022889).
The work of Adel Elmahdy and Soheil Mohajer was supported in part by the National Science Foundation under Grants CCF-1617884 and CCF-1749981.
This article was presented in part at the 2020 Advances in Neural Information Processing Systems Conference (NeurIPS)~\cite{elmahdy2020matrix}.
}
\thanks{
Junhyung Ahn is with the School of Electrical Engineering, Korea Advanced Institute of Science and Technology, Daejeon 34141, South Korea (e-mail: tonyahn96@kaist.ac.kr).
}
\thanks{
Adel Elmahdy and Soheil Mohajer are with the Department of Electrical and Computer Engineering, University of Minnesota, Minneapolis, MN 55455, USA (e-mail: adel@umn.edu, soheil@umn.edu).
}
\thanks{
Changho Suh is with the School of Electrical Engineering, Korea Advanced Institute of Science and Technology, Daejeon 34141, South Korea (e-mail: chsuh@kaist.ac.kr).
}
}

\maketitle

\begin{abstract}
We study the matrix completion problem that leverages hierarchical similarity graphs as side information in the context of recommender systems.
Under a hierarchical stochastic block model that well respects practically-relevant social graphs and a low-rank rating matrix model, we characterize the exact information-theoretic limit on the number of observed matrix entries (i.e., optimal sample complexity) by proving sharp upper and lower bounds on the sample complexity. 
In the achievability proof, we demonstrate that probability of error of the maximum likelihood estimator vanishes for sufficiently large number of users and items, if all sufficient conditions are satisfied.
On the other hand, the converse (impossibility) proof is based on the genie-aided maximum likelihood estimator. Under each necessary condition, we present examples of a genie-aided estimator to prove that the probability of error does not vanish for sufficiently large number of users and items.
One important consequence of this result is that exploiting the hierarchical structure of social graphs yields a substantial gain in sample complexity relative to the one that simply identifies different groups without resorting to the relational structure across them. 
More specifically, we analyze the optimal sample complexity and identify different regimes whose characteristics rely on quality metrics of side information of the hierarchical similarity graph.
Finally, we present simulation results to corroborate our theoretical findings and show that the characterized information-theoretic limit can be asymptotically achieved.
\end{abstract}


\section{Introduction}
\label{sec:intro}
\IEEEPARstart{I}{n} recent years, personalized recommender systems have emerged in an extensive range of Web applications to predict the preferences of its users and provide them with new and relevant items based on the scarce data about the users and/or items~\cite{koren2009matrix}.
There are two major paradigms of recommender systems: (i) content-based filtering systems; (ii) collaborative filtering systems. 
Content-based filtering approach exploits a profile of users' preferences and/or properties of the items to carry out the recommendation task. On the other hand, collaborative filtering approach recommends new items to the users based on similarity measures between users and items. The main advantage of collaborative filtering over content-based filtering is that they do not require domain knowledge since the embeddings are automatically learned, and the more interactions the users have with the items, the more accurate and relevant the new recommendations are.
Inspired by the Netflix challenge, a well-known technique to predict the missing ratings in collaborative filtering frameworks is low-rank matrix completion, which is the main interest of this paper. Given partial observation of a matrix of users by items, the goal is to develop an algorithm to accurately predict the values of the missing ratings.
One of the prime challenges of collaborative filtering systems that rely on user-item interactions is the ``cold start problem'' in which high-quality recommendations are not feasible for the new users/items that bear little or no information.
One prominent technique to overcome the problem with cold start users is to incorporate the community information into the framework of recommender systems in order to enhance the recommendation quality. 
Motivated by the social homophily theory \cite{mcpherson2001birds} that users within the same community are more likely to share similar preferences, socially-aware collaborative filtering approach exploits the social network among the users and provides recommendations based on the similarity measures of the users that have direct or indirect social relationship with a given user.

A plethora of research works have explored the idea of exploiting the information inferred by social graphs to enhance the performance of recommender systems from an algorithmic perspective \cite{koren2009matrix,tang2013social,cai2010graph,jamali2010matrix,li2009relation,ma2011recommender,kalofolias2014matrix,ma2008sorec,ma2009learning,guo2015trustsvd,zhao2017collaborative,chouvardas2017robust,massa2005controversial,golbeck2006filmtrust,jamali2009trustwalker,jamali2009using,yang2012bayesian,yang2012top,monti2017geometric,berg2017graph}. 
However, few works dedicated to developing theoretical insights on the usefulness of graph side-information on the quality of recommendation, and characterizing the maximal achievable gain due to side-information, e.g., \cite{chiang2015matrix,rao2015collaborative}.
Recently, a number of works \cite{ahn2018binary,yoon2018joint,tan2019community,jo2020discrete} have investigated the problem of interest from an information-theoretic perspective.
Ahn et~al. \cite{ahn2018binary} considered a matrix completion problem with $n$ users and $m$ items, and studied a simplified model where there are two clusters of users, and the users of each cluster share the same rating over the items.
A sharp threshold on the sample complexity is derived as a function of the quality of the social graph information, and the gain due to the information provided by social graph is theoretically quantified.
Furthermore, the authors proposed an efficient rating estimation algorithm that provably achieves the minimum sample complexity for reliable recovery of the ground truth rating matrix.
Follow-up works have investigated different models of the matrix completion problem proposed in \cite{ahn2018binary}.
Yoon et~al. \cite{yoon2018joint} considered a general setting where there are $K$ hidden communities of possibly different sizes, and each community is associated to only one feature, and hence the users of each community are assumed to provide the same binary rating over the items. 
Unlike \cite{ahn2018binary,yoon2018joint} where one-sided graph side-information (i.e., user-to-user similarity graph) is considered, Zhang et~al. \cite{tan2019community} studied the benefits of two-sided graph side-information depicted by user-to-user and item-to-item similarity graphs.
Interestingly, the theoretical analysis demonstrates that there is a synergistic effect, under some scenarios, stemmed from considering two pieces of graph side-information. This implies that observing both graphs is necessary in order to slash the sample complexity under those scenarios.
Jo et~al. \cite{jo2020discrete} relaxed the assumption in \cite{ahn2018binary} on the preference matrix whose element at row $i$ and column $j$ denotes the probability that user $i$ likes item $j$, and proposed a new model in which the unknown entries of the preference matrix can take discrete values drawn from a known finite set of probabilities.
While the works of \cite{ahn2018binary,yoon2018joint,tan2019community,jo2020discrete} lay out the theoretical foundation for the problem, they impose a number of strict assumptions on the system model such as the users of the same cluster have same ratings over all items, which limits the practicality of the proposed models for real-world data.

A natural hypothesis in the theory of recommender systems is that the unknown rating matrix has an intrinsic structure of being low rank. This hypothesis is sensible because it is generally believed that only a few factors contribute to one's preference.
Prior works \cite{ahn2018binary,yoon2018joint,tan2019community,jo2020discrete} assume that each cluster is represented by a rank-one matrix, and users within a cluster share the same rating vector over items.
In this work, we relax this assumption and study a more generalized framework where each cluster is represented by a rank-$r$ matrix. More specifically, we consider a matrix completion problem where the users are categorized into $c$ clusters, each of which comprises $g$ sub-clusters, or what we call ``groups'', producing a hierarchical structure in which the features of different groups within a cluster are broadly similar to each other, however, they are different from the features of the groups in other clusters.
The goal is to reliably retrieve the rating matrix under the proposed generalized model, utilizing the information provided by the noisy partial observation of the rating matrix, as well as the hierarchical social graph.

\subsection{Related Works}
\label{subsec:related}
\subsubsection{Connection to Low-Rank Matrix Completion Problem}
The objective of low-rank matrix completion, a recurring problem in collaborative filtering \cite{koren2009matrix}, is to recover an unknown low-rank matrix from partial, and possibly noisy, sampling of its entries \cite{nguyen2019low} .
Since the rank minimization problem is NP-hard, accurate reconstruction is generally ill-posed and computationally intractable. However, exploiting the fact that the structure of the matrix is of low-rank makes the exploration for solution worthwhile.
One direction of research is geared towards studying low-rank matrix completion where the observed subset of matrix entries are exactly known. Under certain conditions, upper bounds on the number of observed entries, which are uniformly drawn at random, are developed to ensure successful reconstruction with high probability \cite{candes2009exact,keshavan2010matrix,candes2010power}. 
A fundamental open question in the literature of low-rank matrix completion with exact observation is how to find a low-rank matrix that is consistent with the partial observation of its entries. This question stems from the fact that the sparse basis of the low-rank matrix is unknown, and that the basis are drawn from a continuous space.
Performance guarantees provided by existing algorithms only hold when certain incoherence assumptions on the singular vectors of the matrix are satisfied. 
By and large, theoretical guarantees on the reconstruction performance are not established even for the rank-one case, and hence, our understanding of the problem is far from complete.
Numerous algorithms for low-rank matrix completion have been proposed over the years. If the rank information of the original low-rank matrix is unknown, various techniques based on nuclear norm minimization are proposed \cite{fazel2002matrix,candes2009exact,keshavan2010matrix,candes2010power,cai2010singular,fornasier2011low,mohan2012iterative}. 
On the other hand, if the rank is known in advance, techniques based on Frobenius norm minimization are proposed \cite{lee2010admira,wang2014rank,tanner2016low,wen2012solving,dai2705set,vandereycken2013low,hu2012fast,gotoh2018dc}.
Another interesting and practical research direction is investigating low-rank matrix completion when the observed entries are contaminated by noise. The objective is to seek a low-rank matrix that best approximates the original matrix, and find an upper bound on the root-mean squared error \cite{candes2010matrix,keshavan2010matrix}.

\subsubsection{Algorithms for Recommender Systems with Graph Side-Information}
The idea of exploring the value of incorporating graph side-information into collaborative filtering approaches has gained a lot of attention from the research community \cite{tang2013social}.
There are two primary approaches of collaborative learning \cite{koren2009matrix}: (i) latent factor approach, and (ii) neighborhood approach.
Latent factor approach learns latent features for users and items from the observed ratings. Most successful realizations of this approach hinge on matrix factorization which characterizes the latent characteristics of users and items by two low-rank user- and item-feature matrices inferred from the rating patterns.
One direction to integrate graph side-information in this approach is through adding some regularization terms in the loss function of the matrix factorization model \cite{cai2010graph,jamali2010matrix,li2009relation,ma2011recommender,kalofolias2014matrix}.
Another direction is to develop matrix factorization frameworks that fuse the user-item rating matrix with the social network of the users \cite{ma2008sorec,ma2009learning,guo2015trustsvd,zhao2017collaborative}.
Moreover, a robust online matrix completion on graphs is designed and analyzed in \cite{chouvardas2017robust} that exploits the graph information to recover the incomplete rating matrix entires in the presence of outlier noise.
On the other hand, for neighborhood approach, the prediction of rating information is based on computing the relationships among items or users. The recommendation accuracy in this approach can be enhanced by incorporating the information provided by the social graphs into the neighborhood definition \cite{massa2005controversial,golbeck2006filmtrust,jamali2009trustwalker,jamali2009using,yang2012bayesian,yang2012top}.
Lately, recent works have proposed novel architectures for graph convolutional neural networks that fully exploit the structure of item/user graphs \cite{monti2017geometric,berg2017graph}.

Few works have provided theoretical insights on the usefulness of side-information for matrix completion problem, e.g., \cite{chiang2015matrix,rao2015collaborative}.
Chiang et~al. \cite{chiang2015matrix} proposed a dirty statistical model to exploit the feature-based side information, yet to be robust to feature noise, in matrix completion applications. They provided theoretical guarantees that the proposed model achieves lower sample complexity than the standard matrix completion (with no graph information) under the condition that the features are not too noisy. 
Rao et~al. \cite{rao2015collaborative} proposed a scalable graph regularized matrix completion, and derived consistency guarantees to demonstrate the gain due to the graph side-information. 
It is worth mentioning that the maximal achievable gain due to graph side-information is not characterized in these works.

\subsubsection{Connection to Community Detection in Stochastic Block Model}
Stochastic Block Model (SBM) \cite{holland1983stochastic} with two communities, which is also known as the planted bisection model, is an extension to Erd\"{o}s-R\'{e}nyi (ER) random graph model that exhibits community structure. It is based on the assumption that agents (nodes) in a network (graph) connect depending on their community assignment, unlike Erd\"{o}s-R\'{e}nyi (ER) model in which agents connect independently.
The goal is to reconstruct the community assignment upon observing the unlabeled graph.
This problem has been well investigated from an information-theoretic perspective where a sharp threshold is characterized for exact recovery of communities \cite{abbe2017community,abbe2015exact,abbe2015community,jog2015information,mossel2015reconstruction,hajek2017information,saad2018community,saad2018exact,saad2018recovering}.
Inspired by applications in physics, computational social science, and machine learning, several generalizations of the classical SBM have been developed to explore the benefits of considering various types of side-information in the task of community recovery.
Generalizations have been made to incorporate infinite number of blocks \cite{kemp2006learning}, probabilistic block membership \cite{airoldi2008mixed,ball2011efficient}, heterogeneous degree within blocks \cite{karrer2011stochastic}, weighted edges between nodes \cite{aicher2013adapting}, and node attributes \cite{yang2013community}.
One crucial and practical generalization to SBM, that we consider in this work, is the hierarchical (nested) organization of blocks, where vertices are partitioned into blocks, each of which is further partitioned into sub-blocks \cite{clauset2008hierarchical,peixoto2014hierarchical,lyzinski2016community,cohen2019hierarchical}. 
The edge probabilities in a hierarchical random graphs are inhomogeneous and determined by the topology structure.

\subsubsection{Connection to Clustering Problem with Side-Information}
There has been some recent works that investigate the clustering 
problem with side-information, where the learner is allowed to interact with a domain expert.
Ashtiani et~al. \cite{ashtiani2016clustering} proved that having access to few pairwise queries leads to more efficient k-means clustering, which is NP-hard in general.
Mazumdar et~al. \cite{mazumdar2017query} explored the benefits of the similarity matrix, that is used to cluster similar points together, to slash the adaptive query complexity.
In both works, information-theoretic lower bounds are proved, and efficient clustering algorithms are designed.
Matrix Completion problem with graph side-information can be seen as a natural extension to clustering problem if we shift our focus to recovering the structure of (hierarchical) clusters instead of reconstructing the rating matrix.

\subsection{Main Contributions}
\label{sebsec:contributions}
The main contributions of this paper are summarized as follows.
We conduct a rigorous theoretical study and characterize an information-theoretic threshold for optimal rating matrix recovery, rather than starting with a certain algorithmic approach.
We establish matching upper and lower bounds on the sample complexity for reliable recovery of the rating matrix, and hence exactly characterize the optimal sample complexity that is a function of the quality of social graph side-information.
In the proof of achievability bound, we show that probability of error of the maximum likelihood estimator vanishes for sufficiently large number of users and items, if all sufficient conditions are met.
On the other hand, the proof of the converse bound is based on the genie-aided maximum likelihood estimator. Under each necessary condition, we present examples of a genie-aided estimator to prove that the probability of error does not vanish for sufficiently large number of users and items.

While numerous low-complexity matrix completion algorithms have been proposed, it remains an open problem to develop optimization algorithms with provable performance guarantees for a generic class of matrices \cite{nguyen2019low}. This work makes substantial progress on this long-standing open problem.
We also emphasize on the fact that this work is a non-trivial extension of \cite{ahn2018binary} and \cite{yoon2018joint}, as will be delineated in the following sections.

A preliminary version of the main results of this paper has been reported in \cite{elmahdy2020matrix} for $(c,g,r,q)=(2,3,2,2)$. In this paper, we characterize the optimal sample complexity result for any $(c,g,r,q)$, and present the complete achievability and converse proofs.

\subsection{Notation}
\label{subsec:notation}
Row vectors and matrices are denoted by lowercase letters (e.g., $v$) and uppercase letters (e.g., $X$), respectively. 
Random matrices are denoted by boldface uppercase letters (e.g., $\mathbf{X}$), while their realizations are denoted by uppercase letters (e.g., $X$).
Sets are denoted by calligraphic letters (e.g., $\mathcal{Z}$). 
Let $\mathbb{F}_{q}$ be a finite field of order $q$ for some prime number $q$.
Let $\mathbf{0}_{n \times m}$ and $\mathbf{1}_{n \times m}$ be all-zero and all-one matrices of dimension $n \times m$, respectively. 
For a matrix $X \in \mathbb{F}_{q}^{n \times m}$, let $X^\intercal$ denote the transpose of $X$.
Let $X(r,t)$ denote the matrix entry at row $r$ and column $t$.
Furthermore, let $X(i,:)$ and $X(:,j)$ denote the $i^{\text{th}}$ row and $j^{\text{th}}$ column of matrix $X$, respectively. Furthermore, for sets $\cI$ and $\cJ$, the submatrix of $X$, that is obtained by rows $i\in\cI$ and columns $j\in\cJ$, is denoted by $X(\cI,\cJ)$.
Let $\numDiffElmnt{X}{Y}$ denote the number of different entries between matrices $X$ and $Y$ for $X,Y \in \mathbb{F}_{q}^{n \times m}$.
For $u,v \in \mathbb{F}_q^{1 \times m}$, we use $[u;\:v]$ to denote $[u^\intercal \ v^\intercal]^\intercal$.
Let $v(t)$ denote the $t^{\text{th}}$ entry of $v$.
Moreover, the Hamming distance between two vectors $u$ and $v$ is denoted by $\hamDist{u}{v}$. It is defined as the number of entries in which $u$ and $v$ differ.
Let $\left\| u \right\|_1$ denote the $\ell_1$ norm of vector $u$.
For an integer $n \geq 1$, let $[n]$ denote the set of integers $\{1,2,\ldots,n\}$. 
For integers $n$ and $m$ where $n \leq m$, define $[n:m]$ as $\{n, n+1, \cdots, m\}$.
Let $\{0,1,\cdots,q-1\}^n$ be the set of all $n$-digit sequences whose digits are drawn from $\mathbb{F}_q$.
Moreover, we use $\indicatorFn{\cdot}$ to denote the indicator function. 
Finally, $\cG=([n], \cE)$ denotes an undirected graph $\cG$ where $[n]$ is the set of $n$ vertices labeled by integers in $[n]$; and $\cE$ is the set of undirected edges. 
The elements of $\cE$ are given by pairs $(a,b)$ for $a,b \in [n]$. 

\subsection{Paper Outline}
\label{subsec:outline}
The remainder of this paper is organized as follows. We first present the problem formulation of the rating matrix and hierarchical similarity graph in Section~\ref{subsec:probForm}. Then, the main results are  presented in Section~\ref{subsec:mainRes}. In Section~\ref{sec:interpretation}, information-theoretic interpretation of the main results is provided, and extensive numerical results are presented.
Next, the achievability proof is presented in Section~\ref{sec:achv}, while the converse proof is presented in Section~\ref{sec:conv}.
In Section~\ref{sec:expResults}, we present simulation results that corroborate our main results.
Finally, the paper is concluded and directions for future research are discussed in Section~\ref{sec:conclusion}.

\section{Problem Formulation}
\label{subsec:probForm}
Consider a rating matrix $X \in \mathbb{F}_{q}^{n \times m}$ where $n$ denotes the number of users and $m$ denotes the number of items.
The ratings of the $r^{\text{th}}$ user over $m$ items are given by the rating vector that corresponds to the $r^{\text{th}}$ row of $X$ for $r \in [n]$. 
The user similarity graphs (e.g., social graphs) are leveraged as side-information during the matrix completion procedure to enhance the item recommendation quality. 
More specifically, we consider a hierarchical similarity graph that consists of $c$ disjoint clusters, and each cluster comprises $g$ disjoint sub-clusters (or that we call ``groups'').
For the sake of tractable mathematical analysis, we assume equal-sized clusters and groups. 
The theoretical guarantees, however, hold as long as the group sizes are order-wise same (See Section~\ref{subsec:mainRes}).
According to social homophily theory~\cite{mcpherson2001birds}, users within the same community (i.e., who are more likely to be connected in a social graph) are more likely to share similar preferences of items. 
This results in a low rank structure of the rating matrix since the rows of the rating matrix associated with such users are highly likely to be similar \cite{nguyen2019low}.
To capture this crucial fact in our model, we make the following assumptions: (i) the rating vectors of the users who belong to the same group are equal, and hence there are $gc$ distinct rating vectors in total; (ii) the rating vectors of the groups of a given cluster are different, yet intimately-related to each other through a linear subspace of $r$ basis vectors for some integer $r \leq g$.
Let $\vecV{i}{x}$ denotes the rating vector of the users in cluster $x$ and group $i$ for $x \in [c]$ and $i \in [g]$. 
Let $\ratingCluster{x} \in \mathbb{F}_{q}^{g \times m}$ denote a matrix whose rows are the rating vectors of the groups in cluster~$x$ for $x \in [c]$.
The set of $g$ rows of $\ratingCluster{x}$ (i.e., set of $g$ rating vectors of the groups in cluster~$x$) is spanned by any subset of $r$ rows of $\ratingCluster{x}$.

Let $\gtMat$ denote the ground truth rating matrix.
Each instance of the problem corresponds to a rating matrix $\gtMat$, which can be represented by a set of rating vectors $\cV_0 = \{ \vecU{i}{x}: x\in [c], i\in [g]\}$ and a user partitioning $\cZ_0$.
For instance, consider a problem with $n=12$ users in $c=2$ clusters and $g=3$ groups. If the rating matrix is given by
\begin{align}
    \gtMat = 
    \begin{bmatrix}
        \vecU{1}{1};\:\: \vecU{2}{1};\:\: 
        \vecU{1}{2};\:\: \vecU{3}{2};\:\: 
        \vecU{3}{1};\:\: \vecU{2}{1};\:\: 
        \vecU{2}{2};\:\: \vecU{1}{2};\:\:
        \vecU{3}{1};\:\: \vecU{3}{2};\:\: 
        \vecU{2}{2};\:\: \vecU{1}{1}
    \end{bmatrix},
\end{align}
then we have the set of rating vectors as
\begin{align}
    \cV_0
    &= 
    \left\{
    \vecU{1}{1},\: \vecU{2}{1},\: \vecU{3}{1},\: \vecU{1}{2},\: \vecU{2}{2},\: \vecU{3}{2}
    \right\},
\end{align}
and the user partitioning as
\begin{align}
    \cZ_0 &=
    \left\{
	\frakz_0(1,1)\!=\!\{1,12\},\: \frakz_0(1,2)\!=\!\{2,6\},\: \frakz_0(1,3)\!=\!\{5,9\},\: \frakz_0(2,1)\!=\!\{3,8\},\: \frakz_0(2,2)\!=\!\{7,11\},\: \frakz_0(2,3)\!=\!\{4,10\}
	\right\}.
\end{align}
Formally, $\cZ_0$ is a family of subsets of $[n]$ that partitions the set of all users $[n]$ into $c$ clusters and $g$ groups (per cluster). That is, 
\begin{align}
	\cZ_0 = 
	\left\{
	\left\{\frakz_0(x,i)\right\}_{x \in [c],\: i \in [g]}: 
	\frakz_0(x,i) \subseteq [n],\:
	\frakz_0(x,i) \cup \frakz_0(y,j) = \varnothing \mbox{ for } (x,i) \neq (y,j), \:
	\bigcup_{x\in[c]} \bigcup_{i\in[g]} \frakz_0(x,i) = [n]
	\right\},
	\label{eq:Z_defn}
\end{align}
where $\frakz_0(x,i)$ denotes the set of users who belong to cluster $x$ and group $i$ for $x \in [c]$ and $i \in [g]$.

The main goal of the problem of interest is to find the best estimate of $\gtMat$ with the knowledge of two types of observations:
\begin{enumerate}
	\item partial and noisy observation $Y$ of $\gtMat$.
	For every $r \in [n]$ and $t \in [m]$, let $Y(r,t) \in \mathbb{F}_q \cup \{*\}$, where $*$ denotes no~observation.
	Let the set of observed entries of $\gtMat$ be denoted by $\Omega=\{(r,t)\in [n]\times [m]: Y(r,t)\neq *\}$. 
	The partial observation is modeled by assuming that each entry of $\gtMat$ is observed with probability $p \in [0,1]$, independently from others.
	Moreover, the potential noise in the observation is modeled by considering a random uniform noise distribution, that is, the noise is not adversarial (i.e., not deterministic). 
	In particular, we assume that each observed entry $\gtMat(r,t)$, for $(r,t) \in \Omega$, can be possibly flipped to any element of the set $\{0,1,\ldots, q\} \setminus \gtMat(r,t)$ with a uniform probability of $\theta/(q-1)$ for $\theta \in [0, (q-1)/q)$. 
	The reasons behind choosing this noise model are: (i) the uniform noise distribution is the worst case distribution in discrete channels; and (ii) this model captures the fact that there may exist a fraction of group members whose ratings are close to the majority ratings, yet they are not exactly identical. Hence, the majority ratings can be considered as the ground truth, while the ratings of this fraction of users can be seen as noisy version of the ground truth.
	Furthermore, since this fraction of users have some ratings that are different from the majority, each of such ratings can take a value that is randomly and uniformly selected from the set of all possible ratings different from that of the majority;
	\item
	user similarity graph ${\mathcal{G}=([n], \mathcal{E})}$.
	A~vertex represents a user, and an edge captures a social connection between two users. 
	The set $[n]$ of vertices is partitioned into $c$ disjoint clusters, each of which is of size $n/c$ users. Each cluster is further partitioned into $g$ disjoint groups, each of which is of size $n/(cg)$ users. 
	The user similarity graph is generated as per the hierarchical stochastic block model (HSBM)~\cite{abbe2017community,lyzinski2016community}, which is a generative model for random graphs exhibiting hierarchical cluster behavior. 
	In this model, each two nodes in the graph are connected by an edge, independent of all other nodes, with probabilities given by
	\begin{align}
		\begin{split}
			\alphaEdge &:= \mathbb{P} \left[(a,b) \in \mathcal{E}: a,b \in \grpG{i}{x} \right], 
			\text{ for } x \in [c],\: i \in [g],
			\\
			\betaEdge &:= \mathbb{P} \left[(a,b) \in \mathcal{E}: a \in \grpG{i}{x}, b \in \grpG{j}{x}\right],
			\text{ for } x \in [c],\: i,j \in [g],\: i \neq j,
			\\
			\gammaEdge &:= \mathbb{P} \left[(a,b) \in \mathcal{E}: a \in \grpG{i}{x}, b \in \grpG{j}{y}\right],
			\text{ for } x,y \in [c],\: x \neq y,\: i,j \in [g].
		\end{split}
		\label{eq:edge_prob_defn}
	\end{align}
	Here, we assume that edge probabilities are scaling with the size of the problem. In particular, we assume 
	\begin{align}
	    \alphaEdge = \alpha \frac{\log n}{n},
	    \quad
	    \betaEdge = \beta \frac{\log n}{n},
	    \quad
	    \gammaEdge = \gamma \frac{\log n}{n}, 
	    \label{eq:edge_prob_assump}
	\end{align}
	where $\alpha$, $\beta$ and $\gamma$ are positive real numbers such that $\alpha \geq \beta \geq \gamma$.
	In other words, there is an edge between two users in the \emph{same group} within a cluster with probability $\alphaConst \frac{\log n}{n}$; there is an edge between two users in \emph{different groups} but within the \emph{same cluster} with probability $\betaConst \frac{\log n}{n}$; and there is an edge between two users in \emph{different clusters} with probability $\gammaConst \frac{\log n}{n}$.
	Note that the considered edge probabilities guarantee the disappearance of isolated vertices (i.e., vertices of degree zero) in the user similarity graph, which is a necessary property for exact recovery in stochastic block model (SBM) \cite{abbe2015community}.
	Furthermore, motivated by the social homophily theory~\cite{mcpherson2001birds}, we study the problem of interest when users within the same group (or cluster) are more likely to be connected than those in different groups (or clusters). That is why we assume that $\alphaConst \geq \betaConst \geq \gammaConst$. 
\end{enumerate}

Let $\psi$ denote an estimator (decoder) that takes as input a pair $(Y,\mathcal{G})$ where $Y$ is the incomplete and noisy rating matrix and $\mathcal{G}$ is the user similarity graph, and outputs a completed rating matrix $X\in \mathbb{F}_{q}^{n \times m}$. 
Note that both the set of rating vectors $\cV$ and the user partitioning\footnote{To be more precise, $\cZ$ can be recovered from $X$ up to a relabeling of the clusters and groups, i.e., we can only identify which users belong to the same group/cluster, but the label associated to a group/cluster cannot be identified.} $\cZ$ can be recovered from the completed rating matrix $X$ and vice versa. Hence, with slightly abuse of notation, we may interchangeably use $X$ or $(\cV, \cZ)$ as the output of the estimator. 
The former notation is adopted when we are interested in the entries of the rating matrix, while the latter notation is used when we are interested in either the set of rating vectors or the user partitioning.

One key parameter, that is instrumental in expressing the main result (see Section~\ref{subsec:mainRes}) as well as proving the main theorem, is the discrepancy between the rating vectors. 
Let $\intraTau$ be the minimum normalized Hamming distance among distinct pairs of rating vectors of groups \emph{within the same cluster}. Let $\interTau$ be the counterpart with respect to distinct pairs of rating vectors \emph{across different clusters}. More formally, $\intraTau$ and $\interTau$ are given by
\begin{align}
\begin{split}
    \intraTau &= \frac{1}{m} 
    \: \min_{x \in [c]} \:
    \min_{\substack{i,j \in [g] \\ i \neq j}} \hamDist{\vecU{i}{x}}{\vecU{j}{x}},
	\\
	\interTau  &= \frac{1}{m}  
	\min_{\substack{i,j \in [g] \\ x,y \in [c], x\neq y}}
	\hamDist{\vecU{i}{x}}{\vecU{j}{y}}.
\end{split}
\end{align}
As will be elaborated in the next section, our result hinges on $\diff \coloneqq (\intraTau, \interTau)$. We provide theoretical guarantees for the recovery of all rating matrices $M$ in which the rating vectors maintain a minimum level of dissimilarity. Formally, we define $\matSet$ to be the set of rating matrices $M=(\cV,\cZ)$ such that the following properties are satisfied:
\begin{itemize}
    \item the set of rating vectors $\cV$ must satisfy the property that the minimum normalized Hamming distance among the rating vectors in different groups within the same cluster and those in different clusters are not smaller that $\intraTau$ and $\interTau$, respectively;
    \item the user partitioning $\cZ$ must satisfy the property that the size of clusters is $c/n$ users, while the size of the groups is $c/(ng)$ users.
\end{itemize}

The performance metric we consider to provide theoretical guarantees on the quality of recommendation is the worst-case probability of error $P_e$. 
In other words, the quality of the estimator is defined by its accuracy of estimation of the \emph{most difficult} ground truth matrix $M=(\cV,\cZ)\in \matSet$. 
Therefore, we apply a minimax optimization approach wherein the objective is to find the estimator that minimizes the maximum risk (i.e., minimizes the worst-case probability of error). This can be expressed as
\begin{align}
\label{eq:errorProb_wc}
	\inf_{\psi} P_e^{(\diff)} (\psi) 
	= 
	\inf_{\psi} \max_{M \in \matSet} 
	\mathbb{P}\left[\psi(Y, \mathcal{G}) \neq M
	\right].
\end{align}
Based on the proposed problem formulation and performance metric, we aim at characterizing the optimal sample complexity (i.e., the minimum number of entries of the rating matrix that is required to be observed), concentrated around $nmp^{\star}$ in the limit of $n$ and $m$, for exact rating matrix recovery.
Here, $p^{\star}$ denotes a sharp threshold on the observation probability such that the following conditions, in the limit of $n$ and $m$, are satisfied:
\begin{itemize}
	\item when $p > p^{\star}$, there exists an estimator such that the error probability can be made arbitrarily close to 0;
	\item when $p < p^{\star}$, the error probability does not converge to zero no matter what and whatsoever.
\end{itemize}

\section{Main Results}
\label{subsec:mainRes}
\begin{theorem}[Optimal Sample Complexity]
\label{thm:p_star}
	Let $m = \omega(\log n)$ and $\log m = o(n)$.
	Let $q,\theta,c,g$ and $r$ be constants such that $q$ is prime, $\theta \in [0,(q-1)/q)$, and $r \leq g$.
	For any constant $\epsilon >0$, if 
	\begin{align}
		p
		\geq\!
		\frac{1}{\left(\!\!\sqrt{1\!-\!\theta} \!-\!\! \sqrt{\frac{\theta}{q-1}}\right)^{\!\!2}} 
		\max \!\left\{\! 
		\frac{gc (1\!+\!\epsilon)}{g\!-\!r\!+\!1} \frac{\log m }{n},  
		\frac{\log n}{\intraTau m} \!\left(\!\!(1\!+\!\epsilon) \!-\! \frac{\left(\sqrt{\alphaConst} \!-\!\! \sqrt{\betaConst}\right)^{\!2}}{gc}\!\right)\!,
		\frac{\log n}{\interTau m} \!\left(\!\!(1\!+\!\epsilon) \!-\! \frac{\left(\sqrt{\alphaConst} \!-\!\! \sqrt{\gammaConst}\right)^{\!2} \!\!+\! (g\!-\!1) \!\left(\sqrt{\betaConst} \!-\!\! \sqrt{\gammaConst}\right)^{\!2}}{gc} \!\right)
		\!\!\right\}\!\!,
		\label{eq:pstar-general_achv}
	\end{align}
	then there exists an estimator $\psi$ that outputs a rating matrix $X \in \matSet$ given $Y$ and ${\cal G}$ such that $\lim_{n \rightarrow \infty} P_e^{(\diff)} (\psi) = 0$; 
	conversely, if
	\begin{align}
		p
		\leq\!
		\frac{1}{\left(\!\!\sqrt{1\!-\!\theta} \!-\!\! \sqrt{\frac{\theta}{q-1}}\right)^{\!\!2}} 
		\max \!\left\{\! 
		\frac{gc (1\!-\!\epsilon)}{g\!-\!r\!+\!1} \frac{\log m }{n},  
		\frac{\log n}{\intraTau m} \!\left(\!\!(1\!-\!\epsilon) \!-\! \frac{\left(\sqrt{\alphaConst} \!-\!\! \sqrt{\betaConst}\right)^{\!2}}{gc}\!\right)\!,
		\frac{\log n}{\interTau m} \!\left(\!\!(1\!-\!\epsilon) \!-\! \frac{\left(\sqrt{\alphaConst} \!-\!\! \sqrt{\gammaConst}\right)^{\!2} \!\!+\! (g\!-\!1) \!\left(\sqrt{\betaConst} \!-\!\! \sqrt{\gammaConst}\right)^{\!2}}{gc} \!\right)
		\!\!\right\}\!\!,
		\label{eq:pstar-general_conv}
	\end{align}
	then $ \lim_{n \rightarrow \infty} P_e^{(\diff)}(\psi) \neq 0$ for any estimator $\psi$.
	Therefore, the optimal observation probability $p^{\star}$ is given by
 	\begin{align}
		p^{\star} 
		=
		\frac{1}{\left(\sqrt{1\!-\!\theta} \!-\! \sqrt{\frac{\theta}{q-1}}\right)^2} 
		\max \!\left\{\! 
		\frac{gc}{g\!-\!r\!+\!1} \frac{\log m }{n},\:  
		\frac{\log n}{\intraTau m} \left(\!1 \!-\! \frac{\left(\sqrt{\alphaConst} \!-\! \sqrt{\betaConst}\right)^2}{gc}\!\right),\:
		\frac{\log n}{\interTau m} \left(\!1-\frac{\left(\sqrt{\alphaConst} \!-\! \sqrt{\gammaConst}\right)^2 \!+\! (g\!-\!1) \!\left(\sqrt{\betaConst} \!-\! \sqrt{\gammaConst}\right)^2}{gc} \!\right)
		\!\!\right\}\!.
		\label{eq:pstar-general}
	\end{align} 
\end{theorem}
\begin{IEEEproof}
    We provide a concise proof sketch of Theorem~\ref{thm:p_star}, along with the technical distinctions with respect to the prior works~\cite{ahn2018binary,yoon2018joint}.
    We defer the complete achievability proof to Section~\ref{sec:achv}, and the converse proof to Section~\ref{sec:conv}.
    
    The achievability proof is based on maximum likelihood estimation (MLE). 
    We first evaluate the likelihood for a given clustering/grouping of users and the corresponding rating matrix.
    Next, we provide an upper bound on the worst-case probability of error, which is given by the probability that the likelihood of the ground truth rating matrix is less than that of a candidate rating matrix.
    Finally, we perform typical and atypical error analyses and demonstrate that if $p > p^\star$, where $p^\star$ is given by \eqref{eq:pstar-general}, the likelihood is maximized only by the ground-truth rating matrix, and hence the worst-case probability of error vanishes, in the limit of $n$ and $m$. This completes the achievability proof.
    
    On the other hand, the converse (impossibility) proof hinges on the genie-aided maximum likelihood estimation.
    We first establish a lower bound on the error probability, and show that it is minimized when employing the maximum likelihood estimator. 
    Next, we prove that if $p$ is smaller than any of the three terms in the RHS of~\eqref{eq:pstar-general}, then there exists another solution that yields a larger likelihood, compared to the ground-truth matrix. 
    More precisely, for any estimator and any ground truth rating matrix, we have the following three cases:
    \begin{itemize}
        \item if $p \leq \frac{(1-\epsilon) gc \log m}{(\sqrt{1-\theta}- \sqrt{\theta / (q-1)})^2 (g-r+1) n}$, there exists a class of matrices that is obtained by replacing one column of the ground truth rating matrix with a carefully chosen sequence, and yields a larger likelihood than the one of the ground truth rating matrix;
        \item if $p \leq \frac{\log n }{ (\sqrt{1-\theta}- \sqrt{\theta / (q-1)})^2 \intraTau m} ((1\!-\!\epsilon) - \frac{(\sqrt{\alphaConst} - \sqrt{\gammaConst})^{2}}{gc})$, there exists a class of rating matrices which is obtained by swapping the rating vectors of two users in the same cluster yet from distinct groups such that the hamming distance between their rating vectors is~$m \intraTau$. We show that the likelihood of any rating matrix from this class is greater than the one of the ground truth rating matrix;
        \item and finally, if $p \leq \frac{\log n }{ (\sqrt{1-\theta}- \sqrt{\theta / (q-1)})^2 \interTau m} ((1\!-\!\epsilon) - \frac{(\sqrt{\alphaConst} - \sqrt{\gammaConst})^{2} + (g-1) (\sqrt{\betaConst} - \sqrt{\gammaConst})^{2}}{gc})$, there exists a class of rating matrices which is obtained by swapping the rating vectors of two users in distinct clusters such that the hamming distance between their rating vectors is~$m \interTau$. We demonstrate that any rating matrix from this class yields a larger likelihood than the one of the ground truth rating matrix.
    \end{itemize}
    For each case, we show that the maximum likelihood estimator will fail in the limit of $n$ and $m$ by selecting one of the rating matrices from the respective class instead of the ground truth rating matrix.
    This completes the converse proof, and concludes the proof of Theorem~\ref{thm:p_star}.
    
    The technical distinctions with respect to the prior works~\cite{ahn2018binary,yoon2018joint} are four folded: (i) the likelihood computation requires more involved combinatorial arguments due to the hierarchical structure of the similarity graph; (ii) sophisticated upper/lower bounding techniques are developed in order to exploit the relational structure across different groups; (iii) novel typical and atypical error analyses are derived for the achievability proof; and (iv) novel failure proof techniques are proposed in the converse proof. 
\end{IEEEproof}
The assumptions $m = \omega(\log n)$ and $\log m = o(n)$ are introduced in order to be able to apply large deviation theories, as will be shown in the proof of Theorem~\ref{thm:p_star}.
These assumptions are also practically relevant as they eliminate the possibility of having extremely tall and wide matrices, respectively.
The following remark demonstrates that the problem setting considered in \cite{elmahdy2020matrix} is a special case of the general setting considered in this paper, and their result is subsumed by our generalized result presented in Theorem~\ref{thm:p_star}. 
\begin{remark}
	Setting $(c,g,r,q)=(2,3,2,2)$, the optimal observation probability $p^{\star}$ reduces to
	\begin{align}
		p^{\star} 
		= 
		\frac{1}{ \left(\sqrt{1- \theta} - \sqrt{\theta} \right)^2}
		\:\max \left \{ 
		3 \:\frac{\log m}{n}, \:
		\frac{\log n}{\intraTau m} \left(1 - \frac{\left(\sqrt{\alphaConst} - \sqrt{\betaConst}\right)^2}{6}\right), \:
		\frac{\log n}{\interTau m} \left(1-
		\frac{\left(\sqrt{\alphaConst} - \sqrt{\gammaConst}\right)^2}{6}
		- \frac{\left(\sqrt{\betaConst} - \sqrt{\gammaConst}\right)^2}{3} \right)
		\right\},
		\label{eq:pstar}
	\end{align}
	which is equal to the shape threshold on $p$ characterized\footnote{The optimal sample complexity for general $(c,g,r,q)$, given by \eqref{eq:pstar-general}, is conjectured by \cite{elmahdy2020matrix}. However, the achievability and converse proofs are provided only for $(c,g,r,q)=(2,3,2,2)$ in \cite{elmahdy2020matrix}. In this paper, we present complete achievability and converse proofs for any $(c,g,r,q)$.} by \cite{elmahdy2020matrix} under the setting of $(c,g,r,q)=(2,3,2,2)$.
\end{remark}

\section{Information-Theoretic Interpretation of Theorem~\ref{thm:p_star}}
\label{sec:interpretation}
\begin{JA_AE}
We investigate the relationship between the optimal sample complexity $nmp^\star$, where $p^\star$ is characterized by \eqref{eq:pstar-general} in Theorem~\ref{thm:p_star}, and different parameters related to the rating matrix as well as the hierarchical user similarity graph.

\subsection{Minimum Hamming Distance}
\label{subsec:min_hamdist}
The optimal sample complexity increases as $\intraTau$ (or $\interTau$) decreases.
This is due to the fact that as the Hamming distance between rating vectors of two users in different groups within the same cluster (or in different clusters) decreases, it becomes harder to distinguish the rating vectors, and hence it leads to imperfect user grouping (or clustering). Thus, one has to sample more entries of the rating matrix in order to exactly recover the rating matrix.

\subsection{Observation Noise}
\label{subsec:obs_noise}
It is evident that the optimal sample complexity increases as $\theta$ increases. 
Furthermore, as $\theta$ approaches $(q-1)/q$, each sampled entry of the rating matrix can take any of the $q$ possible values with a uniform probability of $1/q$, and hence an infinite sample complexity is theoretically required to exactly recover the entries of the rating matrix.

\subsection{Quality of the Hierarchical Similarity Graph}
\label{subsec:quality_graph}
In order to better illustrate the relationship between the optimal sample complexity and the quality of the hierarchical graph, we define the following quality parameters:
\begin{align}
    \Ig \coloneqq \left(\sqrt{\alpha} - \sqrt{\beta}\right)^2,
    \quad
    \IcOne \coloneqq \left(\sqrt{\alpha} - \sqrt{\gamma}\right)^2,
    \quad
    \IcTwo \coloneqq \left(\sqrt{\beta} - \sqrt{\gamma}\right)^2.
\end{align}
Intuitively, as $\Ig$ increases, it becomes easier to distinguish users in different groups within the same cluster. 
On the other hand, larger values of $\IcOne$ and $\IcTwo$ leads to better user clustering.
The optimal sample complexity reads different values depending on the quality parameters of the hierarchical graph. More specifically, we define three regimes as follows: 
\begin{enumerate}
    \item the first term in the RHS of \eqref{eq:pstar-general} is activated when $\Ig, \IcOne$ and $\IcTwo$ are large enough so that the grouping and clustering information are reliable. Hence, this regime is coined as \emph{``perfect clustering/grouping regime''}; 
    \item the second term in the RHS of \eqref{eq:pstar-general} is activated when $\Ig$ is small such the grouping information is not reliable, Therefore, this regime is coined as \emph{``grouping-limited regime''};
    \item the third term in the RHS of \eqref{eq:pstar-general} is activated when $\IcOne$ and $\IcTwo$ are small such that the clustering information is not reliable, and\footnote{It is evident from \eqref{eq:pstar-general} and $\alphaConst \geq \betaConst \geq \gammaConst$ that the third term in the RHS of \eqref{eq:pstar-general} is inactive whenever $\intraTau \leq \interTau$.} $\intraTau > \interTau$. Thus, this regime is coined as \emph{``clustering-limited regime''}.
\end{enumerate}
In what follows, we analyze the optimal sample complexity under each regime, and highlight the novel technical contributions in the achievability proof (See Section~\ref{sec:achv}) as well as the converse proof (See Section~\ref{sec:conv}).
For illustrative simplicity, we focus on the noiseless case where $\theta = 0$.

\subsubsection{Perfect Clustering/Grouping Regime}
The optimal sample complexity reads $(gc / (g-r+1)) m \log m$.
Since the grouping and clustering information are reliable, one can recover the groups and clusters from the similarity graph. However, further increments of the values of these quality parameters do not yield further improvement in the sample complexity, and hence the sample complexity gain from the similarity graph is saturated in this regime. 
Moreover, it should be noted that a naive generalization of~\cite{ahn2018binary, yoon2018joint} requires $c r m \log m$ observations since there are $r$ independent rating vectors to be estimated for each the $c$~clusters, and each rating vector requires $m \log m$ observations under the considered random sampling due to the coupon-collecting effect. 
On the other hand, we leverage the relational structure (i.e, linear dependency) across rating vectors of different group, reflected by the underlying linear MDS code structure (to be detailed in Section~\ref{sec:achv}), and hence this serves to estimate the $rc$~rating vectors more efficiently, precisely by a factor of $r(g-r+1)/g$ improvement, thus yielding $(gc / (g-r+1)) m \log m$.

\subsubsection{Grouping-Limited Regime}
The optimal sample complexity reads 
\begin{align*}
    \frac{1}{\intraTau} \left(1 - \frac{\left(\sqrt{\alphaConst} - \sqrt{\betaConst}\right)^2}{gc}\right) n \log n
    &=
    \frac{1}{\intraTau} \left(1 - \frac{\Ig}{gc}\right) n \log n,
\end{align*}
which is a decreasing function of $\Ig$.
This sample complexity coincides with that of~\cite{yoon2018joint} in which the considered similarity graph consists of only $gc$ clusters. 
This implies that exploiting the relational structure across different groups does not help improving sample complexity when grouping information is not reliable. 
Furthermore, since the clustering information is reliable, clusters can be recovered from the similarity graph. 
However, further increments of $\IcOne$ and $\IcTwo$ do not yield further reduction in the sample complexity, and hence the sample complexity gain from these two quality parameters is saturated in this regime. 

\subsubsection{Clustering-Limited Regime}
The optimal sample complexity reads 
\begin{align*}
    \frac{1}{\interTau} \left(1-\frac{\left(\sqrt{\alphaConst} - \sqrt{\gammaConst}\right)^2 + (g-1) \left(\sqrt{\betaConst} - \sqrt{\gammaConst}\right)^2}{gc} \right) n \log n
    &=
    \frac{1}{\interTau} \left(1-\frac{\IcOne + (g-1) \IcTwo}{gc} \right) n \log n,
\end{align*}
which is a decreasing function of $\IcOne$ and $\IcTwo$.
This is the most challenging scenario which has not been explored by any prior works. 
Since the clustering information is not reliable, it is not possible to recover the groups and clusters from the similarity graph.
Moreover, it should be noted that when $\beta = \gamma$, i.e., groups and clusters are indistinguishable, we have $\Ig = \IcOne$ and $\IcTwo = 0$. As a result, it boils down to a problem setting of $gc$ clusters, and hence the optimal sample complexity reads 
\begin{align*}
    \frac{1}{\interTau} \left(1-\frac{\left(\sqrt{\alphaConst} - \sqrt{\gammaConst}\right)^2}{gc} \right) n \log n
    &=
    \frac{1}{\interTau} \left(1-\frac{\Ig}{gc} \right) n \log n,
\end{align*}
Comparing to the optimal sample complexity expression for the grouping-limited regime, the only distinction appears in the denominator, in which $\intraTau$ is replaced with $\interTau$ due to the fact that $\interTau < \intraTau$.

\begin{figure}
    \centering
    \subfloat[$\intraTau = \frac{1}{3}$ and $\interTau = \frac{1}{6}$.]{\includegraphics[width=0.45\textwidth]{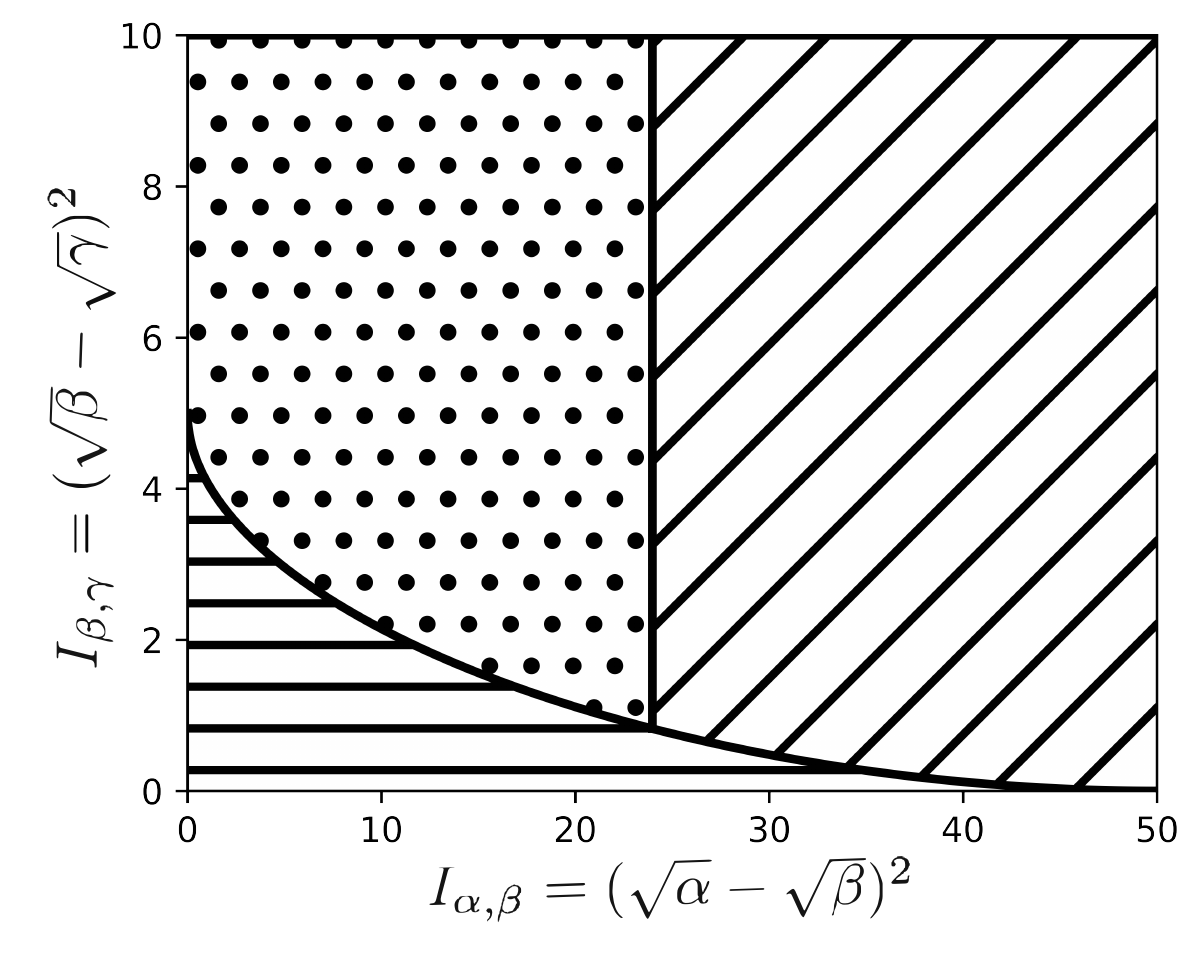}
    \label{fig:3reg}}
    \hfil
    \subfloat[$\intraTau = \frac{1}{6}$ and $\interTau = \frac{1}{3}$.]{\includegraphics[width=0.45\textwidth]{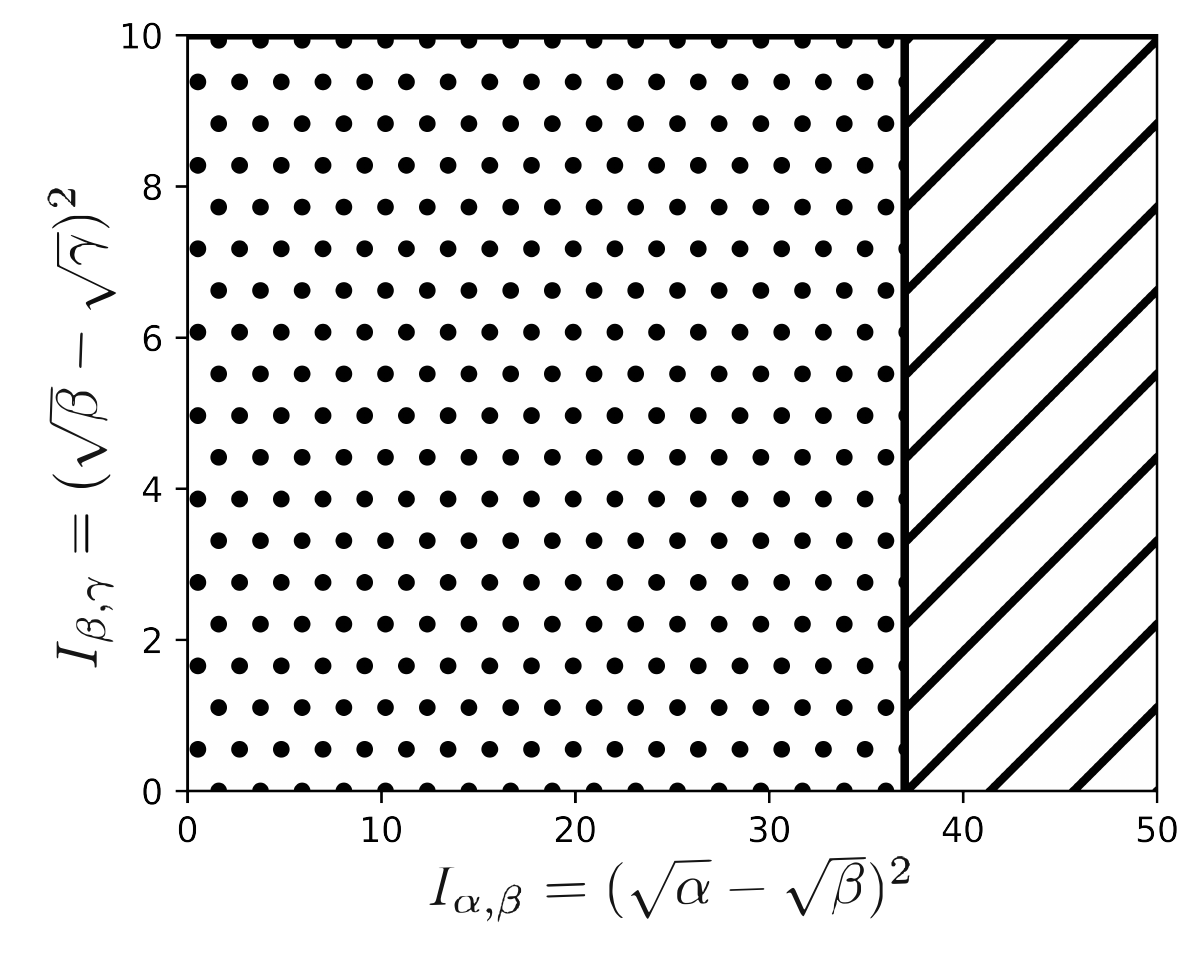}
    \label{fig:2reg}}
    \caption{Let $(n,m,\theta,c,g,r,q)=(4000,500,0,10,5,3,5)$. 
    (a) The different regimes of the optimal sample complexity reported in \eqref{eq:pstar-general} for $\intraTau > \interTau$. 
    (b) The different regimes of the optimal sample complexity reported in \eqref{eq:pstar-general} for $\intraTau < \interTau$. 
    For both sub-figures, diagonal stripes, dots, and horizontal stripes refer to perfect clustering/grouping regime, grouping-limited regime, and clustering-limited regime, respectively.
    }
    \label{fig:regions}
\end{figure}

Consider a problem setting where $n=4000$, $m=500$, $\theta=0$, $c=10$, $g=5$, $r=3$ and $q=5$.
Fig.~\ref{fig:3reg} and Fig.~\ref{fig:2reg} depict the different regimes of the optimal sample complexity as a function of $(\Ig,\IcTwo)$. In Fig.~\ref{fig:3reg}, where $\intraTau = 1/3$ and $\interTau = 1/6$, the region depicted by diagonal stripes corresponds to perfect clustering/grouping regime and the first term in the RHS of \eqref{eq:pstar-general} is active. The graph quality parameters $\Ig$, $\IcTwo$, and consequently $\IcOne$ are large, and graph information is rich enough to perfectly retrieve the clusters and groups. 
The region depicted by dots corresponds to grouping-limited regime, where the second term in the RHS of \eqref{eq:pstar-general} is active. In this regime, graph information suffices to exactly recover the clusters, but we need to rely on rating observation to exactly recover the groups. 
Finally, the third term in the RHS of \eqref{eq:pstar-general} is active in the region captured by horizontal stripes. This indicates the clustering-limited regime, where neither clustering nor grouping is exact without the side information of the rating vectors. 
On the other hand, Fig.~\ref{fig:2reg} depicts the case where $\intraTau = 1/6$ and $\interTau = 1/3$. 
It is worth noting that in practically-relevant systems, we have $\intraTau < \interTau$, i.e., rating vectors of users in the same cluster are expected to be more similar than those in a different cluster. 
Therefore, the third regime, i.e., clustering-limited regime, vanishes in Fig.~\ref{fig:2reg}.

\subsection{Benefit of Hierarchical Graph Structure}
\begin{figure}
    \centering
    \includegraphics[width=0.5\textwidth]{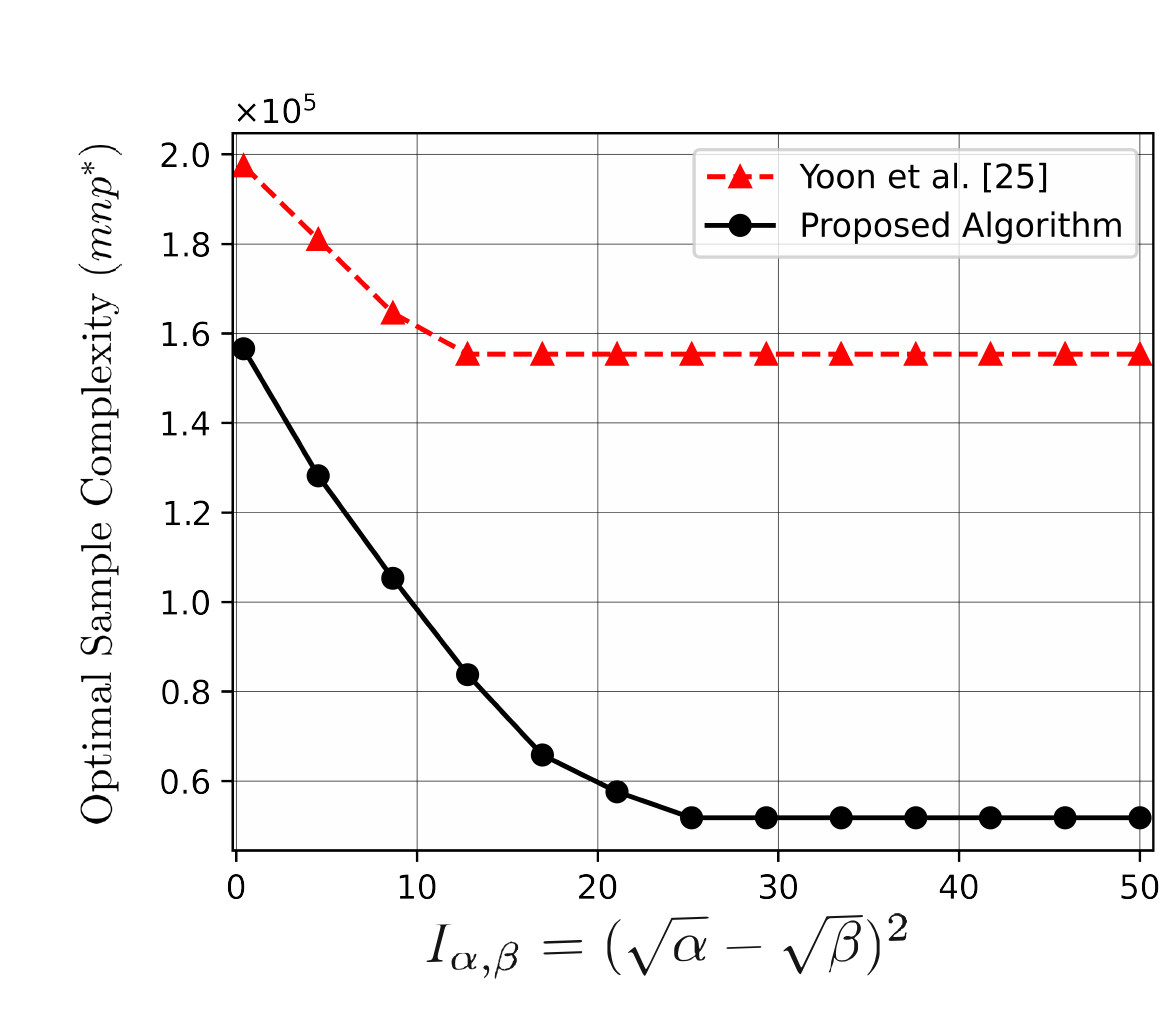}
    \caption{
    Let $(n,m,\theta,c,g,r,q)=(4000,500,0,10,5,3,5)$. 
    Comparison between the sample complexity reported in \eqref{eq:pstar-general} and that of \cite{yoon2018joint} for $\beta=5$, $\gamma = 1$, $\intraTau = 1/3$ and $\interTau = 1/6$.
    }
    \label{fig:baselineComp}
\end{figure}

Consider a problem setting where $n=4000$, $m=500$, $\theta=0$, $c=10$, $g=5$, $r=3$ and $q=5$.
Fig.~\ref{fig:baselineComp} compares the optimal sample complexity, as a function of $\Ig$, between the one reported in~\eqref{eq:pstar-general} and that of \cite{yoon2018joint} for $\intraTau = 1/3$, $\interTau = 1/6$, $\beta = 5$ and $\gamma = 1$. 
It should be noted that \cite{yoon2018joint} leverages neither the hierarchical structure of the graph, nor the linear dependency among the rating vectors. Thus, the problem formulated in Section~\ref{subsec:probForm} will be translated to a graph that consists of $gc$ clusters whose rating vectors are linearly independent in the setting of \cite{yoon2018joint}. Also note that the minimum hamming distance for \cite{yoon2018joint} is $\interTau$. In Fig.~\ref{fig:baselineComp}, we can see that the noticeable gain in the sample complexity of our result in the diagonal parts of the plot (i.e., clustering-limited and grouping-limited regimes on the left side) is due to leveraging the hierarchical graph structure, while the improvement in the sample complexity in the flat part of the plot (i.e, perfect clustering/grouping regime) is a consequence of exploiting the relational structure (i.e., linear dependency) among the rating vectors within each cluster.
\end{JA_AE}

\section{The Achievability proof}
\label{sec:achv}
In this section, we prove the achievability part of Theorem~\ref{thm:p_star}, that is if the condition on $p$ in \eqref{eq:pstar-general_achv} holds, then there exists an estimator $\psi$ such that $\lim_{n \rightarrow \infty} P_e^{(\diff)}(\psi) = 0$. 
 To this end, we prove that $\lim_{n \rightarrow \infty} P_e^{(\diff)}(\psi_{\text{ML}}) = 0$ where $\psi_{\text{ML}}$ is the maximum likelihood (ML) estimator, if all the following inequalities hold: 
\begin{align}
    \frac{g-r+1}{gc}n \Ir &\geq (1+\epsilon) \log m, 
    & & \textbf{(Perfect Clustering/Grouping Regime)}
    \label{eq:suffCond_1}\\
	\intraTau m \Ir + \frac{\Ig}{gc} \log n &\geq (1+\epsilon) \log n,	
	& & \textbf{(Grouping-Limited Regime)}
	\label{eq:suffCond_2}\\
	\interTau m \Ir + \frac{\IcOne}{gc} \log n + \frac{(g-1) \IcTwo }{gc} \log n &\geq (1+\epsilon) \log n,
	& & \textbf{(Clustering-Limited Regime)}
	\label{eq:suffCond_3}
\end{align}
where 
\begin{align}
    \Ir \coloneqq p \left(\sqrt{1-\theta} - \sqrt{\frac{\theta}{q-1}}\right)^2,
    \quad
    \Ig \coloneqq (\sqrt{\alphaConst} - \sqrt{\betaConst})^2,
    \quad
    \IcOne \coloneqq (\sqrt{\alphaConst} - \sqrt{\gammaConst})^2,
    \quad
    \IcTwo \coloneqq (\sqrt{\betaConst} - \sqrt{\gammaConst})^2.
    \label{eq:Ir_I_abg_defn}
\end{align}
Throughout the proof, let $p = \Theta ((\log n) / n)$, and let $q$ and $\theta$ be constants such that $q$ is prime and $\theta \in [0,1]$. 
We first present the structure of the ground truth rating matrix and the underlying linear MDS code structure in Section~\ref{sec:achv_groundTruth}. Next, we introduce a number of auxiliary lemmas in Section~\ref{sec:achv_auxLemma}. Finally, we present the achievability proof of Theorem~\ref{thm:p_star} in Section~\ref{sec:achv_proof_Thm1}.

\subsection{The Structure of Ground Truth Rating Matrix}
\label{sec:achv_groundTruth}
Let the ground truth rating matrix be denoted by $\gtMat = (\cV_0,\cZ_0)$ where $\gtMat \in \matSet$, and
\begin{align}
    \cV_0 &= \left\{\vecU{i}{x}: x\in [c], i\in [g]\right\},
    \qquad
    \cZ_0 = \left\{\left\{\frakz_0(x,i)\right\}_{x \in [c],\: i \in [g]}
    \right\},
\end{align}
where $\cZ_0$ follows the conditions given in \eqref{eq:Z_defn}, but omitted for the sake of brevity.
Let $\gtRatingCluster{x} \in \mathbb{F}_{q}^{g \times m}$ be a matrix obtained by stacking all the rating vectors of cluster $x$ given by $\{\vecU{i}{x}: i\in [g]\}$ for $x \in [c]$.
Consequently, $\gtMat$ is an $n \times m$ matrix where its $r^{\text{th}}$ row equals to $\vecU{i}{x}$ if and only if $r \in \frakz_0(x,i)$.
Furthermore, let the output of an estimator~$\psi$ (i.e., the completed rating matrix) be denoted by $X = (\cV, \cZ)$ where $X \in \mathbb{F}_{q}^{n \times m}$, and
\begin{align}
    \cV &= \left\{\vecV{i}{x}: x\in [c], i\in [g]\right\},
    \qquad
    \cZ = \left\{\left\{\frakz(x,i)\right\}_{x \in [c],\: i \in [g]} 
	\right\},
\end{align}
where $\cZ$ also follows similar conditions listed in \eqref{eq:Z_defn}.

First, we construct a ground truth rating matrix $\gtMat$ that we are supposed to recover using the maximum likelihood estimator $\psi_{\text{ML}}$.
Recall from Section~\ref{subsec:probForm} that the considered hierarchical graph consists of $c$ clusters, and each cluster comprises $g$ equal-sized groups.
The set of $g$ rating vectors of cluster~$x$ is spanned by any subset of $r$ rating vectors for $x \in [c]$.
Without loss of generality, assume that the set of users who belong to cluster~$x$ and group $i$ is given by $\{k + 1, k + 2, \ldots, k + \frac{n}{cg}\}$ for $x \in [c]$, $i \in [g]$ and $k = (x-1)\frac{n}{c} + (i-1)\frac{n}{cg}$.
From the literature of error-correcting codes, a $(g,r)$ linear MDS code in $\mathbb{F}_q$ has a minimum distance of $g-r+1$, and hence reaches the Singleton bound \cite{macwilliams1977theory}.
Furthermore, according to the MDS conjecture \cite{macwilliams1977theory}, \red{there exists a $(g,r)$ MDS code in $\mathbb{F}_q$ if and only if $g \leq q+1$ for all $q$ and $2 \leq r \leq q-1$, except when $q$ is even and $r \in \{3, q-1\}$, in which case $g \leq q + 2$}.
If these conditions are satisfied, then the existence of such an MDS code is guaranteed by the construction of Generalized Reed-Solomon codes \cite{macwilliams1977theory}.
Consider a $(g,r)$ linear MDS code in $\mathbb{F}_q$ where $g$ is the length of the code and $r$ is its dimension.
Let the set of $g$ ground truth rating vectors of the groups in cluster~$x$ be a $(g,r)$ MDS code. 
Hence, the set of $g$ rows of $\gtRatingCluster{x}$
spanned by any subset of $r$ rows of $\gtRatingCluster{x}$.
Let $\iGenerator{x} \in \mathbb{F}_q^{g \times r}$ be a generator matrix of the $(g,r)$ MDS code, and $\iBasis{x} \in \mathbb{F}_q^{r \times m}$ be the basis matrix (with rank $r$), such that 
\begin{align}
	\gtRatingCluster{x} = \iGenerator{x} \iBasis{x},\:
	\text{ for }  x \in [c].
	\label{eq:mds_code}
\end{align}
Without loss of generality, we make the following assumptions:
\begin{itemize}
    \item let the first row of $\gtRatingCluster{1}$ be given by $\gtRatingCluster{1}(1,:)= \mathbf{1}_{1 \times n}$;
    \item for $x \in [c]$, let $\iGenerator{x} = \Phi$ where $\Phi$ is a systematic generator matrix such that $\Phi = \left[I_{r \times r} \ A^\intercal \right]^\intercal$ and $A \in \mathbb{F}_{q}^{(g-r) \times r}$. 
    This ensures that the first $r$ rows of $\iGenerator{x}$ are linearly independent, and hence the first $r$ rows of $\gtRatingCluster{x}$ are linearly independent by \eqref{eq:mds_code}.
\end{itemize}
Based on the aforementioned assumptions, the entries of each column of $\gtMat$ can take values from a set of $q^{cr-1}$ possible column vectors. This is due to the fact that the first row of $\gtMat$ is fixed to all-one vector, and the last $g-r$ rows of $\gtRatingCluster{x}$, for $x \in [c]$, can be constructed by linear combinations of its first $r$ rows. Hence, we have a total of $q^{cg-(1+c(g-r))} = q^{cr-1}$ different choices.
Let the set of column of $\gtMat$ be partitioned into $q^{cr-1}$ sections, where the columns of each section correspond to one choice of the possible $q^{cr-1}$ vectors. Let the number of columns of each section be $\colBlockSize{\ell} m$, where $0 \leq \colBlockSize{\ell} \leq 1$ for $\ell \in \{0,1,\ldots,q-1\}^{cr-1}$ and $\sum_{\ell \in \{0,1,\ldots,q-1\}^{cr-1}} \colBlockSize{\ell} = 1$. 
Let $\colBlock{\ell}$ denote the $\ell^{\text{th}}$ column section of $\gtMat$, and hence $\colBlockSize{\ell} = \vert \colBlock{\ell} \vert / m$.
Accordingly, let each row $\vecU{i}{x}$ of $\gtMat$ be partitioned into $q^{cr-1}$ sections, denoted\footnote{A similar interpretation goes for $\{\vecV{i}{x}(\ell): \ell\in \{0,1,\ldots,q-1\}^{cr-1}\}$.} by $\{\vecU{i}{x}(\ell): \ell\in \{0,1,\ldots,q-1\}^{cr-1}\}$, for $x \in [c]$ and $i \in [g]$.
\red{We assume that the MDS code structure is assumed to be known a priori, and hence the output matrix follows the MDS code structure imposed on the construction of $\gtMat$.}
In the following example, we give an illustrative description of the proposed construction of the ground truth rating matrix $\gtMat$.

\subsubsection{Illustrative Example}
Consider the setting of $(c,g,r,q) = (2,3,2,2)$. Under this setting, the generator matrix and the basis matrix of each cluster are given by
\begin{align}
	\iGenerator{1} &= \iGenerator{2} = \Phi
	= 
	\begin{bmatrix}
		1 & 0 \\
		0 & 1 \\
		1 & 1
	\end{bmatrix}, 
	\nonumber\\
	\iBasis{1}
	&=
	\begin{bmatrix}
    \begin{array}{c|c|c|c|c|c|c|c}
      \mathbf{1}_{1\times \colBlockSize{000} m} & 
      \mathbf{1}_{1\times \colBlockSize{001} m} & 
      \mathbf{1}_{1\times \colBlockSize{010} m} & 
      \mathbf{1}_{1\times \colBlockSize{011} m} & 
      \mathbf{1}_{1\times \colBlockSize{100} m} & 
      \mathbf{1}_{1\times \colBlockSize{101} m} & 
      \mathbf{1}_{1\times \colBlockSize{110} m} & 
      \mathbf{1}_{1\times \colBlockSize{111} m}\\
      \hline
      \mathbf{0}_{1\times \colBlockSize{000} m} & 
      \mathbf{0}_{1\times \colBlockSize{001} m} & 
      \mathbf{0}_{1\times \colBlockSize{010} m} & 
      \mathbf{0}_{1\times \colBlockSize{011} m} & 
      \mathbf{1}_{1\times \colBlockSize{100} m} & 
      \mathbf{1}_{1\times \colBlockSize{101} m} & 
      \mathbf{1}_{1\times \colBlockSize{110} m} & 
      \mathbf{1}_{1\times \colBlockSize{111} m}
    \end{array}
  	\end{bmatrix},
	\nonumber\\
	\iBasis{2}
	&=
	\begin{bmatrix}
    \begin{array}{c|c|c|c|c|c|c|c}
      \mathbf{0}_{1\times \colBlockSize{000} m} & 
      \mathbf{0}_{1\times \colBlockSize{001} m} & 
      \mathbf{1}_{1\times \colBlockSize{010} m} & 
      \mathbf{1}_{1\times \colBlockSize{011} m} & 
      \mathbf{0}_{1\times \colBlockSize{100} m} & 
      \mathbf{0}_{1\times \colBlockSize{101} m} & 
      \mathbf{1}_{1\times \colBlockSize{110} m} & 
      \mathbf{1}_{1\times \colBlockSize{111} m}\\
      \hline
      \mathbf{0}_{1\times \colBlockSize{000} m} & 
      \mathbf{1}_{1\times \colBlockSize{001} m} & 
      \mathbf{0}_{1\times \colBlockSize{010} m} & 
      \mathbf{1}_{1\times \colBlockSize{011} m} & 
      \mathbf{0}_{1\times \colBlockSize{100} m} & 
      \mathbf{1}_{1\times \colBlockSize{101} m} & 
      \mathbf{0}_{1\times \colBlockSize{110} m} & 
      \mathbf{1}_{1\times \colBlockSize{111} m}
    \end{array}
  	\end{bmatrix},
\end{align}
where $0 \leq \colBlockSize{\ell} \leq 1$ for $\ell \in \{0,1\}^3$, and $\sum_{\ell \in \{0,1\}^3} \colBlockSize{\ell} = 1$.
Therefore, from \eqref{eq:mds_code}, we have
\begin{align}
	\gtRatingCluster{1} &= \iGenerator{1} \iBasis{1}
	=
	\begin{bmatrix}
    \begin{array}{c|c|c|c|c|c|c|c}
      \mathbf{1}_{1\times \colBlockSize{000} m} & 
      \mathbf{1}_{1\times \colBlockSize{001} m} & 
      \mathbf{1}_{1\times \colBlockSize{010} m} & 
      \mathbf{1}_{1\times \colBlockSize{011} m} & 
      \mathbf{1}_{1\times \colBlockSize{100} m} & 
      \mathbf{1}_{1\times \colBlockSize{101} m} & 
      \mathbf{1}_{1\times \colBlockSize{110} m} & 
      \mathbf{1}_{1\times \colBlockSize{111} m}\\
      \hline
      \mathbf{0}_{1\times \colBlockSize{000} m} & 
      \mathbf{0}_{1\times \colBlockSize{001} m} & 
      \mathbf{0}_{1\times \colBlockSize{010} m} & 
      \mathbf{0}_{1\times \colBlockSize{011} m} & 
      \mathbf{1}_{1\times \colBlockSize{100} m} & 
      \mathbf{1}_{1\times \colBlockSize{101} m} & 
      \mathbf{1}_{1\times \colBlockSize{110} m} & 
      \mathbf{1}_{1\times \colBlockSize{111} m}\\
      \hline
      \mathbf{1}_{1\times \colBlockSize{000} m} & 
      \mathbf{1}_{1\times \colBlockSize{001} m} & 
      \mathbf{1}_{1\times \colBlockSize{010} m} & 
      \mathbf{1}_{1\times \colBlockSize{011} m} & 
      \mathbf{0}_{1\times \colBlockSize{100} m} & 
      \mathbf{0}_{1\times \colBlockSize{101} m} & 
      \mathbf{0}_{1\times \colBlockSize{110} m} & 
      \mathbf{0}_{1\times \colBlockSize{111} m}\\
    \end{array}
  	\end{bmatrix},
  	\nonumber\\
  	\gtRatingCluster{2} &= \iGenerator{2} \iBasis{2}
	=
	\begin{bmatrix}
    \begin{array}{c|c|c|c|c|c|c|c}
      \mathbf{0}_{1\times \colBlockSize{000} m} & 
      \mathbf{0}_{1\times \colBlockSize{001} m} & 
      \mathbf{1}_{1\times \colBlockSize{010} m} & 
      \mathbf{1}_{1\times \colBlockSize{011} m} & 
      \mathbf{0}_{1\times \colBlockSize{100} m} & 
      \mathbf{0}_{1\times \colBlockSize{101} m} & 
      \mathbf{1}_{1\times \colBlockSize{110} m} & 
      \mathbf{1}_{1\times \colBlockSize{111} m}\\
      \hline
      \mathbf{0}_{1\times \colBlockSize{000} m} & 
      \mathbf{1}_{1\times \colBlockSize{001} m} & 
      \mathbf{0}_{1\times \colBlockSize{010} m} & 
      \mathbf{1}_{1\times \colBlockSize{011} m} & 
      \mathbf{0}_{1\times \colBlockSize{100} m} & 
      \mathbf{1}_{1\times \colBlockSize{101} m} & 
      \mathbf{0}_{1\times \colBlockSize{110} m} & 
      \mathbf{1}_{1\times \colBlockSize{111} m}\\
      \hline
      \mathbf{0}_{1\times \colBlockSize{000} m} & 
      \mathbf{1}_{1\times \colBlockSize{001} m} & 
      \mathbf{1}_{1\times \colBlockSize{010} m} & 
      \mathbf{0}_{1\times \colBlockSize{011} m} & 
      \mathbf{0}_{1\times \colBlockSize{100} m} & 
      \mathbf{1}_{1\times \colBlockSize{101} m} & 
      \mathbf{1}_{1\times \colBlockSize{110} m} & 
      \mathbf{0}_{1\times \colBlockSize{111} m}
    \end{array}
  	\end{bmatrix}.
\end{align}
Consequently, $\gtMat$ is given by
\begin{align}
\label{eq_M0}
  \gtMat =
  \begin{bmatrix}
    \begin{array}{c|c|c|c|c|c|c|c}
      \mathbf{1}_{\frac{n}{6}\times \colBlockSize{000} m} &
      \mathbf{1}_{\frac{n}{6}\times \colBlockSize{001} m} &
      \mathbf{1}_{\frac{n}{6}\times \colBlockSize{010} m} &
      \mathbf{1}_{\frac{n}{6}\times \colBlockSize{011} m} &
      \mathbf{1}_{\frac{n}{6}\times \colBlockSize{100} m} &
      \mathbf{1}_{\frac{n}{6}\times \colBlockSize{101} m} &
      \mathbf{1}_{\frac{n}{6}\times \colBlockSize{110} m} &
      \mathbf{1}_{\frac{n}{6}\times \colBlockSize{111} m}\\
      \hline
      \mathbf{0}_{\frac{n}{6}\times \colBlockSize{000} m} & 
      \mathbf{0}_{\frac{n}{6}\times \colBlockSize{001} m} & 
      \mathbf{0}_{\frac{n}{6}\times \colBlockSize{010} m} & 
      \mathbf{0}_{\frac{n}{6}\times \colBlockSize{011} m} & 
      \mathbf{1}_{\frac{n}{6}\times \colBlockSize{100} m} & 
      \mathbf{1}_{\frac{n}{6}\times \colBlockSize{101} m} & 
      \mathbf{1}_{\frac{n}{6}\times \colBlockSize{110} m} & 
      \mathbf{1}_{\frac{n}{6}\times \colBlockSize{111} m}\\
      \hline
      \mathbf{1}_{\frac{n}{6}\times \colBlockSize{000} m} & 
      \mathbf{1}_{\frac{n}{6}\times \colBlockSize{001} m} & 
      \mathbf{1}_{\frac{n}{6}\times \colBlockSize{010} m} & 
      \mathbf{1}_{\frac{n}{6}\times \colBlockSize{011} m} & 
      \mathbf{0}_{\frac{n}{6}\times \colBlockSize{100} m} & 
      \mathbf{0}_{\frac{n}{6}\times \colBlockSize{101} m} & 
      \mathbf{0}_{\frac{n}{6}\times \colBlockSize{110} m} & 
      \mathbf{0}_{\frac{n}{6}\times \colBlockSize{111} m}\\
      \hline
      \mathbf{0}_{\frac{n}{6}\times \colBlockSize{000} m} & 
      \mathbf{0}_{\frac{n}{6}\times \colBlockSize{001} m} & 
      \mathbf{1}_{\frac{n}{6}\times \colBlockSize{010} m} & 
      \mathbf{1}_{\frac{n}{6}\times \colBlockSize{011} m} & 
      \mathbf{0}_{\frac{n}{6}\times \colBlockSize{100} m} & 
      \mathbf{0}_{\frac{n}{6}\times \colBlockSize{101} m} & 
      \mathbf{1}_{\frac{n}{6}\times \colBlockSize{110} m} & 
      \mathbf{1}_{\frac{n}{6}\times \colBlockSize{111} m}\\
      \hline
      \mathbf{0}_{\frac{n}{6}\times \colBlockSize{000} m} & 
      \mathbf{1}_{\frac{n}{6}\times \colBlockSize{001} m} & 
      \mathbf{0}_{\frac{n}{6}\times \colBlockSize{010} m} & 
      \mathbf{1}_{\frac{n}{6}\times \colBlockSize{011} m} & 
      \mathbf{0}_{\frac{n}{6}\times \colBlockSize{100} m} & 
      \mathbf{1}_{\frac{n}{6}\times \colBlockSize{101} m} & 
      \mathbf{0}_{\frac{n}{6}\times \colBlockSize{110} m} & 
      \mathbf{1}_{\frac{n}{6}\times \colBlockSize{111} m}\\
      \hline
      \mathbf{0}_{\frac{n}{6}\times \colBlockSize{000} m} & 
      \mathbf{1}_{\frac{n}{6}\times \colBlockSize{001} m} & 
      \mathbf{1}_{\frac{n}{6}\times \colBlockSize{010} m} & 
      \mathbf{0}_{\frac{n}{6}\times \colBlockSize{011} m} & 
      \mathbf{0}_{\frac{n}{6}\times \colBlockSize{100} m} & 
      \mathbf{1}_{\frac{n}{6}\times \colBlockSize{101} m} & 
      \mathbf{1}_{\frac{n}{6}\times \colBlockSize{110} m} & 
      \mathbf{0}_{\frac{n}{6}\times \colBlockSize{111} m}
    \end{array}
  \end{bmatrix},
\end{align}
which is the same construction of $\gtMat$ provided in \cite{elmahdy2020matrix} for the special case of $(c,g,r,q) = (2,3,2,2)$.
\hfill$\blacklozenge$

\subsection{The Auxiliary Lemmas}
\label{sec:achv_auxLemma}
We present six auxiliary lemmas that are used to prove the achievability part of Theorem~\ref{thm:p_star}. Before each lemma, we introduce the terminologies and notations needed for the statement of the lemma.

Let $\mathsf{L}(X)$ denotes\footnote{With a slight abuse of notation, we omit the dependence on $(Y,G)$ in the likelihood function for notational compactness.} the negative log-likelihood of a candidate rating matrix $X=(\cV,\mathcal{Z})$ given a fixed input pair $(Y,\mathcal{G})$. 
More formally, we have
\begin{align}
    \mathsf{L}(X) 
    = 
    \left\{
	\begin{array}{cl}
		- \log \mathbb{P}\left[(Y,\mathcal{G}) \given \mathbf{X} = X \right]
		& \textrm{if $X \in \matSet$},
		\vspace{0.5mm}\\
		\infty
		& \textrm{otherwise}.
	\end{array}
	\right.
    \label{eq:likelihood_defn}
\end{align}
We show that the likelihood expression hinges on two factors: (i) the difference between the estimated ratings (entries of $X$) and the observed ratings (elements of $Y$); and (ii) the dissimilarity between the graph induced by the partitioning in $\mathcal{Z}$ and the observed graph $\mathcal{G}$.
As defined in Section~\ref{subsec:notation}, we denote by $\Lambda(X,Y)$ the number of mismatched entries between $X$ and $Y$.
For a user partitioning $\mathcal{Z}$, let $\userPairSet{\alpha}{\mathcal{Z}}$ denote the set of pairs of users within any group; $\userPairSet{\beta}{\mathcal{Z}}$ denote the set of pairs of users in different groups within any cluster; and $\userPairSet{\gamma}{\mathcal{Z}}$ denote the set of pairs of users in different clusters. Formally, we have
\begin{align}
    \begin{split}
    \userPairSet{\alpha}{\mathcal{Z}} 
    &=
    \left\{
    (a,b) : a \in Z(x,i), \: b \in Z(x,i), 
    \mbox{ for } x \in [c], \: i \in [g]
    \right\},
    \\
    \userPairSet{\beta}{\mathcal{Z}}
    &=
    \left\{
    (a,b) : a \in Z(x,i), \: b \in Z(x,j), \:
    \mbox{ for } x \in [c],\: i, j \in [g],\: i\neq j
    \right\},
    \\
    \userPairSet{\gamma}{\mathcal{Z}} 
    &=
    \left\{
    (a,b) : a \in Z(x,i), \: b \in Z(y,j), \:
    \mbox{ for } x,y \in [c],\: x \neq y, \: i, j \in [g]
    \right\}.
    \end{split}
\end{align}
Recall from Section~\ref{subsec:probForm} that the user partitioning induced by any rating matrix in $\matSet$ should satisfy the property that all groups have equal size of $n/(cg)$ users.
This implies that the sizes of $\userPairSet{\alpha}{\mathcal{Z}}$, $\userPairSet{\beta}{\mathcal{Z}}$ and $\userPairSet{\gamma}{\mathcal{Z}}$ are constants and given by 
\begin{align}
    \vert \userPairSet{\alpha}{\mathcal{Z}} \vert = gc \binom{n/(gc)}{2}, 
    \qquad
    \vert \userPairSet{\beta}{\mathcal{Z}} \vert = c \binom{g}{2} \left(n/(gc)\right)^2,
    \qquad
    \vert \userPairSet{\gamma}{\mathcal{Z}} \vert = \binom{c}{2} \left(n/c\right)^2,
\end{align}
for any user partitioning.
Furthermore, for a graph $\mathcal{G}$ and a user partitioning $\mathcal{Z}$, define $\numEdge{\alpha}{\mathcal{G}}{\mathcal{Z}}$ as the number of edges within any group; $\numEdge{\beta}{\mathcal{G}}{\mathcal{Z}}$ as the number of edges across groups within any cluster; and $\numEdge{\gamma}{\mathcal{G}}{\mathcal{Z}}$ as the number of edges across clusters. More formally, we have
\begin{align}
    \numEdge{\mu}{\mathcal{G}}{\mathcal{Z}}
	&=
	\sum_{(a,b) \in \userPairSet{\mu}{\mathcal{Z}}} 
	\indicatorFn{(a,b) \in \mathcal{E}},
	\label{eq:e_mu}
\end{align}
for $\mu \in \{\alphaEdge,\betaEdge,\gammaEdge\}$.
The following lemma gives a precise expression of $\mathsf{L}(X)$.

\begin{lemma}
For a given fixed input pair $(Y,\mathcal{G})$ and any $X \in \matSet$, we have
\begin{align}
    \mathsf{L}(X) 
    =
    \log \left((q-1) \frac{1-\theta}{\theta}\right) \numDiffElmnt{Y}{X} 
	+
	\sum\limits_{\mu\in\left\{\alphaEdge,\betaEdge,\gammaEdge\right\}}
    \left(
    \log \left(\frac{1-\mu}{\mu}\right) \numEdge{\mu}{\mathcal{G}}{\mathcal{Z}}
    -
    \log(1-\mu) \left\vert \userPairSet{\mu}{\mathcal{Z}} \right\vert 
    \right),
    \label{eq:neg_log_likelihood}
\end{align}
where $\alphaEdge$, $\betaEdge$ and $\gammaEdge$ are the edge probabilities defined in \eqref{eq:edge_prob_defn}.
\label{lm:neg_log_likelihood}
\end{lemma}

\begin{IEEEproof}
    We refer to Appendix~\ref{app:neg_log_likelihood} for the proof of Lemma~\ref{lm:neg_log_likelihood}.
\end{IEEEproof}
The following lemma provides an upper bound on the worst-case probability of error $P_e^{(\diff)}(\psi_{\text{ML}})$.

\begin{lemma}
\label{lm:upp_worst_case_err_prob}
    For the maximum likelihood estimator $\psi_{\text{ML}}$, we have
    \begin{align}
        \label{ineq:upp_worst_case_err_prob}
        P_e^{(\diff)}(\psi_{\text{ML}}) 
        \leq
        \sum_{X \neq \gtMat}
        \mathbb{P}\left[\mathsf{L}(\gtMat) \geq \mathsf{L}(X)\right].
    \end{align}
\end{lemma}

\begin{IEEEproof}
    We refer to Appendix~\ref{app:upp_worst_case_err_prob} for the proof of Lemma~\ref{lm:upp_worst_case_err_prob}.
\end{IEEEproof}

For a ground truth rating matrix $\gtMat = (\cV_0,\cZ_0)$; a candidate rating matrix $X = (\cV, \cZ)$; and a tuple $T \in \tupleSetDelta$, define the following disjoint sets:
\begin{itemize}
	\item define $\diffEntriesSet = \diffEntriesSet(\gtMat, X)$ as the set of matrix entries where $X \neq \gtMat$. Formally, we have
	\begin{align}
	    \diffEntriesSet = \left\{(r,t)\in [n] \times [m]: X(r,t) \neq \gtMat(r,t)\right\};
	    \label{eq:N1_defn}
	\end{align}
	\item define $\misClassfSet{\betaEdge}{\alphaEdge} = \misClassfSet{\betaEdge}{\alphaEdge}(\cZ_0, \cZ)$ as the set of pairs of users where the two users of each pair belong to different groups of the same cluster in $\gtMat$ (and therefore they are connected with probability $\betaEdge$), but they are estimated to be in the same group in $X$ (and hence, given the estimator output, the belief for the existence of an edge between these two users is $\alphaEdge$). Formally, we have
	\begin{align}
	    \misClassfSet{\betaEdge}{\alphaEdge} = \left\{
	    (a,b):
	    a \in Z_0(x,i_1) \cap Z(y,j),\: b \in Z_0(x,i_2) \cap Z(y,j),
	    \mbox{ for } x, y \in [c],\:
	    i_1, i_2, j \in [g],\: i_1 \neq i_2
	    \right\}.
	    \label{eq:N2rev_defn}
	\end{align}	
	On the other hand, define $\misClassfSet{\alphaEdge}{\betaEdge} = \misClassfSet{\alphaEdge}{\betaEdge} (\cZ_0, \cZ)$ as
	\begin{align}
	    \misClassfSet{\alphaEdge}{\betaEdge} = \left\{
	    (a,b):
	    a \in Z_0(x,i) \cap Z(y,j_1),\: b \in Z_0(x,i) \cap Z(y,j_2),
	    \mbox{ for } x, y \in [c],\:
	    i, j_1, j_2 \in [g],\: j_1 \neq j_2
	    \right\};
	    \label{eq:N2_defn}
	\end{align}
    \item define $\misClassfSet{\gammaEdge}{\alphaEdge} = \misClassfSet{\gammaEdge}{\alphaEdge}(\cZ_0, \cZ)$ as the set of pairs of users where the two users of each pair belong to different clusters in $\gtMat$ (and therefore they are connected with probability $\gammaEdge$), but they are estimated to be in the same group in $X$ (and hence, given the estimator output, the belief for the existence of an edge between these two users is $\alphaEdge$). Formally, we have
	\begin{align}
	    \misClassfSet{\gammaEdge}{\alphaEdge} = \left\{
	        (a,b) :
	        a \in Z_0(x_1,i_1) \cap Z(y,j),\: b \in Z_0(x_2,i_2) \cap Z(y,j), 
	        \mbox{ for } x_1, x_2, y \in [c],\: x_1 \neq x_2,\:
	        i_1, i_2, j \in [g]
	    \right\}.
	    \label{eq:N3rev_defn}
	\end{align}
	On the other hand, define $\misClassfSet{\alphaEdge}{\gammaEdge} = \misClassfSet{\alphaEdge}{\gammaEdge}(\cZ_0, \cZ)$ as
	\begin{align}
	    \misClassfSet{\alphaEdge}{\gammaEdge} = \left\{
	        (a,b) :
	        a \in Z_0(x,i) \cap Z(y_1,j_1),\: b \in Z_0(x,i) \cap Z(y_2, j_2),
	        \mbox{ for } x, y_1, y_2 \in [c],\: y_1 \neq y_2,\:
	        i, j_1, j_2 \in [g]
	    \right\};
	    \label{eq:N3_defn}
	\end{align}
    \item define $\misClassfSet{\gammaEdge}{\betaEdge} = \misClassfSet{\gammaEdge}{\betaEdge}(\cZ_0, \cZ)$ as the set of pairs of users where the two users of each pair belong to different clusters in $\gtMat$ (and therefore they are connected with probability $\gammaEdge$), but they are estimated to be in different groups of the  same cluster in $X$ (and hence, given the estimator output, the belief for the existence of an edge between these two users is $\betaEdge$). Formally, we have
	\begin{align}
	    \misClassfSet{\gammaEdge}{\betaEdge} &\!=\! \left\{
	        (a,b) \!:
	        a \!\in\! Z_0(x_1,i_1) \!\cap\! Z(y,j_1),\: b \!\in\! Z_0(x_2,i_2) \!\cap\! Z(y,j_2),
	        \!\mbox{ for } x_1, x_2, y \!\in\! [c],\: x_1 \!\neq\! x_2, \:
	        i_1, i_2, j_1, j_2 \!\in\! [g],\: j_1 \!\neq\! j_2
	    \right\}\!.
	    \label{eq:N4rev_defn}
	\end{align}
	On the other hand, define $\misClassfSet{\betaEdge}{\gammaEdge} = \misClassfSet{\betaEdge}{\gammaEdge}(\cZ_0, \cZ)$ as
	\begin{align}
	    \misClassfSet{\betaEdge}{\gammaEdge} &\!=\! \left\{
	        (a,b) \!:
	        a \!\in\! Z_0(x,i_1) \!\cap\! Z(y_1,j_1),\: b \!\in\! Z_0(x, i_2) \!\cap\! Z(y_2, j_2), 
	        \!\mbox{ for } x, y_1, y_2 \!\in\! [c],\: y_1 \!\neq\! y_2, \:
	        i_1, i_2, j_1, j_2 \!\in\! [g],\: i_1 \!\neq\! i_2
	    \right\}\!.
	    \label{eq:N4_defn}
	\end{align}
\end{itemize}
Let $\mathsf{B}_i^{(\sigma)}$ denote the $i^{\text{th}}$~Bernoulli random variable with parameter $\sigma \in \{p, \theta, \frac{1}{q-1}, \alphaEdge, \betaEdge, \gammaEdge\}$. 
Define the following sets of independent Bernoulli random variables:
\begin{align}
	\left\{\mathsf{B}_i^{(p)}: i\in\diffEntriesSet\right\}, \:
	\left\{\mathsf{B}_i^{(\theta)}:i \in \diffEntriesSet \right\}, \:
	\left\{\mathsf{B}_i^{\left(\frac{1}{q-1}\right)}: i\in\diffEntriesSet\right\}, \:
	\left\{
	\mathsf{B}_i^{(\mu)}: 
	i\in \misClassfSet{\mu}{\nu},\:
	\mu, \nu \in \left\{\alphaEdge,\betaEdge,\gammaEdge \right\}, \:
	\mu \neq \nu
	\right\}.
	\label{eq:bern_sets}
\end{align} 
Now, define $\mathbf{B} = \mathbf{B} \left(\diffEntriesSet, \{\misClassfSet{\mu}{\nu} : \mu, \nu \in \{\alphaEdge,\betaEdge,\gammaEdge \}, \: \mu \neq \nu\}\right)$ as
\begin{align}    
	\mathbf{B}
	&\coloneqq
	\log\left((q-1)\frac{1-\theta}{\theta}\right) 
    \sum_{i \in \diffEntriesSet}
	\mathsf{B}_i^{(p)}
	\left(\left(1+\mathsf{B}_i^{\left(\frac{1}{q-1}\right)}\right)\mathsf{B}_i^{(\theta)} - 1\right)
	\nonumber\\
	&\phantom{=}
	+ \left(\log\frac{(1-\betaEdge)\alphaEdge}{(1-\alphaEdge)\betaEdge}\right)
    \left(
    \sum\limits_{i \in \misClassfSet{\betaEdge}{\alphaEdge}} \mathsf{B}_i^{(\betaEdge)}
    - \sum\limits_{i \in \misClassfSet{\alphaEdge}{\betaEdge}} \mathsf{B}_i^{(\alphaEdge)} 
    \right)
    +
    \left(\log\frac{1-\alphaEdge}{1-\betaEdge}\right) 
    \left(
    \left\vert \misClassfSet{\betaEdge}{\alphaEdge}\right\vert
    - \left\vert \misClassfSet{\alphaEdge}{\betaEdge}\right\vert
    \right)
	\nonumber\\
	&\phantom{=}
	+ \left(\log\frac{(1-\gammaEdge)\alphaEdge}{(1-\alphaEdge)\gammaEdge}\right)
    \left(
    \sum\limits_{i \in \misClassfSet{\gammaEdge}{\alphaEdge}} \mathsf{B}_i^{(\gammaEdge)}
    - \sum\limits_{i \in \misClassfSet{\alphaEdge}{\gammaEdge}} \mathsf{B}_i^{(\alphaEdge)} 
    \right)
    +  
	\left(\log\frac{1-\alphaEdge}{1-\gammaEdge}\right) 
    \left(
    \left\vert \misClassfSet{\gammaEdge}{\alphaEdge}\right\vert
    - \left\vert \misClassfSet{\alphaEdge}{\gammaEdge}\right\vert
    \right)
	\nonumber\\
	&\phantom{=}
	+ \left(\log\frac{(1-\gammaEdge)\betaEdge}{(1-\betaEdge)\gammaEdge}\right)
    \left(
    \sum\limits_{i \in \misClassfSet{\gammaEdge}{\betaEdge}} \mathsf{B}_i^{(\gammaEdge)}
    - \sum\limits_{i \in \misClassfSet{\betaEdge}{\gammaEdge}} \mathsf{B}_i^{(\betaEdge)} 
   \right)
    +  
	\left(\log\frac{1-\betaEdge}{1-\gammaEdge}\right)
    \left(
    \left\vert \misClassfSet{\gammaEdge}{\betaEdge}\right\vert
    - \left\vert \misClassfSet{\betaEdge}{\gammaEdge}\right\vert
    \right).
	\label{eq:B_middle}
\end{align}
In the following lemma, we write each summand in \eqref{ineq:upp_worst_case_err_prob} in terms of \eqref{eq:B_middle}.

\begin{lemma}
	\label{lemma:neg_lik_M0_XT}
	For any $X \in \mathcal{X}(T)$ and $T \in \tupleSetDelta$, we have
    \begin{align}
    	\mathbb{P}\left[\mathsf{L}\left(\gtMat\right) \geq \mathsf{L}(X)\right]
    	= 
    	\mathbb{P} \left[\mathbf{B} \geq 0 \right].
    	\label{eq:lemma3}
    \end{align}
\end{lemma}

\begin{IEEEproof}
    We refer to Appendix~\ref{proof:neg_lik_M0_XT} for the proof of Lemma~\ref{lemma:neg_lik_M0_XT}.
\end{IEEEproof}
The following lemma provides an upper bound of the RHS of \eqref{eq:lemma3}. 

\begin{lemma}
	\label{lemma:upper_bound_B}
    For any $\{\misClassfSet{\mu}{\nu} : \mu, \nu \in \{\alphaEdge,\betaEdge,\gammaEdge \}, \: \mu \neq \nu\}$, we have
    \begin{align}
    	&\mathbb{P} \left[ \mathbf{B}
    	\geq 0 
    	\right]
        \leq 
        \exp\left(-\left(1+o(1)\right) \left(\lvert\diffEntriesSet\rvert \Ir 
    	+ \Pleftrightarrow{\alphaEdge}{\betaEdge} \: \Ig \frac{\log n}{n}
    	+ \Pleftrightarrow{\alphaEdge}{\gammaEdge} \:\IcOne \frac{\log n}{n}
    	+ \Pleftrightarrow{\betaEdge}{\gammaEdge} \: \IcTwo \frac{\log n}{n} \right)\right),
        \label{eq:lemma1_2}
    \end{align}
    where
    \begin{align}
        \Pleftrightarrow{\alphaEdge}{\betaEdge} = 
        \frac{\left\vert \misClassfSet{\betaEdge}{\alphaEdge} \right\vert + \left\vert \misClassfSet{\alphaEdge}{\betaEdge} \right\vert}{2},
        \qquad
        \Pleftrightarrow{\alphaEdge}{\gammaEdge} = 
        \frac{\left\vert \misClassfSet{\gammaEdge}{\alphaEdge} \right\vert + \left\vert \misClassfSet{\alphaEdge}{\gammaEdge} \right\vert}{2},
        \qquad
        \Pleftrightarrow{\betaEdge}{\gammaEdge} = 
        \frac{\left\vert \misClassfSet{\gammaEdge}{\betaEdge} \right\vert + \left\vert \misClassfSet{\betaEdge}{\gammaEdge} \right\vert}{2}.
        \label{eq:Pleftrightarrow_defn}
    \end{align}
\end{lemma}

\begin{IEEEproof}
    We refer to Appendix~\ref{proof:upper_bound_B} for the proof of Lemma~\ref{lemma:upper_bound_B}.
\end{IEEEproof}

We show that the interested error event $\{ \mathsf{L}(\gtMat) \geq \mathsf{L}(X) : X \neq \gtMat \}$ in \eqref{ineq:upp_worst_case_err_prob} depends solely on two sets of key parameters which dictate the relationship between $X$ and $\gtMat$:
\begin{enumerate}
	\item the first set includes counters to identify the number of users in cluster $x$ and group $i$ whose rating vector $\vecU{i}{x}$ in $\gtMat$ is changed to the rating vector $\vecV{j}{y}$ of users in cluster $y$ and group $j$ in $X$, for $x,y \in [c]$ and $i,j \in [g]$.
	Formally, we define
	\begin{align}
	    \kUsrs{i}{j}{x}{y} 
	    = 
	    \left\vert \left\{ 
	    r : r \in Z_0(x,i) \cap Z(y,j)
	    \right\}\right\vert, 
	    \:\text{ where }\:
	    0 \leq \kUsrs{i}{j}{x}{y} \leq \frac{n}{gc};
	    \label{eq:kUsrs_defn}
	\end{align}
	\item the second set provides the Hamming distance between $\vecU{i}{x}$ and $\vecV{i}{x}$, for $x \in [c]$ and $i \in [g]$.
	Formally, we define
	\begin{align}
	    \dElmntsGen{i}{j}{x}{y}
	    =
	    \hamDist{\vecU{i}{x}}{\vecV{j}{y}},
	    \:\text{ where }\:
	    0 \leq \dElmntsGen{i}{j}{x}{y} \leq m.
	   \label{eq:dElmntsGen_defn}
	\end{align}
\end{enumerate}
Based on these two parameters, the set of rating matrices $\matSet$ is partitioned into a number of classes of matrices $\mathcal{X}(T)$. Here, each matrix class $\mathcal{X}(T)$ is defined as the set of rating matrices that is characterized by a tuple~$T$ where
\begin{align}
    T
    =
    \left(
    \left\{\kUsrs{i}{j}{x}{y}\right\}_{x,y \in [c], \: i,j \in [g]}, 
    \left\{\dElmntsGen{i}{j}{x}{y}\right\}_{x,y \in [c], \: i,j \in [g]}
    \right).
	\label{eq:tuple_ratMat}
\end{align}
Define $\tupleSetDelta$ as the set of all non-all-zero tuples $T$.
Therefore, we can write $\matSet = \bigcup_{T \in \tupleSetDelta} \mathcal{X}(T)$.

\begin{JA_AE}
Next, we analyze the performance of the ML decoder by comparing the ground truth user partitioning with that of the decoder. 
For a non-negative constant $\relabelConst \in (0,\: (\epsilon\log m - (2+\epsilon)\log(2q)) / (2 (1 + \epsilon)\log m))$, where $\epsilon > \max\{(2 \log 2) / \log n, \: (2(g-r+1) \log 2) / \log (2qm), \: (2\log(2q))/\log(m/2q)\}$, 
define $\sigma(x,i)$ as the set of pairs of cluster and group in $\cZ$ whose number of overlapped users with $\cZ_0(x,i)$ exceeds a $(1-\tau)$ fraction of the group size. Formally, we have
\begin{align}
    \sigma(x,i)
    =
    \left\{(y,j)\in[c]\times[g]: 
    \left| Z_0(x,i) \cap Z(y,j)\right|
    \geq 
    (1-\relabelConst) \frac{n}{gc} \right\}.
   \label{eq:sigma(x,i)}
\end{align}
Note that $\relabelConst < 0.5$, which implies that $|\sigma(x,i)| \leq 1$ since the size of any group is $n/(gc)$ users.
For $|\sigma(x,i)| = 1$, let $\sigma(x,i) = \{(\sigma(x), \sigma(i|x))\}$.
Accordingly, partition the set $\tupleSetDelta$ into two subsets $\TsmallErr$ and $\TlargeErr$ that are defined as follows:
\begin{align}
    \TsmallErr
    &= 
    \left\{ T \in \mathcal{T}^{(\delta)} : 
    \forall (x,i) \in [c]\times [g] \text{ such that }
    \left\vert \sigma(x,i) \right\vert = 1, \:
    \dElmntsGen{i}{\:\sigma(i|x)}{x}{\:\sigma(x)} \leq \tau m \min\{\interTau, \intraTau\}
    \right\},
    \label{eq:TsmallErr}\\
    \TlargeErr
    &= 
    \left\{ T \in \mathcal{T}^{(\delta)} : 
    \exists (x,i) \in [c]\times [g] \text{ such that }
    \left(\left\vert \sigma(x,i) \right\vert = 0\right)
    \right\}
    \nonumber\\
    &\phantom{=}\cup
    \left\{
    T \in \mathcal{T}^{(\delta)} : 
    \forall (x,i) \in [c] \times [g] 
    \text{ such that }
    \left\vert \sigma(x,i) \right\vert = 1,\: 
    \exists (x,i) \in [c] \times [g] 
    \text{ such that }
    \dElmntsGen{i}{\:\sigma(i|x)}{x}{\:\sigma(x)} >  \tau m \min\{\interTau, \intraTau\}
    \right\}.
    \label{eq:TlargeErr}
\end{align}
Intuitively, when $T \in \TsmallErr$, the class of matrices $\mathcal{X}(T)$ corresponds to the typical (i.e., small) error set. On the other hand, when $T \in \TlargeErr$, the class of matrices $\mathcal{X}(T)$ corresponds to the atypical (i.e., large) error set that has negligible probability mass.

The following two lemmas provide an upper bound on the RHS of \eqref{eq:lemma1_2} under different classes of candidate rating matrices, and evaluating the limits as $n$ and $m$ tend to infinity.

\begin{lemma}
\label{lemma:T1_related} 
    For any $\{\misClassfSet{\mu}{\nu} : \mu, \nu \in \{\alphaEdge,\betaEdge,\gammaEdge \}, \: \mu \neq \nu\}$, we have\footnote{As $n$ tends to infinity, $m$ also tends to infinity since $m = \omega(\log n)$.}
    \begin{align}
        & \lim_{n,m\rightarrow \infty}
        \sum\limits_{T \in \TsmallErr} 
	    \sum\limits_{X \in \mathcal{X}(T)}
	    \exp\left(-\left(1+o(1)\right) \left(\lvert\diffEntriesSet\rvert \Ir 
    	+ \Pleftrightarrow{\alphaEdge}{\betaEdge} \: \Ig \frac{\log n}{n}
    	+ \Pleftrightarrow{\alphaEdge}{\gammaEdge} \: \IcOne \frac{\log n}{n}
    	+ \Pleftrightarrow{\betaEdge}{\gammaEdge} \: \IcTwo \frac{\log n}{n} \right)\right) 
    	= 0.
    	\label{eq:lemma_T1_1} 
    \end{align}
\end{lemma}

\begin{IEEEproof}
We refer to Appendix~\ref{proof:T1_related} for the proof of Lemma~\ref{lemma:T1_related}. 
\end{IEEEproof}

\begin{lemma}
\label{lemma:eta_omega(nm)}
    For any $\{\misClassfSet{\mu}{\nu} : \mu, \nu \in \{\alphaEdge,\betaEdge,\gammaEdge \}, \: \mu \neq \nu\}$, we have
    \begin{align}
        & \lim_{n,m\rightarrow \infty }
        \sum\limits_{T \in  \TlargeErr} 
	    \sum\limits_{X \in \mathcal{X}(T)}
	    \exp\left(-\left(1+o(1)\right)  \left(\lvert\diffEntriesSet\rvert \Ir 
    	+ \Pleftrightarrow{\alphaEdge}{\betaEdge} \: \Ig \frac{\log n}{n}
    	+ \Pleftrightarrow{\alphaEdge}{\gammaEdge} \: \IcOne \frac{\log n}{n}
    	+ \Pleftrightarrow{\betaEdge}{\gammaEdge} \: \IcTwo \frac{\log n}{n} \right)\right)
    	= 0.
    	\label{ineq:eta_omega(nm)}
    \end{align}
\end{lemma}

\begin{IEEEproof}
We refer to Appendix~\ref{proof:eta_omega(nm)} for the proof of Lemma~\ref{lemma:eta_omega(nm)}.
\end{IEEEproof}
\end{JA_AE}

\subsection{The Achievability Proof of Theorem~\ref{thm:p_star}}
\label{sec:achv_proof_Thm1}
The worst-case probability of error $P_e^{(\diff)}(\psi_{\text{ML}})$ is upper bounded by
\begin{align}
    &P_e^{(\diff)}(\psi_{\text{ML}}) \nonumber\\
    &\!\leq\!
    \sum_{X \neq \gtMat}
    \mathbb{P}\left[\mathsf{L}(\gtMat) \geq \mathsf{L}(X)\right]
    \label{eq:matrix_enumerate_neg1}\\
    &\!=\!
    \sum_{\substack{X \neq \gtMat, \\ X \in \matSet}}
    \mathbb{P}\left[\mathsf{L}(\gtMat) \geq \mathsf{L}(X)\right]
    \label{eq:matrix_enumerate_0}\\
    &\!=\!
    \sum\limits_{T \in \tupleSetDelta} 
	\sum\limits_{X \in \mathcal{X}(T)}
	\mathbb{P}\left[\mathsf{L}(\gtMat) \geq \mathsf{L}(X)\right]
	\label{eq:matrix_enumerate}\\
	&\!\leq\!  	
	\sum\limits_{T \in \tupleSetDelta} 
	\sum\limits_{X \in \mathcal{X}(T)}
	\exp\left(-\left(1\!+\!o(1)\right) \left(\lvert\diffEntriesSet\rvert \Ir 
    	+ \Pleftrightarrow{\alphaEdge}{\betaEdge} \: \Ig \frac{\log n}{n}
    	+ \Pleftrightarrow{\alphaEdge}{\gammaEdge} \: \IcOne \frac{\log n}{n}
    	+ \Pleftrightarrow{\betaEdge}{\gammaEdge} \: \IcTwo \frac{\log n}{n} \right)\right)
    \label{ineq:upper_bound_M0_XT}\\
    &\!=\!
    \sum\limits_{T \in  \TsmallErr} 
	\sum\limits_{X \in \mathcal{X}(T)}
	\exp\left(-\left(1\!+\!o(1)\right) \left(\lvert\diffEntriesSet\rvert \Ir 
    	+ \Pleftrightarrow{\alphaEdge}{\betaEdge} \: \Ig \frac{\log n}{n}
    	+ \Pleftrightarrow{\alphaEdge}{\gammaEdge} \: \IcOne \frac{\log n}{n}
    	+ \Pleftrightarrow{\betaEdge}{\gammaEdge} \: \IcTwo \frac{\log n}{n} \right)\right)
    \label{partial_sum}\\
    &\phantom{\leq} 
    + 
    \sum\limits_{T \in \TlargeErr} 
	\sum\limits_{X \in \mathcal{X}(T)}
	\exp\left(-\left(1+o(1)\right) \left(\lvert\diffEntriesSet\rvert \Ir 
    	+ \Pleftrightarrow{\alphaEdge}{\betaEdge} \: \Ig \frac{\log n}{n}
    	+ \Pleftrightarrow{\alphaEdge}{\gammaEdge} \: \IcOne \frac{\log n}{n}
    	+ \Pleftrightarrow{\betaEdge}{\gammaEdge} \: \IcTwo \frac{\log n}{n} \right)\right),\nonumber
\end{align}
where \eqref{eq:matrix_enumerate_neg1} follows from Lemma~\ref{lm:upp_worst_case_err_prob}; 
\eqref{eq:matrix_enumerate_0} follows from the definition of negative log-likelihood in \eqref{eq:likelihood_defn}; \eqref{eq:matrix_enumerate} follows from the definition of the tuples characterizing matrix classes in \eqref{eq:tuple_ratMat}; \eqref{ineq:upper_bound_M0_XT} follows from Lemma~\ref{lemma:neg_lik_M0_XT} and Lemma~\ref{lemma:upper_bound_B}; and finally \eqref{partial_sum} follows from the definitions of $\TsmallErr$ and $\TlargeErr$ in \eqref{eq:TsmallErr} and \eqref{eq:TlargeErr}, respectively.

Finally, the limit of the worst-case probability of error $P_e^{(\diff)}(\psi_{\text{ML}})$ in \eqref{partial_sum} as $n$ and $m$ tend to infinity is evaluated as
\begin{align}
    &\lim_{n,m\rightarrow \infty }
    P_e^{(\diff)}(\psi_{\text{ML}})
    \nonumber\\
    &\leq 
    \lim_{n,m\rightarrow \infty }
    \left(
    \sum\limits_{T \in  \TsmallErr} 
	\sum\limits_{X \in \mathcal{X}(T)}
	\exp\left(-\left(1\!+\!o(1)\right) \left(\lvert\diffEntriesSet\rvert \Ir 
    	+ \Pleftrightarrow{\alphaEdge}{\betaEdge} \: \Ig \frac{\log n}{n}
    	+ \Pleftrightarrow{\alphaEdge}{\gammaEdge} \: \IcOne \frac{\log n}{n}
    	+ \Pleftrightarrow{\betaEdge}{\gammaEdge} \: \IcTwo \frac{\log n}{n} \right)\right)
    \right.
    \nonumber\\
    &\phantom{=\lim_{n\rightarrow \infty } \left(\right.}
    \left.
    +\!\sum\limits_{T \in \TlargeErr} 
	\sum\limits_{X \in \mathcal{X}(T)}
	\!\exp\left(-\left(1+o(1)\right) \left(\lvert\diffEntriesSet\rvert \Ir 
    	+ \Pleftrightarrow{\alphaEdge}{\betaEdge} \: \Ig \frac{\log n}{n}
    	+ \Pleftrightarrow{\alphaEdge}{\gammaEdge} \: \IcOne \frac{\log n}{n}
    	+ \Pleftrightarrow{\betaEdge}{\gammaEdge} \: \IcTwo \frac{\log n}{n} \right)\right)
    \right)
    \nonumber\\
    &= 0,
    \label{eq:lim_err_achv}
\end{align}
where \eqref{eq:lim_err_achv} follows from Lemma \ref{lemma:T1_related} and Lemma \ref{lemma:eta_omega(nm)}.
This concludes the achievability proof of Theorem~\ref{thm:p_star}.
\hfill $\blacksquare$

\section{The Converse Proof}
\label{sec:conv}
In this section, we prove the converse part of Theorem~\ref{thm:p_star}, that is if the condition on $p$ in \eqref{eq:pstar-general_conv} holds, then $ \lim_{n \rightarrow \infty} P_e^{(\diff)}(\psi) \neq 0$ for any estimator $\psi$.
To this end, we prove that $\lim_{n\rightarrow \infty} P_e^{(\delta)} (\psi) \neq 0$ for any estimator $\psi$ and any ground truth rating matrix $\gtMat \in \matSet$, if either of the following conditions holds:
\begin{align}
    \frac{g-r+1}{gc}nI_r &\leq (1-\epsilon) \log m, 
    & & \textbf{(Perfect Clustering/Grouping Regime)}
    \label{eq:necCond_1}\\
	\intraTau m I_r + \frac{\Ig}{gc} \log n &\leq (1-\epsilon) \log n, 
	& & \textbf{(Grouping-Limited Regime)}
	\label{eq:necCond_2}\\
	\interTau m I_r + \frac{\IcOne}{gc} \log n + \frac{(g-1) \IcTwo}{gc} \log n &\leq (1-\epsilon) \log n, 
	& & \textbf{(Clustering-Limited Regime)}
	\label{eq:necCond_3}
\end{align}
where $\Ir$, $\Ig$, $\IcOne$ and $\IcTwo$ are defined in \eqref{eq:Ir_I_abg_defn}.
Throughout the proof, let $p = \Theta ((\log n) / n)$, and let $q$ and $\theta$ be constants such that $q$ is prime and $\theta \in [0,1]$.
We first present a number of auxiliary lemmas in Section~\ref{sec:conv_auxLemma}. Then, we present the converse proof of Theorem~\ref{thm:p_star} in Section~\ref{sec:conv_proof_thm1}.

\subsection{The Auxiliary Lemmas}
\label{sec:conv_auxLemma}
We present three auxiliary lemmas that are used to prove the converse part of Theorem~\ref{thm:p_star}. Before each lemma, we introduce the terminologies and notations needed for the statement of the lemma.

First, let $S$ denote the success event that a rating matrix is correctly estimated (i.e., exactly recovered). It is defined as
\begin{align}
	S \coloneqq
	\bigcap_{X \neq \gtMat}
	\left[\mathsf{L}(X) > \mathsf{L}(\gtMat)\right],
	\label{eq:defS}
\end{align}
where $\mathsf{L}(X)$ is the negative log-likelihood of a candidate rating matrix $X$, defined in \eqref{eq:likelihood_defn}. 
The following lemma introduces a lower bound on the infimum of the worst-case probability of error.

\begin{lemma}
\label{lm:inf_worstProbError}
For any estimator $\psi$, we have
\begin{align}
	\inf_{\psi} P_e^{(\delta)} (\psi) 
    \geq
	\mathbb{P}\left[S^c\right].
\end{align} 
\end{lemma}

\begin{IEEEproof}
We refer to Appendix~\ref{app:inf_worstProbError} for the proof of Lemma~\ref{lm:inf_worstProbError}.
\end{IEEEproof}
Next, the following lemma provides, together with Lemma~\ref{lemma:neg_lik_M0_XT}, a lower bound on the probability that $\mathsf{L}(\gtMat)$ is greater than or equal to $\mathsf{L}(X)$.

\begin{lemma} 
\label{lm:lowerB_prob}
    For any $\{\misClassfSet{\mu}{\nu} : \left\vert \misClassfSet{\mu}{\nu} \right\vert = \left\vert \misClassfSet{\nu}{\mu} \right\vert,  \:
    \mu, \nu \in \{\alphaEdge,\betaEdge,\gammaEdge \}, \: \mu \neq \nu\}$, we have
    \begin{align}
    	&\mathbb{P} \left[ \mathbf{B}
    	\geq 0 
    	\right]
        \geq 
        \frac{1}{4}
        \exp\left(-\left(1+o(1)\right) \left(\lvert\diffEntriesSet\rvert \Ir 
    	+ \left\vert \misClassfSet{\betaEdge}{\alphaEdge} \right\vert \Ig \frac{\log n}{n}
    	+ \left\vert \misClassfSet{\gammaEdge}{\alphaEdge} \right\vert \IcOne \frac{\log n}{n}
    	+ \left\vert \misClassfSet{\gammaEdge}{\betaEdge} \right\vert \IcTwo \frac{\log n}{n} \right)\right).
        \label{eq:lowerB_prob}
    \end{align}
    where the random variable $\mathbf{B}$ is defined in \eqref{eq:B_middle}.
\end{lemma}

\begin{IEEEproof}
We refer to Appendix~\ref{app:lowerB_prob_proof} for the proof of Lemma~\ref{lm:lowerB_prob}.
\end{IEEEproof}

We finally present a lemma that guarantees the existence of two subsets of users with specific properties.

\begin{lemma} 
Consider the sets $\grpG{i}{x}$ and $\grpG{j}{y}$ for $x,y \in [c]$, $i,j \in [g]$ and $(x,i) \neq (y,j)$. 
As $n\rightarrow \infty$, with probability approaching $1$, there exists two subsets $\tG{i}{x} \subset \grpG{i}{x}$ and $\tG{j}{y}\subset \grpG{j}{y}$ with cardinalities $\vert \tG{i}{x} \vert \geq \frac{n}{\log^3 n}$ and  $\vert\tG{j}{y}\vert \geq \frac{n}{\log^3 n}$ such that there are no edges between the vertices in $\tG{i}{x} \cup \tG{j}{y}$. That is, 
\begin{align}
\mathcal{E} \cap \left( \left(\tG{i}{x} \cup \tG{j}{y}\right) \times \left(\tG{i}{x} \cup \tG{j}{y}\right)\right) = \varnothing.
\label{eq:noEdgeLemma}
\end{align}
  \label{lm:randGraph}
\end{lemma}

\begin{IEEEproof}
    We refer to Appendix~\ref{app:randGraph} for the proof of Lemma~\ref{lm:randGraph}.
\end{IEEEproof}

\subsection{The Converse Proof of Theorem~\ref{thm:p_star}}
\label{sec:conv_proof_thm1}
In order to prove the converse part of Theorem~\ref{thm:p_star}, we demonstrate that $\lim_{n,m \rightarrow \infty} \mathbb{P} \left[S\right] = 0$ if any of the conditions given by \eqref{eq:necCond_1}, \eqref{eq:necCond_2} or \eqref{eq:necCond_3} holds. 
In the following, we show the claim for each condition in \eqref{eq:necCond_1}, \eqref{eq:necCond_2} or \eqref{eq:necCond_3} separately.

\subsubsection{Failure Proof for the Perfect Clustering/Grouping Regime}
\label{subsec:failureProof1}
In this proof, we introduce a class of rating matrices, where each matrix in this class is obtained by replacing one column of $\gtMat$ with a carefully chosen sequence.
Then, we prove that if \eqref{eq:necCond_1} holds, then with high probability the ML estimator will fail by selecting one of the rating matrices from this class, instead of $\gtMat$. 

Recall the sections of the columns of $\gtMat$ defined in Section~\ref{sec:achv}, and note that there exists (at least) one section $\colBlock{\ell}$ such that $\colBlockSize{\ell} = |\colBlock{\ell}|/m $ is bounded away from zero (i.e., not vanishing with $m$ and $n$).
For each $k \in \colBlock{\ell}$, define $\M{k} \in \mathbb{F}_{q}^{n \times m}$ as a rating matrix that is identical to $\gtMat$ except for its $k^{\text{th}}$, which will be determined below.
Recall from Section~\ref{sec:achv} that $\gtRatingCluster{1} \in \mathbb{F}_{q}^{g \times m}$, the submatrix of $\gtMat$ associated with the first cluster, is obtained by stacking some codeword vectors from a $(g,r)$ MDS code with generator matrix $\iGenerator{1}$. 
Let $w\in \mathbb{F}_q^{g\times 1}$ be another codeword from this MDS code such that 
\begin{align}
	\hamDist{w}{\:\gtRatingCluster{1}\left(:,k\right)} = g-r+1.
	\label{eq:hamDist_Mc}
\end{align}
The existence of such a column vector $w$ is guaranteed due to the fact that the $(g,r)$ MDS code in $\mathbb{F}_q$ has a minimum distance of $g-r+1$.
Consequently, the entries of $\M{k}$ are given by
\begin{align}
	\M{k}(r,t) 
	= 
	\left\{
	\begin{array}{cl}
		w(1) & \textrm{if $r \in Z(1,1)$ and $t=k$},
		\vspace{0.5mm}\\
		w(2) & \textrm{if $r \in Z(1,2)$  and $t=k$},
		\vspace{0.5mm}\\
		\vdots & \qquad\vdots 
		\vspace{0.5mm}\\
		w(g) & \textrm{if $r \in Z(1,g) $  and $t=k$},
		\vspace{0.5mm}\\
		\gtMat(r,t)
		& \textrm{otherwise}.
	\end{array}
	\right.
	\label{eq:Xk_entries}
\end{align}
Furthermore, given $\gtMat$ and $\M{k}$,  we have
\begin{align}
\begin{split}
    &\diffEntriesSet = \left\{(r,k): r \in Z(1,i) \text{ for } i \in [g], \: \M{k}(r,k) \neq \gtMat(r,k)\right\}, 
    \\
    &\misClassfSet{\betaEdge}{\alphaEdge} = \misClassfSet{\alphaEdge}{\betaEdge} = \varnothing,
    \qquad
    \misClassfSet{\gammaEdge}{\alphaEdge} = \misClassfSet{\alphaEdge}{\gammaEdge} = \varnothing,
    \qquad
    \misClassfSet{\gammaEdge}{\betaEdge} = \misClassfSet{\betaEdge}{\gammaEdge} = \varnothing,
\end{split}
\label{eq:failure1_pairs}
\end{align}
according to their definitions in \eqref{eq:N1_defn}--\eqref{eq:N4rev_defn}.
Thus, the cardinalities of the sets in \eqref{eq:failure1_pairs} are given by
\begin{align}
    \left\vert\diffEntriesSet\right\vert = \frac{n}{gc} (g-r+1), 
    \qquad
    \left\vert \misClassfSet{\betaEdge}{\alphaEdge} \right\vert = \left\vert \misClassfSet{\alphaEdge}{\gammaEdge} \right\vert = 0,
    \qquad
    \left\vert \misClassfSet{\gammaEdge}{\alphaEdge} \right\vert = \left\vert \misClassfSet{\alphaEdge}{\gammaEdge} \right\vert = 0,
    \qquad
    \left\vert \misClassfSet{\gammaEdge}{\betaEdge} \right\vert = \left\vert \misClassfSet{\betaEdge}{\gammaEdge} \right\vert = 0.
    \label{eq:failure1_sizePair}
\end{align}

For each $\M{k}$ where $k \in \colBlock{\ell}$, the probability that the negative log-likelihood of $\M{k}$ is greater that that of $\gtMat$ is upper bounded~by
\begin{align}
  \mathbb{P} \left[ \mathsf{L}(\M{k}) > \mathsf{L}(\gtMat) \right]
  &=
  1 - \mathbb{P} \left[ \mathsf{L}(\M{k}) \leq \mathsf{L}(\gtMat)\right]
  \nonumber\\
  &=
  1 - \mathbb{P} 
  \left[
  \log \left((q-1)\frac{1-\theta}{\theta}\right) 
  \sum_{i \in \diffEntriesSet}
  \mathsf{B}_i^{(p)} \left(\left(1+\mathsf{B}_i^{\left(\frac{1}{q-1}\right)}\right)\mathsf{B}_i^{(\theta)}-1\right) 
  \:\geq\: 0
  \right]
  \label{eq:1stTermX_prob_cond1_0}\\
  &=
  1 - \mathbb{P} 
  \left[
  \log \left((q-1)\frac{1-\theta}{\theta}\right) 
  \sum_{i=1}^{\frac{n}{gc} (g-r+1)}
  \mathsf{B}_i^{(p)} \left(\left(1+\mathsf{B}_i^{\left(\frac{1}{q-1}\right)}\right)\mathsf{B}_i^{(\theta)}-1\right)
  \:\geq\: 0\right]
  \label{eq:1stTermX_prob_cond1_eta}
  \\
  &\leq
  1- \frac{1}{4} \exp \left(- (1+o(1))\frac{g-r+1}{gc} n \Ir\right)
  \label{eq:1stTermX_prob_cond1_lm}
  \\
  &\leq
  \exp \left(- \frac{1}{4} \exp \left(- (1+o(1))\frac{g-r+1}{gc} n \Ir\right)\right),
  \label{eq:1stTermX_prob_cond1}
\end{align}
where \eqref{eq:1stTermX_prob_cond1_0} follows from Lemma~\ref{lemma:neg_lik_M0_XT} and \eqref{eq:failure1_pairs}; 
\eqref{eq:1stTermX_prob_cond1_eta} follows from \eqref{eq:failure1_sizePair}; 
and \eqref{eq:1stTermX_prob_cond1_lm}~is an immediate consequence of  Lemma~\ref{lm:lowerB_prob}.

Finally, since $\colBlockSize{\ell}$ is bounded away from zero, the probability of exact rating matrix recovery is upper bounded by 
\begin{align}
  \mathbb{P}[S] 
  &\leq \mathbb{P} \left[\bigcap_{k\in \colBlock{\ell}}
  \left(\mathsf{L}(\M{k}) > \mathsf{L}(\gtMat)\right)\right]
  \label{eq:1stTerm_cond1_0}\\
  & =
  \prod_{k\in \colBlock{\ell}} \mathbb{P} \left[ \mathsf{L}(\M{k}) > \mathsf{L}(\gtMat) \right]
  \label{eq:1stTerm_cond1_indp}
  \\
  & \leq 
  \left(\exp \left(- \frac{1}{4}  \exp \left(- (1+o(1)) \frac{g-r+1}{gc} n \Ir\right)\right)\right)^{\colBlockSize{\ell} m}
  \label{eq:1stTerm_cond1_ident}
  \\
  & = 
  \exp \left(- \frac{1}{4} \colBlockSize{\ell} \exp \left(- (1+o(1)) \frac{g-r+1}{gc} n \Ir + \log m\right)\right)
  \nonumber\\
  & \leq 
  \exp \left(- \frac{1}{4} \colBlockSize{\ell}
  \exp \Bigl( -\bigl( (1+o(1)) (1-\epsilon) - 1 \bigr) \log m \Bigr)
  \right)
  \label{eq:1stTerm_cond1_cond}\\
  & \leq 
  \exp \left(- \frac{1}{4} \colBlockSize{\ell}
  \exp \Bigl( \bigl( \epsilon - o(1) (1-\epsilon)  \bigr) \log m \Bigr)
  \right),
\end{align}
where \eqref{eq:1stTerm_cond1_0} follows from the definition in \eqref{eq:defS}; \eqref{eq:1stTerm_cond1_indp} holds since the events $\{\mathsf{L}(\M{k}) > \mathsf{L}(\gtMat) : k \in \colBlock{\ell}\}$ are mutually independent due to the fact that each event corresponds to a different column $k$ within $\colBlock{\ell}$; \eqref{eq:1stTerm_cond1_ident}~follows from \eqref{eq:1stTermX_prob_cond1}; and \eqref{eq:1stTerm_cond1_cond} follows from the condition in \eqref{eq:necCond_1}. Therefore, we obtain 
\begin{align}
  \lim_{n,m \rightarrow \infty} \mathbb{P} \left[S\right] 
  &\leq
  \lim_{n,m \rightarrow \infty}  \exp \left(- \frac{1}{4} \colBlockSize{\ell}
  \exp \Bigl( \bigl( \epsilon - o(1) (1-\epsilon)  \bigr) \log m \Bigr)
  \right) = 0,
  \label{eq:lim_Ps_cond1}
\end{align}
which shows that if the condition in \eqref{eq:necCond_1} holds, then the ML estimator will fail in finding $\gtMat$ with high probability. 

\subsubsection{Failure Proof for the Grouping-Limited Regime}
\label{subsec:failureProof2}
Without loss of generality, assume $\intraTau m = \hamDist{\vecU{1}{1}}{\vecU{2}{1}}$, i.e., the rating vectors of groups $1$ and $2$ in cluster~$1$ have the minimum 
Hamming distance among distinct pairs of rating vectors of groups within the same cluster.
In this proof, we introduce a class of rating matrices, which are obtained by switching two users between groups $1$ and $2$ in cluster~$1$. 
Then, we prove that if \eqref{eq:necCond_2} holds, then with high probability the ML estimator will fail by selecting one of the rating matrices from this class, instead of $\gtMat$. 

Set $(x,i)=(1,1)$ and $(y,j)=(1,2)$ in Lemma~\ref{lm:randGraph}. 
Thus, there exist subsets $\tG{1}{1} \subset \grpG{1}{1}$ and $\tG{2}{1}\subset \grpG{2}{1}$ with $|\tG{1}{1}| =|\tG{2}{1}|=\frac{n}{\log^3 n}$, such that the subgraph induced by the vertices in $\tG{1}{1} \cup \tG{2}{1}$ is edge-free. 
Define $\M{a,b} \in \mathbb{F}_{q}^{n \times m}$, for $a\in \tG{1}{1}$ and $b\in \tG{2}{1}$, as a rating matrix that is identical to $\gtMat$ except for its $a^{\text{th}}$ and $b^{\text{th}}$ rows, which are swapped.
More formally, the entries of $\M{a,b}$ are given by
\begin{align}
	\M{a,b}(r,:) = \left\{
	\begin{array}{cc}
		\gtMat(b,:)=\vecU{2}{1} & \textrm{if $r=a$},\\
		\gtMat(a,:)=\vecU{1}{1} & \textrm{if $r=b$},\\
		\gtMat(r,:) & \textrm{otherwise}.
	\end{array}
	\right.
	\label{eq:Xab_defn}
\end{align}
The user partitioning $\mathcal{Z}_{\langle a,b\rangle}$ induced by $\M{a,b}$ is given by
\begin{align}
	\Z{a,b}(x,i)
	= 
	\left\{
	\begin{array}{cl}
		\grpG{1}{1} \cup \{b\} \setminus\{a\}
		& \textrm{if $(x,i) = (1,1)$},
		\vspace{1mm}\\
		\grpG{2}{1} \cup \{a\} \setminus\{b\} 
		& \textrm{if $(x,i) = (1,2)$},
		\vspace{0.7mm}\\
		\grpG{i}{x}
		& \textrm{otherwise}.
	\end{array}
	\right.
	\label{eq:Zab_defn}
\end{align}
Furthermore, given $\gtMat$ and $\M{a,b}$, we have
\begin{align}
\begin{split}
    \diffEntriesSet &= \left\{(r,t): r \in \{a,b\}, t \in \left[m\right], \: \M{a,b}(r,t) \neq \gtMat(r,t)\right\}, 
    \\
    \misClassfSet{\betaEdge}{\alphaEdge} &= 
    \left\{(a,h): 
    h\in \grpG{1}{2}\setminus\{b\} 
    \right\}
    \:\cup\:
    \left\{(b,h):
    h\in \grpG{1}{1}\setminus\{a\} 
    \right\},
    \\
    \misClassfSet{\alphaEdge}{\betaEdge} &= \left\{(a,h): 
    h\in \grpG{1}{1}\setminus\{a\} 
    \right\}
    \:\cup\:
    \left\{(b,h):
    h\in \grpG{2}{1}\setminus\{b\} 
    \right\},
    \\
    \misClassfSet{\gammaEdge}{\alphaEdge} &= \misClassfSet{\alphaEdge}{\gammaEdge} = \varnothing,
    \qquad
    \misClassfSet{\gammaEdge}{\betaEdge} = \misClassfSet{\betaEdge}{\gammaEdge} = \varnothing,
\end{split}
\label{eq:failure2_pairs}
\end{align}
according to their definitions in \eqref{eq:N1_defn}--\eqref{eq:N4rev_defn}.
Thus, the cardinalities of the sets in \eqref{eq:failure2_pairs} are given by
\begin{align}
\begin{split}
    \left\vert\diffEntriesSet\right\vert 
    &= \hamDist{\M{a,b}(a,:)}{\gtMat(a,:)}
    + 
    \hamDist{\M{a,b}(b,:)}{\gtMat(b,:)}
    \\
    &= 
    \hamDist{\gtMat(b,:)}{\gtMat(a,:)}
    + 
    \hamDist{\gtMat(a,:)}{\gtMat(b,:)}
    \\
    &=
    2\intraTau m,
    \\
    \left\vert \misClassfSet{\betaEdge}{\alphaEdge} \right\vert &= \left\vert \misClassfSet{\alphaEdge}{\gammaEdge} \right\vert = 2\left(\frac{n}{cg}-1\right),
    \\
    \left\vert \misClassfSet{\gammaEdge}{\alphaEdge} \right\vert &= \left\vert \misClassfSet{\alphaEdge}{\gammaEdge} \right\vert = 0,
    \qquad
    \left\vert \misClassfSet{\gammaEdge}{\betaEdge} \right\vert = \left\vert \misClassfSet{\betaEdge}{\gammaEdge} \right\vert = 0.
    \label{eq:failure2_sizePair}
\end{split}
\end{align}

For each $\M{a,b}$ where $a\in \tG{1}{1}$ and $b\in \tG{2}{1}$, we have 
\begin{align}
	\mathsf{L}(\gtMat) - \mathsf{L}\left(\M{a,b}\right) 
	&= 
	\log \left((q-1)\frac{1-\theta}{\theta}\right) 
	\sum_{i \in \diffEntriesSet}
	\mathsf{B}_i^{(p)} \left(\left(1+\mathsf{B}_i^{\left(\frac{1}{q-1}\right)}\right)\mathsf{B}_i^{(\theta)}-1\right)
	+ \log\left(\frac{(1-\betaEdge)\alphaEdge}{(1-\alphaEdge)\betaEdge}\right) \sum_{j \in \misClassfSet{\betaEdge}{\alphaEdge}}
	\left( \mathsf{B}_{j}^{(\betaEdge)} - \mathsf{B}_{j}^{(\alphaEdge)}\right)
	\label{eq:c2_diff_1}\\
	&= 
	\log \left((q-1)\frac{1-\theta}{\theta}\right) 
	\sum_{i=1}^{2\intraTau m}
	\mathsf{B}_i^{(p)} \left(\left(1+\mathsf{B}_i^{\left(\frac{1}{q-1}\right)}\right)\mathsf{B}_i^{(\theta)}-1\right)
	+ \log\left(\frac{(1-\betaEdge)\alphaEdge}{(1-\alphaEdge)\betaEdge}\right) \sum_{j=1}^{2(\frac{n}{cg}-1)}
	\left( \mathsf{B}_{j}^{(\betaEdge)} - \mathsf{B}_{j}^{(\alphaEdge)}\right),
	\label{eq:c2_diff_2}
\end{align}
where \eqref{eq:c2_diff_1} follows from Lemma~\ref{lemma:neg_lik_M0_XT} and \eqref{eq:failure2_pairs}; and \eqref{eq:c2_diff_2} follows from \eqref{eq:failure2_sizePair}.
Therefore, the probability that the negative log-likelihood of $\M{a,b}$ is greater that that of $\gtMat$ is upper bounded by
\begin{align}
	\mathbb{P} \left[
	\mathsf{L}\left(\M{a,b}\right) > \mathsf{L}(\gtMat)\right]
	&=
	1 - \mathbb{P}\left[ 
	\mathsf{L}(\gtMat) - \mathsf{L}\left(\M{a,b}\right) 
	\geq 0\right] 
	\nonumber
	\\
	&\leq 
	1 - \frac{1}{4} \exp \left( - (1 + o(1)) \left(2 \intraTau m \Ir + 2\left(\frac{n}{cg}-1\right) \Ig \frac{\log n}{n} \right) \right)
	\label{eq:2nd-L-diff_0}\\
	&\leq
	\exp
	\left(
	- \frac{1}{4} \exp \left( - (1+o(1)) \left(2\intraTau m \Ir + 2\left(\frac{n}{cg}-1\right) \Ig \frac{\log n}{n} \right) \right)
	\right),
	\label{eq:2nd-L-diff}
\end{align}
where \eqref{eq:2nd-L-diff_0} follows from \eqref{eq:c2_diff_2} and Lemma~\ref{lm:lowerB_prob}.

Finally, the probability of exact rating matrix recovery is upper bounded by 
\begin{align}
  \mathbb{P}[S] 
  &\leq 
  \mathbb{P} \left[\bigcap_{\substack{a\in \tG{1}{1} \\ b\in \tG{2}{1}}}
  \left(\mathsf{L}\left(\M{a,b}\right) > \mathsf{L}(\gtMat)\right)\right]
  \label{eq:2nd-success-0}\\
  &=
  \prod_{\substack{a\in \tG{1}{1} \\ b\in \tG{2}{1}}} \mathbb{P} \left[ \mathsf{L}\left(\M{a,b}\right) > \mathsf{L}(\gtMat) \right]
  \label{eq:2nd-indep}
  \\
  & \leq 
  \left(\exp
  \left(
  - \frac{1}{4} \exp \left( - (1+o(1)) \left(2\intraTau m \Ir + 2\left(\frac{n}{cg}-1\right) \Ig \frac{\log n}{n} \right) \right)
  \right) \right)^{\left|\tG{1}{1}\right|\cdot \left|\tG{2}{1}\right|}
  \label{eq:2nd-eval}
  \\
  & = 
  \exp
  \left( -\frac{n^2}{4\log^6 n}
    \exp \left( - (1+o(1)) \left(2\intraTau m \Ir + 2\left(\frac{n}{cg}-1\right) \Ig \frac{\log n}{n} \right) \right)
  \right) 
  \label{eq:2nd-SetSize}\\
  & \leq 
  \exp
  \left( -\frac{n^2}{4 \log^6 n}
   \exp \left( - 2(1+o(1)) (1-\epsilon) \log n \right) \right)
   \label{eq:2d-BadCond}\\
  & \leq 
  \exp \left(-\frac{n^{2(\epsilon - o(1)(1-\epsilon))}}{4 \log^6 n }
  \right),
\end{align}
where \eqref{eq:2nd-success-0} follows from the definition in \eqref{eq:defS}; \eqref{eq:2nd-indep} holds since the events $\{\mathsf{L}\left(\M{a,b}\right) > \mathsf{L}(\gtMat) : a\in \tG{1}{1}, b\in \tG{2}{1}\}$ are mutually independent due to the fact that there are no edges among the vertices in $\tG{1}{1} \cup \tG{2}{1}$, as per Lemma~\ref{lm:randGraph}; 
\eqref{eq:2nd-eval} follows from \eqref{eq:2nd-L-diff}; \eqref{eq:2nd-SetSize} holds since $|\tG{1}{1}| =|\tG{2}{1}|=\frac{n}{\log^3 n}$; and  
\eqref{eq:2d-BadCond} follows from the condition in \eqref{eq:necCond_2}. Therefore, we obtain 
\begin{align*}
\lim_{n,m \rightarrow \infty } \mathbb{P}[S] \leq 
\lim_{n,m \rightarrow \infty } 
\exp \left( -\frac{n^{2(\epsilon - o(1)(1-\epsilon))}}{4 \log^6 n}
  \right) =0,
\end{align*}
which shows that if the condition in \eqref{eq:necCond_2} holds, then the ML estimator will fail in finding $\gtMat$ with high probability.

\subsubsection{Failure Proof for the Clustering-Limited Regime}
\label{subsec:failureProof3}
The proof follows the same structure as that presented in Section~\ref{subsec:failureProof2} where the condition in \eqref{eq:necCond_2} holds.
Without loss of generality, assume that the rating vectors of group $1$ in cluster $1$ and group $2$ in cluster $2$ have the minimum Hamming distance among distinct pairs of rating vectors across different clusters, i.e., $ \hamDist{\vecU{1}{1}}{ \vecU{2}{2}} = \interTau m$. 
Note that the corresponding groups defined by such rating vectors belong to different clusters, as opposed to the same cluster in Section~\ref{subsec:failureProof2}.
In this proof, we introduce a class of rating matrices, which are obtained by switching two users between group $1$ in cluster $1$ and group $2$ in cluster $2$. 
Then, we prove that if \eqref{eq:necCond_3} holds, then with high probability the ML estimator will fail by selecting one of the rating matrices from this class, instead of~$\gtMat$. 

Set $(x,i)=(1,1)$ and $(y,j)=(2,2)$ in Lemma~\ref{lm:randGraph}. 
Hence, there exist subsets $\tG{1}{1} \subset \grpG{1}{1}$ and $\tG{2}{2}\subset \grpG{2}{2}$ with $|\tG{1}{1}| =|\tG{2}{2}|=\frac{n}{\log^3 n}$, such that the subgraph induced by the vertices in $\tG{1}{1} \cup \tG{2}{2}$ is edge-free. 
Similar to \eqref{eq:Xab_defn} and \eqref{eq:Zab_defn} in Section~\ref{subsec:failureProof2}, define $\M{a,b} \in \mathbb{F}_{q}^{n \times m}$, for $a\in \tG{1}{1}$ and $b\in \tG{2}{2}$, as 
\begin{align}
	\M{a,b}(r,:) = \left\{
	\begin{array}{cc}
		\gtMat(b,:)=\vecU{2}{2} & \textrm{if $r=a$},\\
		\gtMat(a,:)=\vecU{1}{1} & \textrm{if $r=b$},\\
		\gtMat(r,:) & \textrm{otherwise}.
	\end{array}
	\right.
	\label{eq:Xab_defn_2}
\end{align}
The corresponding user partitioning $\mathcal{Z}_{\langle a,b\rangle}$ is given by 
\begin{align}
	\Z{a,b}(x,i)
	= 
	\left\{
	\begin{array}{cl}
		\grpG{1}{1} \cup \{b\} \setminus\{a\}
		& \textrm{if $(x,i) = (1,1)$},
		\vspace{1mm}\\
		\grpG{2}{2} \cup \{a\} \setminus\{b\} 
		& \textrm{if $(x,i) = (2,2)$},
		\vspace{0.7mm}\\
		\grpG{i}{x}
		& \textrm{otherwise}.
	\end{array}
	\right.
	\label{eq:Zab_defn_2}
\end{align}
Furthermore, given $\gtMat$ and $\M{a,b}$, we have
\begin{align}
\begin{split}
    \diffEntriesSet &= \left\{(r,t): r \in \{a,b\}, t \in \left[m\right], \: \M{a,b}(r,t) \neq \gtMat(r,t)\right\}, 
    \\
    \misClassfSet{\betaEdge}{\alphaEdge} &= \misClassfSet{\alphaEdge}{\betaEdge} = \varnothing,
    \\
    \misClassfSet{\gammaEdge}{\alphaEdge} &= 
    \left\{(a,h): 
    h\in \grpG{2}{2}\setminus\{b\} 
    \right\}
    \:\cup\:
    \left\{(b,h):
    h\in \grpG{1}{1}\setminus\{a\} 
    \right\},
    \\
    \misClassfSet{\alphaEdge}{\gammaEdge} &= \left\{(a,h): 
    h\in \grpG{1}{1}\setminus\{a\} 
    \right\}
    \:\cup\:
    \left\{(b,h):
    h\in \grpG{2}{2}\setminus\{b\} 
    \right\},
    \\
    \misClassfSet{\gammaEdge}{\betaEdge} &=
    \left\{(a,h): 
    h \in \bigcup_{i \in [g] \setminus \{2\}} \grpG{i}{2}
    \right\}
    \:\cup\:
    \left\{(b,h):
    h \in \bigcup_{i \in [g] \setminus \{1\}} \grpG{i}{1}
    \right\},
    \\
    \misClassfSet{\betaEdge}{\gammaEdge} &= \left\{(a,h): 
    h \in \bigcup_{i \in [g] \setminus \{1\}} \grpG{i}{1}
    \right\}
    \:\cup\:
    \left\{(b,h):
    h \in \bigcup_{i \in [g] \setminus \{2\}} \grpG{i}{2}
    \right\},
\end{split}
\label{eq:failure3_pairs}
\end{align}
according to their definitions in \eqref{eq:N1_defn}--\eqref{eq:N4rev_defn}.
Thus, the cardinalities of the sets in \eqref{eq:failure3_pairs} are given by
\begin{align}
\begin{split}
    \left\vert\diffEntriesSet\right\vert 
    &= \hamDist{\M{a,b}(a,:)}{\gtMat(a,:)}
    + 
    \hamDist{\M{a,b}(b,:)}{\gtMat(b,:)}
    \\
    &= 
    \hamDist{\gtMat(b,:)}{\gtMat(a,:)}
    + 
    \hamDist{\gtMat(a,:)}{\gtMat(b,:)}
    \\
    &=
    2\interTau m,
    \\
    \left\vert \misClassfSet{\betaEdge}{\alphaEdge} \right\vert &= \left\vert \misClassfSet{\alphaEdge}{\gammaEdge} \right\vert = 0,
    \\
    \left\vert \misClassfSet{\gammaEdge}{\alphaEdge} \right\vert &= \left\vert \misClassfSet{\alphaEdge}{\gammaEdge} \right\vert = 2 \left(\frac{n}{cg}-1\right),
    \\
    \left\vert \misClassfSet{\gammaEdge}{\betaEdge} \right\vert &= \left\vert \misClassfSet{\betaEdge}{\gammaEdge} \right\vert = 2 \left(\frac{g-1}{gc}\right) n.
\end{split}
\label{eq:failure3_sizePair}
\end{align}

For each $\M{a,b}$ where $a\in \tG{1}{1}$ and $b\in \tG{2}{2}$, we have 
\begin{align}
	\mathsf{L}(\gtMat) - \mathsf{L}\left(\M{a,b}\right) 
	&= 
	\log \left((q-1)\frac{1-\theta}{\theta}\right) 
	\sum_{i \in \diffEntriesSet}
	\mathsf{B}_i^{(p)} \left(\left(1+\mathsf{B}_i^{\left(\frac{1}{q-1}\right)}\right)\mathsf{B}_i^{(\theta)}-1\right)
	\nonumber\\
	&\phantom{=}
	+ \left(\log\frac{(1-\gammaEdge)\alphaEdge}{(1-\alphaEdge)\gammaEdge}\right)
	\sum_{i \in \misClassfSet{\gammaEdge}{\alphaEdge}} \left(\mathsf{B}_{i}^{(\gammaEdge)} - \mathsf{B}_{i}^{(\alphaEdge)}\right) 
	+ \left(\log\frac{(1-\gammaEdge)\betaEdge}{(1-\betaEdge)\gammaEdge}\right)
	\sum_{i \in \misClassfSet{\gammaEdge}{\betaEdge}} \left(\mathsf{B}_{i}^{(\gammaEdge)} - \mathsf{B}_{i}^{(\betaEdge)}\right) 
	\label{eq:c1_diff_0}\\
	&= 
	\log \left((q-1)\frac{1-\theta}{\theta}\right) 
	\sum_{i = 1}^{2\interTau m}
	\mathsf{B}_i^{(p)} \left(\left(1+\mathsf{B}_i^{\left(\frac{1}{q-1}\right)}\right)\mathsf{B}_i^{(\theta)}-1\right)
	\nonumber\\
	&\phantom{=}
	+ \left(\log\frac{(1-\gammaEdge)\alphaEdge}{(1-\alphaEdge)\gammaEdge}\right)
	\sum_{i = 1}^{2 \left(\frac{n}{cg}-1\right)} \left(\mathsf{B}_{i}^{(\gammaEdge)} - \mathsf{B}_{i}^{(\alphaEdge)}\right) 
	+ \left(\log\frac{(1-\gammaEdge)\betaEdge}{(1-\betaEdge)\gammaEdge}\right)
	\sum_{i = 1}^{2 \left(\frac{g-1}{gc}\right) n} \left(\mathsf{B}_{i}^{(\gammaEdge)} - \mathsf{B}_{i}^{(\betaEdge)}\right),
	\label{eq:c1_diff_1}
\end{align}
where \eqref{eq:c1_diff_0} follows from Lemma~\ref{lemma:neg_lik_M0_XT} and \eqref{eq:failure3_pairs}; and \eqref{eq:c1_diff_1} follows from \eqref{eq:failure3_sizePair}.
Therefore, the probability that the negative log-likelihood of $\M{a,b}$ is greater that that of $\gtMat$ is upper bounded by
\begin{align}
	\mathbb{P} \left[
	\mathsf{L}\left(\M{a,b}\right) > \mathsf{L}(\gtMat)\right]
	&=
	1 - \mathbb{P}\left[ 
	B\left(2\interTau m,\: 0,\: 2 \left(\frac{n}{cg}-1\right),\: 2 \left(\frac{g-1}{gc}\right) n \right)
	\geq 0
	\right] 
	\nonumber
	\\
	&\leq
	1 - \frac{1}{4} \exp \left( - (1+o(1)) \left(2\interTau m \Ir + 2\left(\frac{n}{cg}-1\right) \IcOne \frac{\log n}{n} + 2 \left(\frac{g-1}{gc}\right) n \IcTwo \frac{\log n}{n} \right) \right)
	\label{eq:3rd-L-diff}\\
	&\leq
	\exp
	\left(
	- \frac{1}{4} \exp \left( - (1+o(1)) \left(2\interTau m \Ir + 2\left(\frac{n}{cg}-1\right) \IcOne \frac{\log n}{n} + 2 \left(\frac{g-1}{gc}\right) n \IcTwo \frac{\log n}{n} \right) \right)
	\right),
\end{align}
where \eqref{eq:3rd-L-diff} follows from \eqref{eq:c1_diff_1} and Lemma~\ref{lm:lowerB_prob}.

Finally, the probability of exact rating matrix recovery is upper bounded by 
\begin{align}
	\mathbb{P}[S] 
	&\leq
	\mathbb{P} \left[\bigcap_{\substack{a\in \tG{1}{1} \\ b\in \tG{2}{2}}}
    \left(\mathsf{L}\left(\M{a,b}\right) > \mathsf{L}(\gtMat)\right)\right]
    \label{eq:3rd-success-0}\\
	&= \prod_{\substack{a\in \tG{1}{1} \\ b\in \tG{2}{2}}} \mathbb{P} \left[
	\mathsf{L}\left(\M{a,b}\right) > \mathsf{L}(\gtMat)\right]\label{eq:3rd-indep}\\
	&\leq   \left(\exp
	\left(
	- \frac{1}{4} \exp \left( - (1+o(1)) \left(2\interTau m \Ir + 2\left(\frac{n}{cg}-1\right) \IcOne \frac{\log n}{n} + 2\left(\frac{g-1}{gc}\right) n \IcTwo \frac{\log n}{n} \right) \right)
	\right)\right)^{\left|\tG{1}{1}\right|\cdot \left|\tG{2}{2}\right|}
	\label{eq:3rd-eval}
	\\
	& \leq 
	\exp
	\left( -\frac{n^2}{4 \log^6 n}
	\exp \left( - 2(1+o(1)) (1-\epsilon) \log n \right) \right)
	\label{eq:3rd-BadCond}\\
	& \leq 
	\exp \left(-\frac{n^{2(\epsilon - o(1)(1-\epsilon))}}{4 \log^6 n}
  \right),
\end{align}
where \eqref{eq:3rd-success-0} follows from the definition in \eqref{eq:defS}; \eqref{eq:3rd-indep} holds since the events $\{\mathsf{L}\left(\M{a,b}\right) > \mathsf{L}(\gtMat) : a\in \tG{1}{1}, b\in \tG{2}{2}\}$ are mutually independent due to the fact that there are no edges among the vertices in $\tG{1}{1} \cup \tG{2}{2}$, as per Lemma~\ref{lm:randGraph}; \eqref{eq:3rd-eval} follows from \eqref{eq:3rd-L-diff}; and \eqref{eq:3rd-BadCond} follows from the condition in \eqref{eq:necCond_3}, and $|\tG{1}{1}| =|\tG{2}{1}|=\frac{n}{\log^3 n}$. 
Thus, we obtain
\begin{align*}
\lim_{n,m\rightarrow \infty } \mathbb{P}[S] 
= \lim_{n,m\rightarrow \infty } \exp \left(-\frac{n^{2(\epsilon - o(1)(1-\epsilon))}}{4 \log^6 n}
  \right)
=0,
\end{align*} 
which shows that if the condition in \eqref{eq:necCond_3} holds, then the ML estimator will fail in finding $\gtMat$ with high probability.

Since $\lim_{n,m \rightarrow \infty} \mathbb{P} \left[S\right] = 0$ is proved under each of the three conditions stated in \eqref{eq:necCond_1}, \eqref{eq:necCond_2} and \eqref{eq:necCond_3}, the converse proof of Theorem~\ref{thm:p_star} is concluded.
\hfill $\blacksquare$

\section{Simulation Results}
\label{sec:expResults}

We conduct Monte Carlo experiments\footnote{The proposed achievable scheme is based on maximum likelihood estimation whose computational complexity increases dramatically with the problem size (i.e., as $n$ and $m$ increase). Therefore, we use the computationally efficient matrix completion algorithm, proposed in \cite{elmahdy2020matrix}.} to corroborate Theorem~\ref{thm:p_star}. 
We consider a problem setting in which we have $c=2$ clusters, $g=3$ groups per cluster, finite field of order $q=2$, and $r=2$ basis vectors per group. 
The MDS code structure is given $\vecU{3}{x} = \vecU{1}{x} + \vecU{2}{x}$ for $x\in[2]$. 
Furthermore, the parameters of observation noise, graph and rating vectors are set to $\theta=0.1$, $(\beta,\gamma)=(10,0.5)$ and $(\intraTau,\interTau)=(0.5,0.5)$, respectively.
Finally, the synthetic data is generated as per the model in Section~\ref{subsec:probForm}.

\begin{figure*}
    \centering
    \subfloat[]{\includegraphics[width=0.45\textwidth]{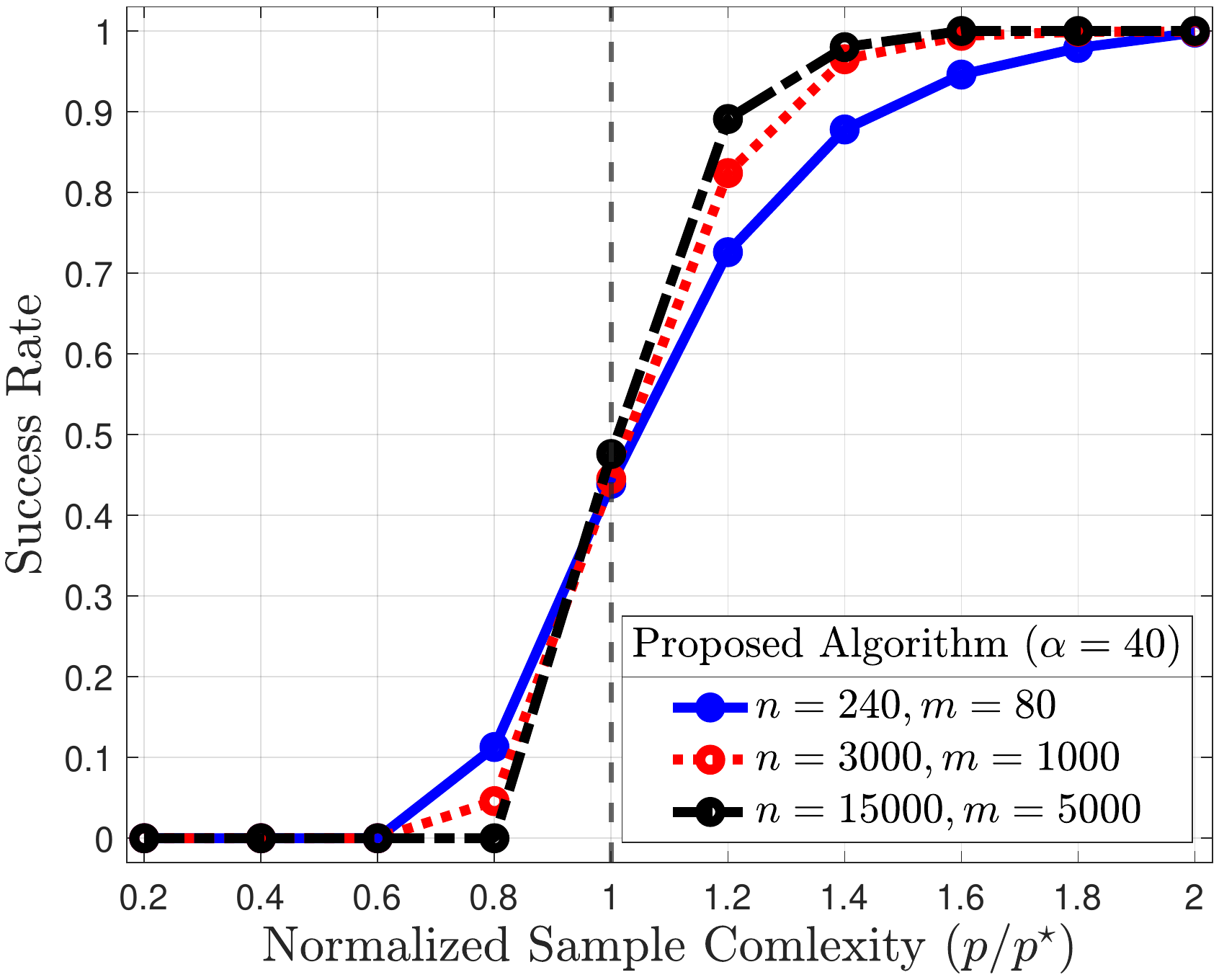}
    \label{fig:synth_alph40}}
    \hfill
    \subfloat[]{\includegraphics[width=0.45\textwidth]{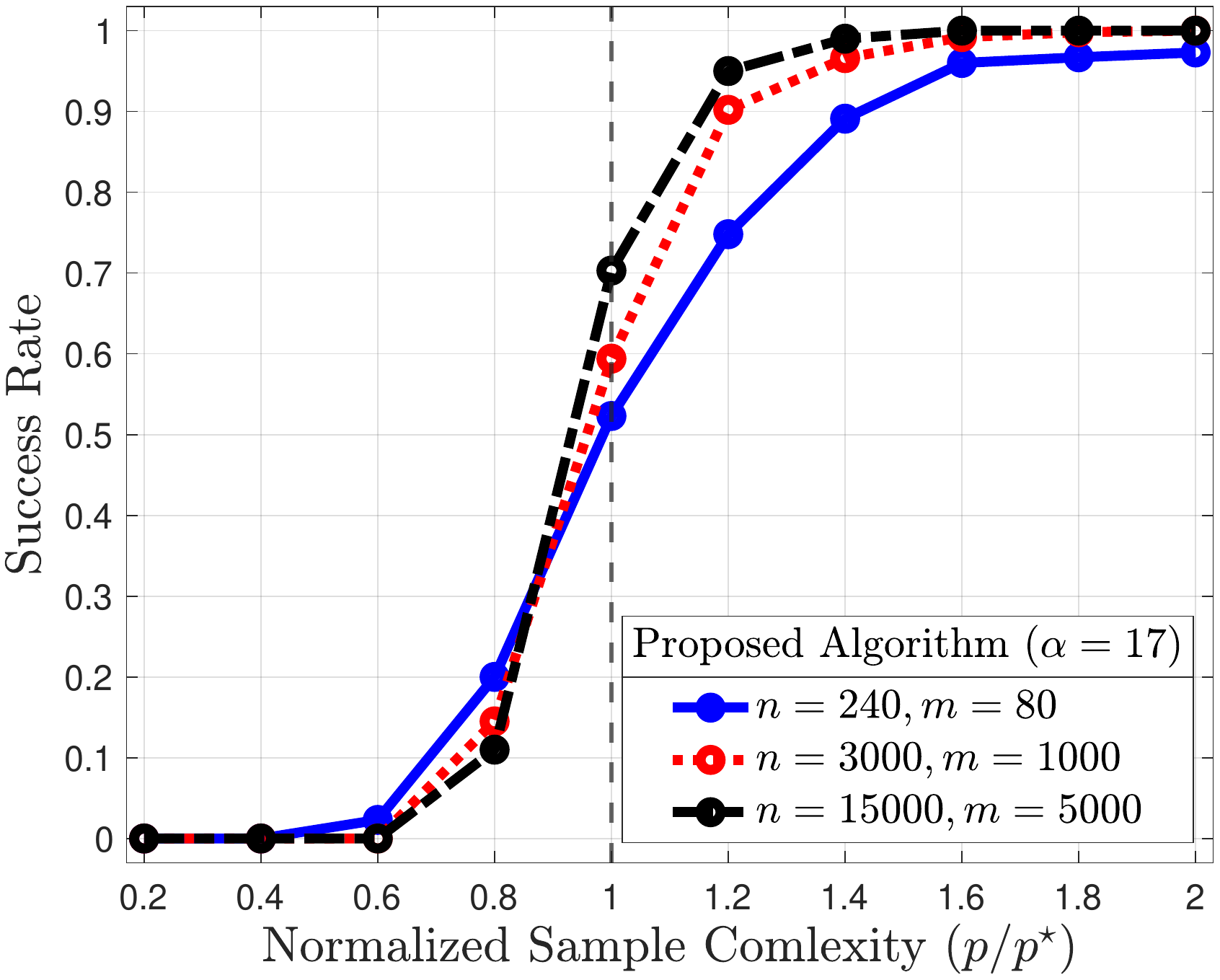}
    \label{fig:synth_alph17}}
    \hfill
    \caption{
    The success rate of the proposed algorithm as a function of $p/p^\star$ for under different values of $n$, $m$ and $\alpha$.
    The considered problem setting is characterized by $(c,g,q,r)=(2,3,2,2)$, $\vecU{3}{x} = \vecU{1}{x} + \vecU{2}{x}$ for $x\in[2]$, $\theta=0.1$, $(\beta,\gamma)=(10,0.5)$, and $(\intraTau,\interTau)=(0.5,0.5)$.
    We study two cases: (a) perfect clustering/grouping regime where $\alpha = 40$; and (b) grouping-limited regime where $\alpha = 17$. 
    }
\end{figure*}

In Figs. \ref{fig:synth_alph40} and \ref{fig:synth_alph17}, we evaluate the performance of the proposed algorithm, and quantify the empirical success rate as a function of the normalized sample complexity over $10^3$~randomly drawn realizations of rating vectors and hierarchical graphs. We vary $n$ and $m$, preserving the ratio $n/m=3$. 
Fig.~\ref{fig:synth_alph40} depicts the case of $\alpha = 40$ which corresponds to perfect clustering/grouping regime, while Fig.~\ref{fig:synth_alph17} illustrates the case of $\alpha = 17$ which corresponds to grouping-limited regime.
In both figures, we observe a phase transition\footnote{The transition is ideally a step function at $p = p^\star$ as $n$ and $m$ tend to infinity.} in the success rate at $p = p^\star$, and the phase transition gets sharper as $n$ and $m$ increase. Figs. \ref{fig:synth_alph40} and \ref{fig:synth_alph17} corroborate Theorem~\ref{thm:p_star} in different regimes when the graph side information is not scarce.

\section{Conclusion}
\label{sec:conclusion}
In this paper, we consider a rating matrix that consists of $n$ users and $m$ items, and a hierarchical similarity graph that consists of $c$ disjoint clusters, and each cluster comprises $g$ disjoint groups. The rating vectors of the groups of a given cluster are different, but related to each other through a linear subspace of $r$ basis vectors.
We characterize the optimal sample complexity to jointly recover the hierarchical structure of the similarity graph as well as the rating matrix entries. 
We propose a matrix completion algorithm that is based on the maximum likelihood estimation, and achieve the characterized sample complexity. 
The optimality of the proposed achievable scheme was demonstrated through a matching converse proof. 
We demonstrate that the optimal sample complexity hinge on the quality of side information of the hierarchical similarity graph. We also highlighted the fact that leveraging the graph side information enables us to achieve significant gain in sample complexity, compared to existing schemes that identifies different groups without taking into consideration the hierarchical structure across them. 

One potential research direction is to develop a computationally efficient algorithm to achieve the sharp threshold on the optimal sample complexity characterized in this paper.  
Another research direction is to characterize the optimal sample complexity for a more general case of $c$ clusters, each of which comprise arbitrary number of groups of possibly different number of users.

\bibliographystyle{IEEEtran}
\bibliography{myRef_IT}

\appendices
\section{Proof of Lemma~\ref{lm:neg_log_likelihood}}
\label{app:neg_log_likelihood}
From the definition in \eqref{eq:likelihood_defn}, the negative log-likelihood of a candidate rating matrix $X=(\cV,\mathcal{Z})$, for $X \in \matSet$, given a fixed input pair $(Y,\mathcal{G})$ can be written as
\begin{align}
    \mathsf{L}(X) 
    &= 
    - \log \mathbb{P}\left[(Y,\mathcal{G}) \given \mathbf{X} = X \right]
    \nonumber\\
    &=
    - \log \left(\mathbb{P}\left[Y \given \mathbf{X} = X \right]  
    \:\mathbb{P} \left[\mathcal{G} \given \mathbf{X} = X \right]\right)
    \nonumber\\
    &=
    - \log \mathbb{P}\left[Y \given \mathbf{X} = X \right] 
    - \log \mathbb{P}\left[\mathcal{G} \given \mathbf{X} = X \right],
\end{align}
where
\begin{align}
	\mathbb{P}\left[Y \given \mathbf{X} = X \right] 
	&= 
	p^{|\Omega|}\:
	(1-p)^{nm-|\Omega|}
	\left(\frac{\theta}{q-1}\right)^{\numDiffElmnt{Y}{X}}
	(1-\theta)^{|\Omega| - \numDiffElmnt{Y}{X}}, 
	\\
	\mathbb{P}\left[\mathcal{G} \given \mathbf{X} = X \right]
	&=
	\alphaEdge^{\numEdge{\alpha}{\mathcal{G}}{Z}} 
	(1-\alphaEdge)^{ \vert \userPairSet{\alpha}{\mathcal{Z}} \vert - \numEdge{\alpha}{\mathcal{G}}{Z}} \:
	\betaEdge^{\numEdge{\beta}{\mathcal{G}}{Z}} 
	(1-\betaEdge)^{\vert \userPairSet{\beta}{\mathcal{Z}} \vert - \numEdge{\beta}{\mathcal{G}}{Z}} \:
	\gammaEdge^{\numEdge{\gamma}{\mathcal{G}}{Z}} 
	(1-\gammaEdge)^{ \vert \userPairSet{\gamma}{\mathcal{Z}} \vert - \numEdge{\gamma}{\mathcal{G}}{Z}}.
\end{align}
Consequently, $\mathsf{L}(X)$ is given by
\begin{align}
	\mathsf{L}(X) 
	=  
	\log \left((q-1) \frac{1-\theta}{\theta}\right) \numDiffElmnt{Y}{X} 
	+
	\sum\limits_{\mu\in\left\{\alphaEdge,\betaEdge,\gammaEdge\right\}}
    \left(
    \log \left(\frac{1-\mu}{\mu}\right) \numEdge{\mu}{\mathcal{G}}{\mathcal{Z}}
    -
    \log(1-\mu) \left\vert \userPairSet{\mu}{\mathcal{Z}} \right\vert 
    \right).
	\label{eq:neg_log_lik}
\end{align}
This completes the proof of Lemma~\ref{lm:neg_log_likelihood}.
\hfill $\blacksquare$

\section{Proof of Lemma~\ref{lm:upp_worst_case_err_prob}}
\label{app:upp_worst_case_err_prob}
The worst-case probability of error $P_e^{(\diff)}(\psi_{\text{ML}})$ for the maximum likelihood estimator $\psi_{\text{ML}}$ is upper bounded by
\begin{align}
    P_e^{(\diff)}(\psi_{\text{ML}})
    & = 
    \max_{M \in \matSet } 
    \mathbb{P}\left[\psi_{\text{ML}}(Y, \mathcal{G}) \neq M \right]
    \nonumber\\
    & =
    \mathbb{P}\left[\psi_{\text{ML}}(Y, \mathcal{G}) \neq \gtMat \:\vert\: \mathbf{M} = \gtMat \right]
    \label{eq:achv_lm_probErr_upper_1}\\
    & =
    \mathbb{P}\left[
    \bigcup_{X \neq \gtMat}
    \mathsf{L}(X) \leq \mathsf{L}(\gtMat)  
    \right]
    \label{eq:achv_lm_probErr_upper_2}\\
    & \leq
    \sum_{X \neq \gtMat}
    \mathbb{P}\left[
    \mathsf{L}(X) \leq \mathsf{L}(\gtMat)  
    \right],
    \label{eq:achv_lm_probErr_upper_3}
\end{align}
where \eqref{eq:achv_lm_probErr_upper_1} holds since $\gtMat \in \matSet$ by the construction of $\gtMat$ presented in Section~\ref{sec:achv}, and the error event $\left\{\psi_{\text{ML}}(Y, \mathcal{G}) \neq M\right\}$ is statistically identical over all $M \in \matSet$; \eqref{eq:achv_lm_probErr_upper_2} follows from the fact that the output of the maximum likelihood estimator is different from the ground truth rating matrix $\gtMat$ only if there exists a candidate rating matrix $X$ whose negative log-likelihood 
is less than or equal to that of $\gtMat$; and \eqref{eq:achv_lm_probErr_upper_3} follows from the union bound.
This completes the proof of Lemma~\ref{lm:upp_worst_case_err_prob}.
\hfill $\blacksquare$

\section{Proof of Lemma~\ref{lemma:neg_lik_M0_XT}}
\label{proof:neg_lik_M0_XT}
By Lemma~\ref{lm:neg_log_likelihood}, the LHS of \eqref{eq:lemma3} can be written as
\begin{align}
  \mathsf{L}\left(\gtMat\right) - \mathsf{L}(X)
  &=
  \log\left(\!(q-1)\frac{1-\theta}{\theta}\right) 
  \underbrace{\left(\numDiffElmnt{Y}{\gtMat} - \numDiffElmnt{Y}{X}\right)}_{\mathsf{Term_1}}
  +\!\! 
  \underbrace{
  \sum\limits_{\mu\in\left\{\alphaEdge,\betaEdge,\gammaEdge\right\}}
  \!\!\left(\!
  \log \left(\frac{1-\mu}{\mu}\right)
  \left(
  \numEdge{\mu}{\mathcal{G}}{\partitionSet_0}
  - \numEdge{\mu}{\mathcal{G}}{\partitionSet}
  \right)\!
  \right)}_{\mathsf{Term_2}}
  \nonumber\\ 
  &\phantom{=}
  + 
  \underbrace{
  \sum\limits_{\mu \in \left\{\alphaEdge,\betaEdge,\gammaEdge\right\}}
  \left(
  \log(1-\mu) 
  \left(
  \left\vert \userPairSet{\mu}{\partitionSet} \right\vert
  -
  \left\vert  \userPairSet{\mu}{\partitionSet_0} \right\vert
  \right)
  \right)}_{\mathsf{Term_3}}.
  \label{eq:LX_LM0}
\end{align}
In what follows, we evaluate each of the three terms in \eqref{eq:LX_LM0}.

Recall from Section~\ref{subsec:notation} that $\numDiffElmnt{A}{B}$ denotes the number of different entries between matrices $A_{n \times m}$ and $B_{n \times m}$.
Therefore, $\mathsf{Term_1}$ can be expanded as
\begin{align}
    \mathsf{Term_1}
    &= 
	\numDiffElmnt{Y}{\gtMat} - \numDiffElmnt{Y}{X}
	\nonumber\\
	&= 
	\sum_{(r,t)\in \Omega} 
	\left(\indicatorFn{Y(r,t) \neq \gtMat(r,t)}\right)
	- \sum_{(r,t)\in \Omega} 
	\left(\indicatorFn{Y(r,t) \neq X(r,t)}\right)
	\nonumber\\ 
	&=
	\sum_{(r,t)\in \Omega} 
	\left(nm - \indicatorFn{Y(r,t) = \gtMat(r,t)} \right)
	- \left(nm - \indicatorFn{Y(r,t) = X(r,t)} \right)
	\nonumber\\
	&= 
	\sum_{\substack{(r,t) \in \Omega : \\ X(r,t) \neq \gtMat(r,t)}} 
	\indicatorFn{Y(r,t) = X(r,t)} 
	- \indicatorFn{Y(r,t) = \gtMat(r,t)}
	\label{eq:phiX_0}\\ 
	&= \sum_{i \in \lvert\left\{(r,t)\in [n] \times [m] \::\: X(r,t) \neq \gtMat(r,t)\right\}\rvert} 
	\mathsf{B}^{(p)}_i \mathsf{B}_i^{\left(\frac{\theta}{q-1}\right)} 
	- \mathsf{B}^{(p)}_i\left(1-\mathsf{B}_i^{(\theta)}\right)
	\label{eq:phiX_00}\\ 
	&=
	\sum_{i \in \lvert\left\{(r,t)\in [n] \times [m] \::\: X(r,t) \neq \gtMat(r,t)\right\}\rvert}
	\mathsf{B}^{(p)}_i 
	\left(
	\mathsf{B}_i^{\left(\theta\right)} \mathsf{B}_i^{\left(\frac{1}{q-1}\right)}
	- \left(1-\mathsf{B}_i^{(\theta)}\right)
	\right)
	\label{eq:phiX_1}\\
	&= \sum_{i \in \diffEntriesSet}
	\mathsf{B}_i^{(p)}
	\left(\left(1+\mathsf{B}_i^{\left(\frac{1}{q-1}\right)}\right)\mathsf{B}_i^{(\theta)} - 1\right), 
	\label{eq:phiX_3}
\end{align}
where \eqref{eq:phiX_0} follows since $\indicatorFn{Y(i,j) = X(i,j)} = \indicatorFn{Y(i,j) = \gtMat(i,j)}$ if $X(i,j) = \gtMat(i,j)$; the first term of each summand in \eqref{eq:phiX_00} follows since the probability that the observed rating matrix entry is $X(i,j)$, which is not equal to $\gtMat(i,j)$, is $p(\theta/(q-1))$, while the second term of each summand in \eqref{eq:phiX_00} follows since the probability that the observed rating matrix entry is $\gtMat(i,j)$ is $p(1-\theta)$, for every $(i,j)\in [n] \times [m]$; \eqref{eq:phiX_1} follows since $\mathsf{B}_l^{\left(\theta/(q-1)\right)} = \mathsf{B}_l^{\left(\theta\right)} \mathsf{B}_l^{\left(1/(q-1)\right)}$; and finally \eqref{eq:phiX_3} follows from \eqref{eq:N1_defn}.

Next, we evaluate $\mathsf{Term_2}$. 
We first evaluate the quantity $\numEdge{\alpha}{\mathcal{G}}{\partitionSet_0} - \numEdge{\alpha}{\mathcal{G}}{Z}$ as
\begin{align}
    &\numEdge{\alpha}{\mathcal{G}}{\partitionSet_0} - \numEdge{\alpha}{\mathcal{G}}{Z} 
    \nonumber\\
    &=
     \left\vert
    \left\{
    (a,b) \in \mathcal{E}: 
    a \in Z_0(x,i) \cap Z(y,j_1),\: b \in Z_0(x,i) \cap Z(y,j_2),
    \mbox{ for } x, y \in [c],\: i, j_1, j_2 \in [g],\: j_1 \neq j_2
    \right\}
    \right\vert
    \nonumber\\
    &\phantom{=} +
    \left\vert
    \left\{
    (a,b) \in \mathcal{E}: 
    a \in Z_0(x,i) \cap Z(y_1,j_1),\: b \in Z_0(x,i) \cap Z(y_2, j_2),
    \mbox{ for } x, y \in [c],\: y_1 \neq y_2, \:
    i, j_1, j_2 \in [g]
    \right\}
    \right\vert
    \nonumber\\
    &\phantom{=}
    -\left\vert
    \left\{
    (a,b) \in \mathcal{E}: 
    a \in Z_0(x,i_1) \cap Z(y,j),\: b \in Z_0(x,i_2) \cap Z(y,j),
    \mbox{ for } x, y \in [c],\: i_1, i_2, j \in [g],\: i_1 \neq i_2
    \right\}
    \right\vert
    \nonumber\\
    &\phantom{=} - 
    \left\vert
    \left\{
    (a,b) \in \mathcal{E}: 
    a \in Z_0(x_1,i_1) \cap Z(y,j),\: b \in Z_0(x_2,i_2) \cap Z(y,j),
    \mbox{ for } x_1, x_2, y \in [c],\: x_1 \neq x_2, \:
    i_1, i_2, j \in [g]
    \right\}
    \right\vert
    \label{eq:diff_edge0}\\
    &=
    \sum\limits_{i=1}^{\misClassfSet{\alphaEdge}{\betaEdge}} \mathsf{B}_i^{(\alphaEdge)}
    +\sum\limits_{i=1}^{\misClassfSet{\alphaEdge}{\gammaEdge}} \mathsf{B}_i^{(\alphaEdge)}
    -\sum\limits_{i=1}^{\misClassfSet{\betaEdge}{\alphaEdge}} \mathsf{B}_i^{(\betaEdge)} 
    - \sum\limits_{i=1}^{\misClassfSet{\gammaEdge}{\alphaEdge}} \mathsf{B}_i^{(\gammaEdge)},
    \label{eq:diff_edge1}
\end{align}
where \eqref{eq:diff_edge0} holds since the edges that remain after the subtraction are: (i) edges that exist in the same group in $\partitionSet_0$, but are estimated to be in different groups within the same cluster in $\partitionSet$; (ii) edges that exist in the same group in $\partitionSet_0$, but are estimated to be in different clusters in $\partitionSet$; (iii) edges that exist in different groups within the same cluster in $\partitionSet_0$, but are estimated to be in the same group in $\partitionSet$; (iv) edges that exist in different clusters in $\partitionSet_0$, but are estimated to be in the same group in $\partitionSet$; and finally \eqref{eq:diff_edge1} follows from \eqref{eq:N2_defn}--\eqref{eq:N3rev_defn}.
In a similar way, one can evaluate the following quantities:
\begin{align}
    \numEdge{\betaEdge}{\mathcal{G}}{\partitionSet_0} - \numEdge{\betaEdge}{\mathcal{G}}{\partitionSet}
    &=
    \sum\limits_{i=1}^{\misClassfSet{\betaEdge}{\alphaEdge}} \mathsf{B}_i^{(\betaEdge)}
    +\sum\limits_{i=1}^{\misClassfSet{\betaEdge}{\gammaEdge}} \mathsf{B}_i^{(\betaEdge)}
    -\sum\limits_{i=1}^{\misClassfSet{\alphaEdge}{\betaEdge}} \mathsf{B}_i^{(\alphaEdge)} 
    - \sum\limits_{i=1}^{\misClassfSet{\gammaEdge}{\betaEdge}} \mathsf{B}_i^{(\gammaEdge)},
    \label{eq:diff_edge2}\\
    \numEdge{\gammaEdge}{\mathcal{G}}{\partitionSet_0} - \numEdge{\gammaEdge}{\mathcal{G}}{\partitionSet}
    &=
    \sum\limits_{i=1}^{\misClassfSet{\gammaEdge}{\alphaEdge}} \mathsf{B}_i^{(\gammaEdge)}
    +\sum\limits_{i=1}^{\misClassfSet{\gammaEdge}{\betaEdge}} \mathsf{B}_i^{(\gammaEdge)}
    -\sum\limits_{i=1}^{\misClassfSet{\alphaEdge}{\gammaEdge}} \mathsf{B}_i^{(\alphaEdge)} 
    - \sum\limits_{i=1}^{\misClassfSet{\betaEdge}{\gammaEdge}} \mathsf{B}_i^{(\betaEdge)}.
    \label{eq:diff_edge3}
\end{align}
Consequently, $\mathsf{Term_2}$ can be written as
\begin{align}
    \mathsf{Term_2} 
    &=
    \sum\limits_{\mu\in{\alphaEdge,\betaEdge,\gammaEdge}}
    \left(
    \log \left(\frac{1-\mu}{\mu}\right)
    \left(
    \numEdge{\mu}{\mathcal{G}}{\partitionSet_0}
    - \numEdge{\mu}{\mathcal{G}}{\partitionSet}
    \right)
    \right)
    \nonumber\\
    &=
    \log \left(\frac{1-\alphaEdge}{\alphaEdge}\right)
    \left(
	\sum_{i \in \misClassfSet{\alphaEdge}{\betaEdge}} 
	\mathsf{B}_i^{(\alphaEdge)}
	+ \sum_{i \in \misClassfSet{\alphaEdge}{\gammaEdge}} 
	\mathsf{B}_i^{(\alphaEdge)}
	- \sum_{i \in \misClassfSet{\betaEdge}{\alphaEdge}} 
	\mathsf{B}_i^{(\betaEdge)}
	- \sum_{i \in \misClassfSet{\gammaEdge}{\alphaEdge}} 
	\mathsf{B}_i^{(\gammaEdge)}
    \right)
    \nonumber\\
    &\phantom{=}
    + \log \left(\frac{1-\betaEdge}{\betaEdge}\right)
    \left(
	\sum_{i \in \misClassfSet{\betaEdge}{\alphaEdge}} 
	\mathsf{B}_i^{(\betaEdge)}
	+ \sum_{i \in \misClassfSet{\betaEdge}{\gammaEdge}} 
	\mathsf{B}_i^{(\betaEdge)}
    - \sum_{i \in \misClassfSet{\alphaEdge}{\betaEdge}} 
	\mathsf{B}_i^{(\alphaEdge)}
	- \sum_{i \in \misClassfSet{\gammaEdge}{\betaEdge}} 
	\mathsf{B}_i^{(\gammaEdge)}
    \right)
    \nonumber\\
    &\phantom{=}
    + \log \left(\frac{1-\gammaEdge}{\gammaEdge}\right)
    \left(
	\sum_{i \in \misClassfSet{\gammaEdge}{\alphaEdge}} 
	\mathsf{B}_i^{(\gammaEdge)}
	+ \sum_{i \in \misClassfSet{\gammaEdge}{\betaEdge}} 
	\mathsf{B}_i^{(\gammaEdge)}
    - \sum_{i \in \misClassfSet{\alphaEdge}{\gammaEdge}} 
	\mathsf{B}_i^{(\alphaEdge)}
	- \sum_{i \in \misClassfSet{\betaEdge}{\gammaEdge}} 
	\mathsf{B}_i^{(\betaEdge)}
    \right)
    \label{eq:term2_L(x)_0}\\
    &=
    \left(\log\frac{(1-\betaEdge)\alphaEdge}{(1-\alphaEdge)\betaEdge}\right)
    \left(
    \sum\limits_{i \in \misClassfSet{\betaEdge}{\alphaEdge}} \mathsf{B}_i^{(\betaEdge)}
    - \sum\limits_{i \in \misClassfSet{\alphaEdge}{\betaEdge}} \mathsf{B}_i^{(\alphaEdge)} 
    \right)
    + 
    \left(\log\frac{(1-\gammaEdge)\alphaEdge}{(1-\alphaEdge)\gammaEdge}\right)
    \left(
    \sum\limits_{i \in \misClassfSet{\gammaEdge}{\alphaEdge}} \mathsf{B}_i^{(\gammaEdge)}
    - \sum\limits_{i \in \misClassfSet{\alphaEdge}{\gammaEdge}} \mathsf{B}_i^{(\alphaEdge)} 
    \right)
    \nonumber\\
    &\phantom{=}
    +
    \left(\log\frac{(1-\gammaEdge)\betaEdge}{(1-\betaEdge)\gammaEdge}\right)
     \left(
    \sum\limits_{i \in \misClassfSet{\gammaEdge}{\betaEdge}} \mathsf{B}_i^{(\gammaEdge)}
    - \sum\limits_{i \in \misClassfSet{\betaEdge}{\gammaEdge}} \mathsf{B}_i^{(\betaEdge)} 
   \right),
   \label{eq:term2_L(x)_1}
\end{align}
where \eqref{eq:term2_L(x)_0} follows from \eqref{eq:diff_edge1}--\eqref{eq:diff_edge3}.

Finally, $\mathsf{Term_3}$ is evaluated as follows:
\begin{align}
    \mathsf{Term_3} 
    &=
    \sum\limits_{\mu \in \left\{\alphaEdge,\betaEdge,\gammaEdge\right\}}
    \left(
    \log(1-\mu) 
    \left(
    \left\vert \userPairSet{\mu}{\partitionSet} \right\vert
    -
    \left\vert  \userPairSet{\mu}{\partitionSet_0} \right\vert
    \right)
    \right)
    \nonumber\\
    &=
    \sum\limits_{\mu \in \left\{\alphaEdge,\betaEdge,\gammaEdge\right\}}
    \left(
    \log(1-\mu) 
    \left\vert 
    \bigcup_{\nu \in \left\{\alphaEdge, \betaEdge, \gammaEdge\right\}} \misClassfSet{\nu}{\mu}
    \right\vert
    \right)
    -
    \sum\limits_{\mu \in \left\{\alphaEdge,\betaEdge,\gammaEdge\right\}}
    \left(
    \log(1-\mu)
    \left\vert  
    \bigcup_{\nu \in \left\{\alphaEdge, \betaEdge, \gammaEdge\right\}} \misClassfSet{\mu}{\nu}
    \right\vert
    \right)
    \label{eq:term3_L(x)_0}\\
    &=
    \sum\limits_{\mu \in \left\{\alphaEdge,\betaEdge,\gammaEdge\right\}}
    \left(
    \log(1-\mu) 
    \left( 
    \left\vert\misClassfSet{\alphaEdge}{\mu}\right\vert + \left\vert\misClassfSet{\betaEdge}{\mu}\right\vert + \left\vert\misClassfSet{\gammaEdge}{\mu}\right\vert
    \right)
    \right)
    -
    \sum\limits_{\mu \in \left\{\alphaEdge,\betaEdge,\gammaEdge\right\}}
    \left(
    \log(1-\mu)
    \left( 
    \left\vert\misClassfSet{\mu}{\alphaEdge}\right\vert + \left\vert\misClassfSet{\mu}{\betaEdge}\right\vert + \left\vert\misClassfSet{\mu}{\gammaEdge}\right\vert
    \right)
    \right)
    \label{eq:term3_L(x)_1}\\
    &=
    \log(1-\alphaEdge) 
    \left( 
    \left\vert\misClassfSet{\betaEdge}{\alphaEdge}\right\vert + \left\vert\misClassfSet{\gammaEdge}{\alphaEdge}\right\vert
    \right)
    -
    \log(1-\alphaEdge)
    \left( 
    \left\vert\misClassfSet{\alphaEdge}{\betaEdge}\right\vert + \left\vert\misClassfSet{\alphaEdge}{\gammaEdge}\right\vert
    \right)
    + \log(1-\betaEdge) 
    \left( 
    \left\vert\misClassfSet{\alphaEdge}{\betaEdge}\right\vert +  \left\vert\misClassfSet{\gammaEdge}{\betaEdge}\right\vert
    \right)
    \nonumber\\
    &\phantom{=}
    -
    \log(1-\betaEdge)
    \left( 
    \left\vert\misClassfSet{\betaEdge}{\alphaEdge}\right\vert +  \left\vert\misClassfSet{\betaEdge}{\gammaEdge}\right\vert
    \right)
    + \log(1-\gammaEdge) 
    \left( 
    \left\vert\misClassfSet{\alphaEdge}{\gammaEdge}\right\vert + \left\vert\misClassfSet{\betaEdge}{\gammaEdge}\right\vert 
    \right)
    -
    \log(1-\gammaEdge)
    \left( 
    \left\vert\misClassfSet{\gammaEdge}{\alphaEdge}\right\vert + \left\vert\misClassfSet{\gammaEdge}{\betaEdge}\right\vert 
    \right)
    \nonumber\\
    &=
    \left(\log\frac{1-\alphaEdge}{1-\betaEdge}\right) 
    \left(
    \left\vert \misClassfSet{\betaEdge}{\alphaEdge}\right\vert
    - \left\vert \misClassfSet{\alphaEdge}{\betaEdge}\right\vert
    \right)
	+  
	\left(\log\frac{1-\alphaEdge}{1-\gammaEdge}\right) 
    \left(
    \left\vert \misClassfSet{\gammaEdge}{\alphaEdge}\right\vert
    - \left\vert \misClassfSet{\alphaEdge}{\gammaEdge}\right\vert
    \right)
	+  
	\left(\log\frac{1-\betaEdge}{1-\gammaEdge}\right)
    \left(
    \left\vert \misClassfSet{\gammaEdge}{\betaEdge}\right\vert
    - \left\vert \misClassfSet{\betaEdge}{\gammaEdge}\right\vert
    \right),
    \label{eq:term3_L(x)}
\end{align}
where \eqref{eq:term3_L(x)_0} holds since
\begin{align}
    \userPairSet{\mu}{\partitionSet_0} 
    = \bigcup_{\nu \in \left\{\alphaEdge, \betaEdge, \gammaEdge\right\}} \misClassfSet{\mu}{\nu},
    \qquad
    \userPairSet{\mu}{\partitionSet} 
    = \bigcup_{\nu \in \left\{\alphaEdge, \betaEdge, \gammaEdge\right\}} \misClassfSet{\nu}{\mu};
\end{align}
and \eqref{eq:term3_L(x)_1} follows since the sets $\{\misClassfSet{\mu}{\nu} : \mu, \nu \in \{\alphaEdge,\betaEdge,\gammaEdge \}, \: \mu \neq \nu\}$ are disjoint.

By \eqref{eq:phiX_3}, \eqref{eq:term2_L(x)_1} and \eqref{eq:term3_L(x)}, the LHS of \eqref{eq:LX_LM0} is given by
\begin{align}
    \mathsf{L}\left(\gtMat\right) - \mathsf{L}(X)
    &=
    \log\left((q-1)\frac{1-\theta}{\theta}\right) 
    \sum_{i \in \diffEntriesSet}
	\mathsf{B}_i^{(p)}
	\left(\left(1+\mathsf{B}_i^{\left(\frac{1}{q-1}\right)}\right)\mathsf{B}_i^{(\theta)} - 1\right)
	\nonumber\\
	&\phantom{=}
	+ \left(\log\frac{(1-\betaEdge)\alphaEdge}{(1-\alphaEdge)\betaEdge}\right)
    \left(
    \sum\limits_{i \in \misClassfSet{\betaEdge}{\alphaEdge}} \mathsf{B}_i^{(\betaEdge)}
    - \sum\limits_{i \in \misClassfSet{\alphaEdge}{\betaEdge}} \mathsf{B}_i^{(\alphaEdge)} 
    \right)
    +
    \left(\log\frac{1-\alphaEdge}{1-\betaEdge}\right) 
    \left(
    \left\vert \misClassfSet{\betaEdge}{\alphaEdge}\right\vert
    - \left\vert \misClassfSet{\alphaEdge}{\betaEdge}\right\vert
    \right)
	\nonumber\\
	&\phantom{=}
	+ \left(\log\frac{(1-\gammaEdge)\alphaEdge}{(1-\alphaEdge)\gammaEdge}\right)
    \left(
    \sum\limits_{i \in \misClassfSet{\gammaEdge}{\alphaEdge}} \mathsf{B}_i^{(\gammaEdge)}
    - \sum\limits_{i \in \misClassfSet{\alphaEdge}{\gammaEdge}} \mathsf{B}_i^{(\alphaEdge)} 
    \right)
    +  
	\left(\log\frac{1-\alphaEdge}{1-\gammaEdge}\right) 
    \left(
    \left\vert \misClassfSet{\gammaEdge}{\alphaEdge}\right\vert
    - \left\vert \misClassfSet{\alphaEdge}{\gammaEdge}\right\vert
    \right)
	\nonumber\\
	&\phantom{=}
	+ \left(\log\frac{(1-\gammaEdge)\betaEdge}{(1-\betaEdge)\gammaEdge}\right)
    \left(
    \sum\limits_{i \in \misClassfSet{\gammaEdge}{\betaEdge}} \mathsf{B}_i^{(\gammaEdge)}
    - \sum\limits_{i \in \misClassfSet{\betaEdge}{\gammaEdge}} \mathsf{B}_i^{(\betaEdge)} 
   \right)
    +  
	\left(\log\frac{1-\betaEdge}{1-\gammaEdge}\right)
    \left(
    \left\vert \misClassfSet{\gammaEdge}{\betaEdge}\right\vert
    - \left\vert \misClassfSet{\betaEdge}{\gammaEdge}\right\vert
    \right),
\end{align}
which immediately proves \eqref{eq:lemma3}.
This completes the proof of Lemma~\ref{lemma:neg_lik_M0_XT}.
\hfill $\blacksquare$

\section{Proof of Lemma~\ref{lemma:upper_bound_B}}
\label{proof:upper_bound_B}
We first define three random variables (indexed by $i$) before proceeding with the proof.
Recall from Section~\ref{sec:achv} that $\mathsf{B}_i^{(\sigma)}$ denotes the $i^{\text{th}}$~Bernoulli random variable with parameter $\sigma \in \{p, \theta, \frac{1}{q-1}, \alphaEdge, \betaEdge, \gammaEdge\}$, that is $\mathbb{P}[\mathsf{B}_i^{(\sigma)}=1] = 1- \mathbb{P}[\mathsf{B}_i^{(\sigma)}=0] = \sigma$. 
For $p=\Theta\left(\frac{\log n}{n}\right)$ and a constant $\theta \in[0,1]$, we define the first random variable $\mathbf{U}_i = \mathbf{U}_i(p,\theta,q)$ as
\begin{align}
	\mathbf{U}_i(p,\theta,q) 
	&= \log \left((q-1)\frac{1-\theta}{\theta}\right) \mathsf{B}_i^{(p)} \left( \left(1+\mathsf{B}_i^{\left(\frac{1}{q-1}\right)}\right)\mathsf{B}_i^{(\theta)}-1\right)
	\nonumber\\
	&= \left\{
	\begin{array}{ll}
	-\log \left((q-1)\frac{1-\theta}{\theta}\right) & \textrm{w.p. } \: p(1-\theta),\\
	0 & \textrm{w.p. } \: (1-p) + p \theta \left(1-\frac{1}{q-1}\right),\\
	\log \left((q-1)\frac{1-\theta}{\theta}\right)  & \textrm{w.p. } \: p\theta\frac{1}{q-1}.
	\end{array}
	\right.
	\label{eq:app_x(p,t,q)}
\end{align}
The moment generating function $M_{\mathbf{U}_i(p,\theta,q)}\left(t\right)$ of $\mathbf{U}_i(p,\theta,q)$ at $t=1/2$ is evaluated as
\begin{align}
    M_{\mathbf{U}_i(p,\theta,q)}\left(\frac{1}{2}\right)
    &= 
    \mathbbm{E}\left[\exp\left(\frac{1}{2} \mathbf{U}_i(p,\theta,q) \right) \right] 
    \nonumber\\
    &=
    \left[
    p(1\!-\!\theta)  \exp\left(\!-\frac{1}{2}\log\left(\!(q\!-\!1)\frac{1\!-\!\theta}{\theta}\right)\!\!\!\right)\!
    \right]
    + 
    \left[
    1 \!-\! p + p \theta \left(\!1-\frac{1}{q\!-\!1}\right)\!
    \right]
    + 
    \left[
    \frac{p\theta}{q\!-\!1} \exp\left(\frac{1}{2}\log\left(\!(q-1) \frac{1\!-\!\theta}{\theta}\right)\!\!\!\right)\!
    \right]
    \nonumber \\
    &=
    p\sqrt{\frac{\theta(1-\theta)}{q-1}} 
    + 1-p+p\theta\left(1-\frac{1}{q-1}\right)  
    + p\sqrt{\frac{\theta(1-\theta)}{q-1}}   \nonumber \\
    &=
    1-p\left(1-\theta
    -2\sqrt{\frac{\theta(1-\theta)}{q-1}}
    +\frac{\theta}{q-1}\right)   \nonumber \\
    &=
    1-p\left(\sqrt{1-\theta}-\sqrt{\frac{\theta}{q-1}} \right)^2, 
    \label{eq:M-X}
\end{align}
and hence we have 
\begin{align}
    -\log M_{\mathbf{U}_i(p,\theta,q)}\left(\frac{1}{2}\right) 
    &= - \log \left(
    1-p\left(\sqrt{1-\theta}-\sqrt{\frac{\theta}{q-1}} \right)^2
    \right)
    \nonumber\\
    &= p\left(\sqrt{1-\theta}-\sqrt{\frac{\theta}{q-1}} \right)^2
    + O\left(p^2\right)
    \label{eq:taylor_Mx}\\
    &= (1+o(1)) \left(\sqrt{1-\theta} - \sqrt{\frac{\theta}{q-1}}\right)^2 p
    \nonumber\\
    &= (1+o(1)) \Ir,
    \label{eq:log-M-X}
\end{align}
where \eqref{eq:taylor_Mx} follows from Taylor series of $\log (1-x)$, for $x = p\left(\sqrt{1-\theta}-\sqrt{\frac{\theta}{q-1}} \right)^2$, which converges since $p=\Theta\left(\frac{\log n}{n}\right)$.
Next, for $\mu,\nu = \Theta\left(\frac{\log n}{n}\right)$, define the second random variable $\mathbf{V}_i = \mathbf{V}_i(\mu,\nu)$ as 
\begin{align}
	\mathbf{V}_i(\mu,\nu) 
	&=
	\left(\log\frac{(1-\mu)\nu}{(1-\nu)\mu}\right) \left(\mathsf{B}_i^{(\mu)} - \mathsf{B}_i^{(\nu)}\right)
    =
	\left\{
	\begin{array}{ll}
	-\log\frac{(1-\mu)\nu}{(1-\nu)\mu} & \textrm{w.p. }\: (1-\mu)\nu,\\
	0 & \textrm{w.p. }\: (1-\mu)(1-\nu) + \mu\nu,\\
	\log\frac{(1-\mu)\nu}{(1-\nu)\mu} & \textrm{w.p. }\: \mu(1-\nu).
	\end{array}
	\right.
	\label{eq:app_z(m,n)}
\end{align}
The moment generating function $M_{\mathbf{V}_i(\mu,\nu)}\left(t\right)$ of $\mathbf{V}_i(\mu,\nu)$ at $t=1/2$ is evaluated as
\begin{align}
    M_{\mathbf{V}_i(\mu,\nu)}\left(\frac{1}{2}\right)
    &= 
    \mathbb{E}\left[\exp\left(\frac{1}{2}\mathbf{V}_i(\mu,\nu)\right)\right] 
    \nonumber\\ 
    &= 
    \left[
    (1-\mu)\nu \exp\left(-\frac{1}{2}\log\frac{(1-\mu)\nu}{(1-\nu)\mu} \right) 
    \right]
    + \left[ 
    (1-\mu)(1-\nu) + \mu\nu 
    \right]
    + 
    \left[
    \mu(1-\nu)\exp\left(\frac{1}{2}\log\frac{(1-\mu)\nu}{(1-\nu)\mu} \right) 
    \right]
    \nonumber\\
    &= 
    (1-\mu)\nu \sqrt{\frac{(1-\nu)\mu}{(1-\mu)\nu}}
    + (1-\mu)(1-\nu) + \mu\nu 
    + (1-\nu)\mu \sqrt{\frac{(1-\mu)\nu}{(1-\nu)\mu}}
    \nonumber\\
    &= 
    \mu\nu 
    + 2 \sqrt{(1-\mu)(1-\nu)\mu\nu}
    + (1-\mu)(1-\nu) 
    \nonumber\\
    &= \left(\sqrt{\mu\nu}+\sqrt{(1-\mu)(1-\nu)}\right)^2,
    \label{eq:M-Y}
\end{align}
and thus we have 
\begin{align}
    -\log M_{\mathbf{V}_i(\mu,\nu)}\left(\frac{1}{2}\right) 
    &= -2\log \left(\sqrt{\mu\nu}+\sqrt{(1-\mu)}\sqrt{(1-\nu)}\right)
    \nonumber\\
    &= -2\log \left(
    \sqrt{\mu\nu}
    + \left(1-\frac{1}{2}\mu+O\left(\mu^2\right)\right) \left(1-\frac{1}{2}\nu+O\left(\nu^2\right)\right)
    \right)
    \label{eq:log-Mz_1}\\
    &= -2\log \left(
    \sqrt{\mu\nu}
    + \left(1-\frac{1}{2}\mu-\frac{1}{2}\nu+O\left(\mu^2+\nu^2\right)\right)
    \right)
    \nonumber\\
    &=
    -2\log \left(1
    - \left(\frac{1}{2}\mu+\frac{1}{2}\nu-\sqrt{\mu\nu}+O\left(\mu^2+\nu^2\right)\right)
    \right)
    \nonumber\\
    &=
    \left(\sqrt{\mu} - \sqrt{\nu}\right)^2 + O\left(\mu^2+\nu^2\right)
    \label{eq:log-Mz_2}\\
    &= (1+o(1)) \left(\sqrt{\mu} - \sqrt{\nu}\right)^2
    =
    \left\{
	\begin{array}{ll}
	(1+o(1)) \Ig \frac{\log n}{n} & \textrm{if }\: \mu = \betaEdge,\: \nu = \alphaEdge,\\
	(1+o(1)) \IcOne \frac{\log n}{n}& \textrm{if }\: \mu = \gammaEdge,\: \nu = \alphaEdge,\\
	(1+o(1)) \IcTwo \frac{\log n}{n}& \textrm{if }\: \mu = \gammaEdge,\: \nu = \betaEdge,
	\end{array}
	\right.
    \label{eq:log-M-Y}
\end{align}
where \eqref{eq:log-Mz_1} follows form Taylor series of $\sqrt{1-\mu}$ and $\sqrt{1-\nu}$, which both converge since $\mu,\nu = \Theta\left(\frac{\log n}{n}\right)$; and \eqref{eq:log-Mz_2} follows from Taylor series of $\log (1-x)$, for $x = \frac{1}{2}\mu+\frac{1}{2}\nu-\sqrt{\mu\nu}+O\left(\mu^2+\nu^2\right)$, which also converges for $\mu,\nu = \Theta\left(\frac{\log n}{n}\right)$.
Finally, for $\mu,\nu = \Theta\left(\frac{\log n}{n}\right)$, define the third random variable $\mathbf{W}_i = \mathbf{W}_i (\mu,\nu)$ as 
\begin{align}
	\mathbf{W}_i (\mu,\nu) 
	&=
	\left(\log\frac{1-\nu}{1-\mu}\right) + \left(\log\frac{(1-\mu)\nu}{(1-\nu)\mu}\right) \mathsf{B}_i^{(\mu)}
    =
	\left\{
	\begin{array}{ll}
	\log\frac{\nu}{\mu} & \textrm{w.p. }\: \mu,\\
	\log\frac{1-\nu}{1-\mu} & \textrm{w.p. }\: (1-\mu).
	\end{array}
	\right.
	\label{eq:app_w(m,n)}
\end{align}
The moment generating function $M_{\mathbf{W}_i (\mu,\nu)}\left(t\right)$ of $\mathbf{W}_i (\mu,\nu)$ at $t=1/2$ is evaluated as
\begin{align}
    M_{\mathbf{W}_i(\mu,\nu)}\left(\frac{1}{2}\right)
    &= 
    \mathbb{E}\left[\exp\left(\frac{1}{2}\mathbf{W}_i(\mu,\nu)\right)\right] 
    \nonumber\\ 
    &= 
    \left[
    \mu \exp\left(\frac{1}{2}\log\frac{\nu}{\mu} \right) 
    \right]
    +
    \left[
    (1-\mu)\exp\left(\frac{1}{2}\log\frac{1-\nu}{1-\mu} \right) 
    \right]
    \nonumber\\
    &= \sqrt{\mu\nu}+\sqrt{(1-\mu)(1-\nu)},
    \label{eq:M-W}
\end{align}
and hence we have 
\begin{align}
    -\log M_{\mathbf{W}_i(\mu,\nu)}\left(\frac{1}{2}\right) 
    &= -\log \left(\sqrt{\mu\nu}+\sqrt{(1-\mu)}\sqrt{(1-\nu)}\right)
    \nonumber \\
    &= \frac{1}{2} (1+o(1)) \left(\sqrt{\mu} - \sqrt{\nu}\right)^2
    =
    \left\{
	\begin{array}{ll}
	\frac{1}{2} (1+o(1)) \Ig \frac{\log n}{n} & \textrm{if }\: \mu = \betaEdge,\: \nu = \alphaEdge,\\
	\frac{1}{2} (1+o(1)) \IcOne \frac{\log n}{n} & \textrm{if }\: \mu = \gammaEdge,\: \nu = \alphaEdge,\\
	\frac{1}{2} (1+o(1)) \IcTwo \frac{\log n}{n} & \textrm{if }\: \mu = \gammaEdge,\: \nu = \betaEdge,
	\end{array}
	\right.
	\label{eq:log-M-W}
\end{align}
where \eqref{eq:log-M-W} follows from \eqref{eq:log-M-Y}.
Next, based on the random variables defined in \eqref{eq:app_z(m,n)} and \eqref{eq:app_w(m,n)}, we present the following proposition that will be used in the proof of Lemma~\ref{lemma:upper_bound_B}. The proof of the proposition is presented at the end of this appendix.
\begin{prop}
\label{prop:A_moment}
For $\mu,\nu = \Theta\left(\frac{\log n}{n}\right)$, let $\mathbf{A} = \mathbf{A}(\mu,\nu)$ be a random variable that is defined as
\begin{align}
    \mathbf{A}(\mu,\nu) =
    \left(\log\frac{(1-\mu)\nu}{(1-\nu)\mu}\right)
    \left(
    \sum\limits_{i \in \misClassfSet{\mu}{\nu}} \mathsf{B}_i^{(\mu)}
    - \sum\limits_{i \in \misClassfSet{\nu}{\mu}} \mathsf{B}_i^{(\nu)} 
    \right)
    +
    \left(\log\frac{1-\nu}{1-\mu}\right) 
    \left(
    \left\vert \misClassfSet{\mu}{\nu}\right\vert
    - \left\vert \misClassfSet{\nu}{\mu}\right\vert
    \right),
    \label{eq:prop_A}
\end{align}
where $\{\mathsf{B}_i^{(\mu)}: i \in \misClassfSet{\nu}{\mu}\}$  and $\{\mathsf{B}_i^{(\nu)}: i \in \misClassfSet{\nu}{\mu}\}$ are sets of independent and identically distributed Bernoulli random variables.
The moment generating function $M_{\mathbf{A}(\mu,\nu)}\left(t\right)$ of $\mathbf{A}(\mu,\nu)$ at $t=1/2$ is given by
\begin{align}
    M_{\mathbf{A}(\mu,\nu)}\left(t\right)
    &=
    \exp \left( - (1+o(1)) \frac{\left\vert \misClassfSet{\mu}{\nu} \right\vert + \left\vert \misClassfSet{\nu}{\mu} \right\vert}{2} \left(\sqrt{\mu} - \sqrt{\nu}\right)^2 \right).
    \\
    &=
    \left\{
	\begin{array}{ll}
	 \exp \left( - (1+o(1)) \Pleftrightarrow{\alphaEdge}{\betaEdge} \: \Ig \frac{\log n}{n} \right)
	& \textrm{if }\: \mu = \betaEdge,\: \nu = \alphaEdge,\\
	 \exp \left( - (1+o(1)) \Pleftrightarrow{\alphaEdge}{\gammaEdge} \: \IcOne \frac{\log n}{n} \right)
	& \textrm{if }\: \mu = \gammaEdge,\: \nu = \alphaEdge,\\
	 \exp \left( - (1+o(1)) \Pleftrightarrow{\betaEdge}{\gammaEdge} \: \IcTwo \frac{\log n}{n} \right)
	& \textrm{if }\: \mu = \gammaEdge,\: \nu = \betaEdge.
	\end{array}
	\right.
	\label{eq:prop_A_momentGen}
\end{align}
\end{prop}

Let $\left\{\mathbf{U}_i (p,\theta,q): i\in \diffEntriesSet \right\}$, and $\{\mathbf{A}(\betaEdge,\alphaEdge), \mathbf{A}(\gammaEdge,\alphaEdge), \mathbf{A}(\gammaEdge,\betaEdge)\}$ be sets of independent and identically distributed random variables defined as per \eqref{eq:app_x(p,t,q)}, and \eqref{eq:prop_A} in Proposition~\ref{prop:A_moment}.
Note that the sets $\{\misClassfSet{\mu}{\nu} : \mu, \nu \in \{\alphaEdge,\betaEdge,\gammaEdge \}, \: \mu \neq \nu\}$ are disjoint as per their definitions given by \eqref{eq:N1_defn}--\eqref{eq:N4rev_defn}.
Consequently, the LHS of \eqref{eq:lemma1_2} is upper bounded by
\begin{align}
    \mathbb{P}\left[ \mathbf{B} \geq 0\right]
	&= 
	\mathbb{P}\left[
	\left(\sum_{i\in\diffEntriesSet} 
	\mathbf{U}_i(p,\theta,q) \right)
	+ \mathbf{A}(\betaEdge,\alphaEdge)
	+ \mathbf{A}(\gammaEdge,\alphaEdge)
	+ \mathbf{A}(\gammaEdge,\betaEdge)
	\geq 0\right]
	\nonumber\\
	&\leq
	\left(M_{\mathbf{U}_i(p,\theta,q) }\left(\frac{1}{2}\right)\right)^{\lvert\diffEntriesSet\rvert}
	\left(M_{\mathbf{A}(\betaEdge,\alphaEdge)}\left(\frac{1}{2}\right)\right)
	\left(M_{\mathbf{A}(\gammaEdge,\alphaEdge)}\left(\frac{1}{2}\right)\right)
	\left(M_{\mathbf{A}(\gammaEdge,\betaEdge)}\left(\frac{1}{2}\right)\right)
	\label{eq:app_chernoff_1}\\
	&=
	\exp\left(-\left(1+o(1)\right) \left(\lvert\diffEntriesSet\rvert \Ir 
	+ \Pleftrightarrow{\alphaEdge}{\betaEdge} \: \Ig \frac{\log n}{n}
	+ \Pleftrightarrow{\alphaEdge}{\gammaEdge} \: \IcOne \frac{\log n}{n}
	+ \Pleftrightarrow{\betaEdge}{\gammaEdge} \: \IcTwo \frac{\log n}{n} \right)\right), 
	\label{eq:app_chernoff_2}
\end{align}
where \eqref{eq:app_chernoff_1} follows from the Chernoff bound, and the independence of the random variables $\left\{\mathbf{U}_i (p,\theta,q): i\in \diffEntriesSet \right\}$, and $\{\mathbf{A}(\betaEdge,\alphaEdge), \mathbf{A}(\gammaEdge,\alphaEdge), \mathbf{A}(\gammaEdge,\betaEdge)\}$;
and finally \eqref{eq:app_chernoff_2} follows from \eqref{eq:log-M-X}, and \eqref{eq:prop_A_momentGen} in Proposition~\ref{prop:A_moment}.
This completes the proof of Lemma~\ref{lemma:upper_bound_B}.
\hfill $\blacksquare$

It remains to prove Proposition~\ref{prop:A_moment}. The proof is presented as follows.
\begin{IEEEproof}[Proof of Proposition~\ref{prop:A_moment}]
First, consider the case of $\left\vert \misClassfSet{\mu}{\nu} \right\vert \geq \left\vert \misClassfSet{\nu}{\mu} \right\vert$. Therefore, the random variable $\mathbf{A}(\mu,\nu)$ can be expressed as
\begin{align}
    \mathbf{A}(\mu,\nu)
    &=
    \sum\limits_{i \in \misClassfSet{\nu}{\mu}} 
    \left(
    \left(\log\frac{(1-\mu)\nu}{(1-\nu)\mu}\right)
    \left(
    \mathsf{B}_i^{(\mu)} - \mathsf{B}_i^{(\nu)} 
    \right)
    \right)
    +
    \sum\limits_{i \in \misClassfSet{\mu}{\nu} \setminus \misClassfSet{\nu}{\mu}} 
    \left(
    \left(\log\frac{1-\nu}{1-\mu}\right) 
    +
    \left(\log\frac{(1-\mu)\nu}{(1-\nu)\mu}\right)
    \mathsf{B}_i^{(\mu)}
    \right)
    \label{eq:app_A_defn_0}\\
    &=
    \sum\limits_{i \in \misClassfSet{\nu}{\mu}} 
    \mathbf{V}_i
    +
    \sum\limits_{i \in \misClassfSet{\mu}{\nu} \setminus \misClassfSet{\nu}{\mu}} 
    \mathbf{W}_i,
    \label{eq:app_A_defn}
\end{align}
where \eqref{eq:app_A_defn_0} holds since the sets $\misClassfSet{\nu}{\mu}$ and $\misClassfSet{\mu}{\nu}$ are disjoint; and \eqref{eq:app_A_defn} follows from \eqref{eq:app_z(m,n)} and \eqref{eq:app_w(m,n)}.
\begin{align}
    M_{\mathbf{A}(\mu,\nu)}\left(\frac{1}{2}\right)
    &=
    \mathbb{E}\left[\exp\left(\frac{1}{2}\mathbf{A}(\mu,\nu)\right)\right] 
    \nonumber\\ 
    &= 
    \mathbb{E}\left[
    \left(
    \prod_{i \in \misClassfSet{\nu}{\mu}}  
    \exp \left( \frac{1}{2} \mathbf{V}_i(\mu,\nu) \right)
    \right) 
    \left(
    \prod_{i \in \misClassfSet{\mu}{\nu} \setminus \misClassfSet{\nu}{\mu}} 
    \exp \left( \frac{1}{2} \mathbf{W}_i(\mu,\nu) \right)
    \right)
    \right]
    \nonumber\\
    &= 
    \left(
    \prod_{i \in \misClassfSet{\nu}{\mu}}  
    \mathbb{E}\left[
    \exp \left( \frac{1}{2} \mathbf{V}_i(\mu,\nu) \right)
    \right] 
    \right) 
    \left(
    \prod_{i \in \misClassfSet{\mu}{\nu} \setminus \misClassfSet{\nu}{\mu}} 
    \mathbb{E}\left[
    \exp \left( \frac{1}{2} \mathbf{W}_i(\mu,\nu) \right)
    \right] 
    \right)
    \label{eq:prop_A_0}\\
    &=
    \left(
    \prod_{i \in \misClassfSet{\nu}{\mu}}  
    M_{\mathbf{V}_i(\mu,\nu)}\left(\frac{1}{2}\right)
    \right) 
    \left(
    \prod_{i \in \misClassfSet{\mu}{\nu} \setminus \misClassfSet{\nu}{\mu}} 
    M_{\mathbf{W}_i(\mu,\nu)}\left(\frac{1}{2}\right)
    \right)
    \nonumber\\
    &=
    \left(
    \exp \left( - (1+o(1)) \left(\sqrt{\mu} - \sqrt{\nu}\right)^2 \right)
    \right)^{\left\vert \misClassfSet{\nu}{\mu} \right\vert} 
    \left(
    \exp \left( - \frac{1}{2} (1+o(1)) \left(\sqrt{\mu} - \sqrt{\nu}\right)^2 \right)
    \right)^{\left\vert \misClassfSet{\mu}{\nu} \right\vert - \left\vert \misClassfSet{\nu}{\mu} \right\vert}
    \label{eq:prop_A_2}\\
    &=
    \exp \left( - (1+o(1)) \frac{\left\vert \misClassfSet{\mu}{\nu} \right\vert + \left\vert \misClassfSet{\nu}{\mu} \right\vert}{2} \left(\sqrt{\mu} - \sqrt{\nu}\right)^2 \right),
    \label{eq:prop_A_3}
\end{align}
where \eqref{eq:prop_A_0} holds since the random variables $\{\mathbf{V}_i : i \in \misClassfSet{\nu}{\mu}\}$ and $\{\mathbf{W}_i : i \in \misClassfSet{\mu}{\nu} \setminus \misClassfSet{\nu}{\mu}\}$ are independent; and \eqref{eq:prop_A_2} follows from \eqref{eq:log-M-Y} and \eqref{eq:log-M-W}.

Next, consider the case of $\left\vert \misClassfSet{\mu}{\nu} \right\vert \leq \left\vert \misClassfSet{\nu}{\mu} \right\vert$. In a similar way, the random variable $\mathbf{A}$ can be written as
\begin{align}
    \mathbf{A}(\mu,\nu)
    &=
    \sum\limits_{i \in \misClassfSet{\mu}{\nu}} 
    \left(
    \left(\log\frac{(1-\mu)\nu}{(1-\nu)\mu}\right)
    \left(
    \mathsf{B}_i^{(\mu)} - \mathsf{B}_i^{(\nu)} 
    \right)
    \right)
    +
    \sum\limits_{i \in \misClassfSet{\nu}{\mu} \setminus \misClassfSet{\mu}{\nu}} 
    \left(
    \left(\log\frac{1-\mu}{1-\nu}\right) 
    +
    \left(\log\frac{(1-\nu)\mu}{(1-\mu)\nu}\right)
    \mathsf{B}_i^{(\nu)}
    \right)
    \nonumber\\
    &=
    \sum\limits_{i \in \misClassfSet{\mu}{\nu}} 
    \mathbf{V}_i
    +
    \sum\limits_{i \in \misClassfSet{\nu}{\mu} \setminus \misClassfSet{\mu}{\nu}} 
    \mathbf{W}_i.
\end{align}
Following the same procedure presented in the previous case, one can show that $M_{\mathbf{A}(\mu,\nu)}\left(\frac{1}{2}\right)$ is also given by \eqref{eq:prop_A_3} in this case.
This completes the proof of Proposition~\ref{prop:A_moment}.
\end{IEEEproof}

\section{Proof of Lemma~\ref{lemma:T1_related}}
\label{proof:T1_related}
\begin{JA_AE}
The LHS of \eqref{eq:lemma_T1_1} is given by
\begin{align}
    & \lim_{n,m\rightarrow \infty}
    \sum\limits_{T \in \TsmallErr} 
	\sum\limits_{X \in \mathcal{X}(T)}
	\exp\left(-\left(1+o(1)\right) \left(\lvert\diffEntriesSet\rvert \Ir 
    + \Pleftrightarrow{\alphaEdge}{\betaEdge} \: \Ig \frac{\log n}{n}
    + \Pleftrightarrow{\alphaEdge}{\gammaEdge} \: \IcOne \frac{\log n}{n}
    + \Pleftrightarrow{\betaEdge}{\gammaEdge} \: \IcTwo \frac{\log n}{n} \right)\right) 
    \nonumber \\
    & \qquad =
    \lim_{n,m\rightarrow \infty}
    \sum\limits_{T \in \TsmallErr} 
	\underbrace{\left\vert \mathcal{X}(T) \right\vert}_{\mathsf{Term_1}}
	\:\:
	\underbrace{\exp\left(-\left(1+o(1)\right) \left(\lvert\diffEntriesSet\rvert \Ir 
    + \Pleftrightarrow{\alphaEdge}{\betaEdge} \: \Ig \frac{\log n}{n}
    + \Pleftrightarrow{\alphaEdge}{\gammaEdge} \: \IcOne \frac{\log n}{n}
    + \Pleftrightarrow{\betaEdge}{\gammaEdge} \: \IcTwo \frac{\log n}{n} \right)\right)}_{\mathsf{Term_2}}.
    \label{eq:largeErr_LHS_app}
\end{align}
In what follows, we derive upper bounds on $\mathsf{Term_1}$ and $\mathsf{Term_2}$ for a fixed non-all-zero tuple $T \in \TsmallErr$ given by
\begin{align}
    T
    =
    \left(
    \left\{\kUsrs{i}{j}{x}{y}\right\}_{x,y \in [c], \: i,j \in [g]}, 
    \left\{\dElmntsGen{i}{j}{x}{y}\right\}_{x,y \in [c], \: i,j \in [g]}
    \right),
\end{align}
according to \eqref{eq:tuple_ratMat}.

\noindent \underline{(1) Upper Bound on $\mathsf{Term_1}$:} 
The cardinality of the set $\mathcal{X}(T)$ is given by
\begin{align}
    \mathsf{Term_1}
    &= 
    \underbrace{
    \left\vert 
    \mathcal{X}
    \left(
    \left\{\kUsrs{i}{j}{x}{y} : x,y \in [c], i,j \in [g] \right\}, \:
    \left\{\dElmntsGenHat{i}{j}{x}{y} = 0 : x,y \in [c], i,j \in [g]\right\}
    \right)
    \right\vert
    }_{\mathsf{Term_{1,1}}}
    \nonumber \\
    &\phantom{=} 
    \times
    \underbrace{
    \left\vert 
    \mathcal{X}
    \left(
    \left\{\kUsrsHat{i}{j}{x}{y} = 0 : x,y \in [c], i,j \in [g] \right\}, \:
    \left\{\dElmntsGen{i}{j}{x}{y} : x,y \in [c], i,j \in [g] \right\}
    \right)
    \right\vert
    }_{\mathsf{Term_{1,2}}},
    \label{eq:term1_11_12}
\end{align}
which follows from the fact that the number of ways of counting the rating matrices subject to $\{\kUsrs{i}{j}{x}{y} : x,y \in [c], i,j \in [g]\}$, and subject to $\{\dElmntsGen{i}{j}{x}{y} : x,y \in [c], i,j \in [g] \}$ are independent.
Next, we provide upper bounds on $\mathsf{Term_{1,1}}$ and $\mathsf{Term_{1,2}}$.

\noindent \underline{(1-1) Upper Bound on $\mathsf{Term_{1,1}}$:} 
An upper bound on $\mathsf{Term_{1,1}}$ is given by
\begin{align}
    \mathsf{Term_{1,1}}
    &=
    \left\vert 
    \mathcal{X}
    \left(
    \left\{\kUsrs{i}{j}{x}{y} : x,y \in [c], i,j \in [g] \right\}, \:
    \left\{\dElmntsGenHat{i}{j}{x}{y} = 0 : x,y \in [c], i,j \in [g]\right\}
    \right)
    \right\vert
    \nonumber\\
    &=
    \prod_{x \in [c]}
    \prod_{i \in [g]} 
    \binom{n/(gc)}
    {\kUsrs{i}{1}{x}{1}, \kUsrs{i}{2}{x}{1}, \cdots, \kUsrs{i}{g}{x}{1}, \kUsrs{i}{1}{x}{2}, \kUsrs{i}{2}{x}{2}, \cdots, \kUsrs{i}{g}{x}{c}}
    \label{eq:term11_00}\\
    &\leq
    \prod_{x \in [c]}
    \prod_{i \in [g]} 
    \left(
    \frac{n}{gc}
    \right)^{
    \textstyle
    \sum_{(y,j) \neq (\sigma(x), \sigma(i|x))} 
    \kUsrs{i}{j}{x}{y}
    }
    \label{eq:term11_0}\\
    &\leq
    \prod_{x \in [c]}
    \prod_{i \in [g]} 
    \exp
    \left(
    \left(
    \sum_{(y,j) \neq (\sigma(x), \sigma(i|x))} 
    \kUsrs{i}{j}{x}{y}
    \right)
    \log n
    \right)
    \nonumber\\
    &=
    \exp
    \left(
    \log n
    \left(
    \sum_{x \in [c]} 
    \sum_{i \in [g]} 
    \sum_{(y,j) \neq (\sigma(x), \sigma(i|x))} 
    \kUsrs{i}{j}{x}{y}
    \right)
    \right),
    \label{eq:term11_1}
\end{align}
where \eqref{eq:term11_00}  follows from the definitions in \eqref{eq:kUsrs_defn} and \eqref{eq:tuple_ratMat}; \eqref{eq:term11_0} follows from the definition of a multinomial coefficient, and the fact that $\binom{n}{k} \leq n^k$.

\noindent \underline{(1-2) Upper Bound on $\mathsf{Term_{1,2}}$:} 
An upper bound on $\mathsf{Term_{1,2}}$ is given by
\begin{align}
    \mathsf{Term_{1,2}}
    &=
    \left\vert 
    \mathcal{X}
    \left(
    \left\{\kUsrsHat{i}{j}{x}{y} = 0 : x,y \in [c], i,j \in [g] \right\}, \:
    \left\{\dElmntsGen{i}{j}{x}{y} : x,y \in [c], i,j \in [g] \right\}
    \right)
    \right\vert
    \nonumber\\
    &\leq
    \left\vert 
    \mathcal{X}
    \left(
    \left\{\kUsrsHat{i}{j}{x}{y} = 0 : x,y \in [c], i,j \in [g] \right\}, \:
    \left\{\dElmntsGen{i}{\sigma(i|x)}{x}{\sigma(x)} : x \in [c], i \in [g] \right\}, \:
    \right.
    \right.
    \nonumber\\
    &\phantom{\leq \vert \mathcal{X} \left( \right.}
    \left.
    \left.
    \left\{\dElmntsGenHat{i}{j}{x}{y} =  t: 
    0 \leq t \leq m, \:
    x,y \in [c] ,\: i,j \in [g],\: (y,j) \neq (\sigma(x),\sigma(i|x)) \right\}
    \right)
    \right\vert
    \nonumber\\
    &\leq
    \prod_{z \in [c]}
    \left\vert 
    \mathcal{X}
    \left(
    \left\{\kUsrsHat{i}{j}{x}{y} = 0 : x,y \!\in\! [c], i,j \!\in\! [g] \right\}\!,
    \left\{\dElmntsGen{i}{\sigma(i|z)}{z}{\sigma(z)} : i \!\in\! [g] \right\}\!,
    \left\{\dElmntsGenHat{i}{\sigma(i|x)}{x}{\sigma(x)} = t : 0 \leq t \leq m,\: x \!\in\! [c] \!\setminus\! \{z\},\: i \!\in\! [g] \right\}\!,
    \right.
    \right.
    \nonumber\\
    &\phantom{= \prod_{z \in [c]}
    \left\vert 
    \mathcal{X}
    \left(
    \right. \right.}
    \left.
    \left.
    \left\{\dElmntsGenHat{i}{j}{x}{y} =  t: 
    0 \leq t \leq m, \:
    x,y \in [c] ,\: i,j \in [g],\: (y,j) \neq (\sigma(x),\sigma(i|x)) \right\}
    \right)
    \right\vert.
    \label{eq:term12_upperB}
\end{align}
    
Recall from Section~\ref{subsec:probForm} that $\gtRatingCluster{x} \in \mathbb{F}_{q}^{g \times m}$ denotes a matrix that is obtained by stacking all the rating vectors of cluster $x$ given by $\{\vecU{i}{x}: i\in [g]\}$ for $x \in [c]$, and whose columns are elements of $(g,r)$ MDS code. 
Similarly, define $\XRatingCluster{x} \in \mathbb{F}_{q}^{g \times m}$ as a matrix that is obtained by stacking all the rating vectors of cluster $x$ given by $\{\vecU{i}{x}: i\in [g]\}$ for $x \in [c]$, and whose columns are also elements of $(g,r)$ MDS code. 
Furthermore, define the binary matrix $\gtRatingClusterHat{x} \in \mathbb{F}_{q}^{g \times m}$ as follows:
\begin{align}
    \gtRatingClusterHat{x}(i,t)
    = 
    \indicatorFn{\gtRatingCluster{x}(i,t) \neq \XRatingCluster{x}(i,t)}, \: \text{for}\: x\in[c], i\in[g], t\in[m]. 
\end{align}
Note that $\gtRatingClusterHat{x}(i,t) = 1$ when there is an error in estimating the rating of the users in cluster $x$ and group $i$ for item $t$ for $x\in[c]$, $i\in[g]$ and $t\in[m]$.
Let any non-zero column of $\gtRatingClusterHat{x}$ be denoted as an ``error column''.

Then, for a given cluster $x \in [c]$, we enumerate all possible matrices $\gtRatingClusterHat{x}$ subject to a given number of error columns.
To this end, define $\totNumErrCol{x}$ as the total number of error columns of $\gtRatingClusterHat{x}$.
Moreover, define $\numErrColConfig$ as the number of possible configurations of an error column.
Let $\{\colVecRhat{k} : k \in [\numErrColConfig]\}$ be the set of all possible error columns. 
Note that this set only depends on the problem setting and the MDS code structure.
For instance, for $(c,g,q,r)=(2,3,2,2)$ and $\vecU{3}{x} = \vecU{1}{x} + \vecU{2}{x}$ for $x\in[2]$,
we have $\numErrColConfig = 3$ since the possible configurations of an error column are given by
\begin{align}
    \colVecRhat{1} = 
    \begin{bmatrix}
    \begin{array}{c}
        1 \\ 0 \\ 1
    \end{array}
    \end{bmatrix}, \:\:
    \colVecRhat{2} = 
    \begin{bmatrix}
    \begin{array}{c}
        1 \\ 1 \\ 0
    \end{array}
    \end{bmatrix}, \:\:
    \colVecRhat{\numErrColConfig} = \colVecRhat{3} = 
    \begin{bmatrix}
    \begin{array}{c}
        0 \\ 1 \\ 1
    \end{array}
    \end{bmatrix}.
\end{align}
Let $\numErrColElemnt{k}{x}$ denote the number of columns of $\gtRatingClusterHat{x}$ that are equal to $\colVecRhat{k}$ for $x \in [c]$ and $k \in [\numErrColConfig]$. Note that $0 \leq \numErrColElemnt{k}{x} \leq \totNumErrCol{x}$ and $\sum_{k=1}^{\numErrColConfig} \numErrColElemnt{k}{x} = \totNumErrCol{x}$.
For cluster $x \in [c]$, let $\ratingVecSet{x}(\totNumErrCol{x},\{\colVecRhat{k} : k \in [\numErrColConfig]\})$ denote the set of matrices $\XRatingCluster{x} \in \mathbb{F}_{q}^{g \times m}$ characterized by $\totNumErrCol{x}$ and $\{\colVecRhat{k} : k \in [\numErrColConfig]\}$.
Consequently, an upper bound on  $\vert\ratingVecSet{x}(\totNumErrCol{x}, \{\colVecRhat{k} : k \in [\numErrColConfig]\})\vert$ is given by
\begin{align}
    \left\vert \ratingVecSet{x}\left(\totNumErrCol{x},\{\colVecRhat{k} : k \in [\numErrColConfig]\}\right)\right\vert 
    &\leq
    \binom{m}{\totNumErrCol{x}}  
    \binom{\totNumErrCol{x}+\numErrColConfig-1}{\numErrColConfig-1} 
    \displaystyle (q-1)^{g\totNumErrCol{x}}
    \label{ineq:Xf_1}\\
    &\leq 
    \displaystyle m^{\totNumErrCol{x}} \: 
    2^{\totNumErrCol{x}+\numErrColConfig-1} \: 
    q^{g \totNumErrCol{x}} 
    \label{ineq:Xf_2}\\
    &\leq
    2^{q^g-1}q^g (2qm)^{\totNumErrCol{x}},
    \label{ineq:Xf_3} 
\end{align}
where 
\begin{itemize}
    \item \eqref{ineq:Xf_1} follows by first choosing $\totNumErrCol{x}$ columns from $m$ columns to be error columns, then counting the number of integer solutions of $\sum_{k=1}^{\numErrColConfig} \numErrColElemnt{k}{x} = \totNumErrCol{x}$, and lastly counting the number of estimation error combination within the $g$ entries of each of the $\totNumErrCol{x}$ error columns;
    \item \eqref{ineq:Xf_2} follows from bounding the first binomial coefficient by $\binom{a}{b} \leq a^b$, and the second binomial coefficient by $\binom{a}{b} \leq \sum_{i=1}^{a} \binom{a}{i} =  2^a$, for $a \geq b$;
    \item and finally \eqref{ineq:Xf_3} follows from $\numErrColConfig \leq q^g$ which is due to the fact that each entry of a rating matrix column can take one of $q$ values.
\end{itemize}
    
Next, for a given cluster $x \in [c]$, we evaluate the maximum number of error columns among all candidate matrices $\XRatingCluster{x}$.
On one hand, row-wise counting of the error entries in $\XRatingCluster{x}$, compared to $\gtRatingCluster{x}$, yields
\begin{align}
    \sum_{i\in[g]} \dElmntsGen{i}{\sigma(i|x)}{x}{\sigma(x)}.
    \label{eq:term12_rowCount}
\end{align}
On the other hand, column-wise counting of the error entries in $\gtRatingClusterHat{x}$ (i.e., number of ones) yields
\begin{align}
    \sum_{k\in[\numErrColConfig]}
    \left\| \colVecRhat{k} \right\|_{1} \numErrColElemnt{k}{x}.
    \label{eq:term12_columnCount}
\end{align}
From \eqref{eq:term12_upperB}, we are interested in the class of candidate rating matrices where the clustering and grouping are done correctly without any errors in user associations to their respective clusters and groups.
Therefore, the expression given by \eqref{eq:term12_rowCount} and \eqref{eq:term12_columnCount} are counting the elements of the same set, and hence we obtain
\begin{align}
    \sum_{i\in[g]} \dElmntsGen{i}{\sigma(i|x)}{x}{\sigma(x)} 
    &= 
    \sum_{k\in[\numErrColConfig]}
    \left\| \colVecRhat{k} \right\|_{1} \numErrColElemnt{k}{x}
    \nonumber\\
    &\geq 
    (g-r+1) 
    \sum_{k\in[\numErrColConfig]} \numErrColElemnt{k}{x} 
    \label{eq:mds_code_1} \\ 
    &=
    (g-r+1) \totNumErrCol{x} 
    \label{eq:mds_code_2} ,
\end{align}
where \eqref{eq:mds_code_1} follows since the MDS code structure is known at the decoder side, and the fact that minimum distance between any two codewords in a  $(g,r)$ linear MDS code is $g-r+1$.
Therefore, by \eqref{eq:mds_code_2}, we get
\begin{align}
    \max \totNumErrCol{x} 
    =
    \frac{1}{g-r+1} \sum_{i\in[g]} \dElmntsGen{i}{\sigma(i|x)}{x}{\sigma(x)}.
    \label{eq:mds_code_3} 
\end{align}

Finally, by \eqref{eq:term12_upperB} and \eqref{eq:mds_code_3}, $\mathsf{Term_{1,2}}$ can be further upper bounded by
\begin{align}
    \mathsf{Term_{1,2}}
    &\leq
    \prod_{z\in[c]}
    \sum_{\ell=1}^{\max \totNumErrCol{z}}
    \left|\ratingVecSet{z}\left(\totNumErrCol{z}= \ell,\{\colVecRhat{k} : k \in [\numErrColConfig]\} \right)\right|
    \nonumber\\
    &\leq
    \prod_{z\in[c]}
    \sum_{\ell=1}^{\max \totNumErrCol{z}}
    2^{q^g-1} \: q^g \: (2qm)^{\ell}
    \label{ineq:term_12_0}\\
    &\leq
    \prod_{z\in[c]}
    2^{q^g-1} \: q^g \: m^{\max \totNumErrCol{z}}
    \sum_{\ell=1}^{\max \totNumErrCol{z}}
    (2q)^{\ell}
    \nonumber\\
    &\leq 
    \prod_{z\in[c]}
    2^{q^g-1} \:q^g \: m^{\max \totNumErrCol{z}} \:
    (2q)^{\max \totNumErrCol{z} + 1}
    \label{ineq:term_12_1} \\ 
    &=
    \prod_{z\in[c]}
    2^{q^g} \: q^{g+1} \: (2qm)^{\max \totNumErrCol{z}}
    \nonumber\\ 
    &=
    \left(2^{q^g}q^{g+1}\right)^c 
    (2qm)^{\sum\limits_{x\in[c]} \max \totNumErrCol{x}}
    \nonumber\\ 
    &=
    c_0 \exp\left(\frac{\log(c_1m)}{g-r+1}\sum_{x\in[c]}\sum_{i\in[g]} \dElmntsGen{i}{\sigma(i|x)}{x}{\sigma(x)} \right),
    \label{ineq:term_12_4}
\end{align}
where \eqref{ineq:term_12_0} follows from \eqref{ineq:Xf_3}; 
\eqref{ineq:term_12_1} follows from $\sum_{\ell=1}^{\max \totNumErrCol{x}} (2q)^\ell \leq \sum_{\ell=0}^{\max \totNumErrCol{x}} (2q)^\ell \leq (2q)^{\max \totNumErrCol{x}+1}$; 
and \eqref{ineq:term_12_4} follows by setting $c_0 = \left(2^{q^g}q^{g+1}\right)^c \geq 1$ and $c_1 = 2q \geq 1$.

Substituting \eqref{eq:term11_1} and \eqref{ineq:term_12_4} into \eqref{eq:term1_11_12}, an upper bound on $\mathsf{Term_1}$ is thus given by
\begin{align}
    \mathsf{Term_{1}}
    &\leq
    c_0
    \exp
    \left(
    \log n
    \left(
    \sum_{x \in [c]} 
    \sum_{i \in [g]} 
    \sum_{(y,j) \neq  (\sigma(x), \sigma(i|x))} 
    \kUsrs{i}{j}{x}{y}
    \right)
    +
    \frac{\log (c_1m)}{g-r+1}
    \left(
    \sum\limits_{x\in[c]}\sum\limits_{i\in[g]}\dElmntsGen{i}{\sigma(i|x)}{x}{\sigma(x)}
    \right)
    \right). \label{eq:term1_lemma_large}
\end{align}

\noindent \underline{(2) Upper Bound on $\mathsf{Term_{2}}$:} 
To this end, we derive lower bounds on the cardinalities of different sets in the exponent of $\mathsf{Term_2}$. 
Recall from \eqref{eq:TsmallErr} that
\begin{align}
    \TsmallErr
    &= 
    \left\{ T \in \mathcal{T}^{(\delta)} : 
    \forall (x,i) \in [c]\times [g] \text{ such that }
    \left\vert \sigma(x) \right\vert = 1, \:
    \left\vert \sigma(i|x) \right\vert = 1, \:\:
    \dElmntsGen{i}{\:\sigma(i|x)}{x}{\:\sigma(x)} \leq \tau m \min\{\intraTau, \interTau\}
    \right\},
    \nonumber\\
    &=
    \left\{ T \in \mathcal{T}^{(\delta)} : 
    \forall (x,i) \in [c]\times [g] \text{ such that }
    \kUsrs{i}{\sigma(i|x)}{x}{\sigma(x)} \geq (1-\relabelConst) \frac{n}{gc},\:\:
    \dElmntsGen{i}{\:\sigma(i|x)}{x}{\:\sigma(x)} \leq \tau m \min\{\intraTau, \interTau\}
    \right\},
    \label{eq:Tlarge_complement_0}
\end{align}
where \eqref{eq:Tlarge_complement_0} follows from \eqref{eq:sigma(x,i)}.

\noindent \underline{(2-1) Lower Bound on $\lvert\diffEntriesSet\rvert$:} 
For $T\in \TsmallErr$, a lower bound on  $\lvert\diffEntriesSet\rvert$ is given by
\begin{align}
    \lvert\diffEntriesSet\rvert
	&=
	\sum\limits_{x \in [c]}
	\sum\limits_{i \in [g]}
	\sum\limits_{y \in [c]}
	\sum\limits_{j \in [g]}
	\kUsrs{i}{j}{x}{y} \dElmntsGen{i}{j}{x}{y} 
	\label{ineq:Lambda_1}\\
	&=
	\left[
	\sum\limits_{x \in [c]}
	\sum\limits_{i \in [g]}
	\kUsrs{i}{\sigma(i|x)}{x}{\sigma(x)} 
	\dElmntsGen{i}{\sigma(i|x)}{x}{\sigma(x)}
	\right]
	+
	\left[
	\sum\limits_{x \in [c]}
	\sum\limits_{i \in [g]}
	\sum\limits_{j \in [g]\setminus \sigma(i|x)}
	\kUsrs{i}{j}{x}{\sigma(x)} 
	\dElmntsGen{i}{j}{x}{\sigma(x)} 
	\right]
	+
	\left[
	\sum\limits_{x \in [c]}
	\sum\limits_{y \in [c] \setminus \sigma(x)}
	\sum\limits_{i \in [g]}
	\sum\limits_{j \in [g]}
	\kUsrs{i}{j}{x}{y}
	\dElmntsGen{i}{j}{x}{y}
	\right]
	\nonumber\\
    &\geq
	\left(
	\sum\limits_{x \in [c]} \sum\limits_{i \in [g]} 
	\dElmntsGen{i}{\sigma(i|x)}{x}{\sigma(x)}
	\left(
	(1-\relabelConst) \frac{n}{gc}
	\right)
	\right)
	+
    \left(
    \sum_{x\in[c]}\sum_{i\in[g]}\sum_{j\in[g] \setminus \sigma(i|x)} \kUsrs{i}{j}{x}{\sigma(x)}
    \left(\hamDist{\vecV{\sigma(i|x)}{\sigma(x)}}{\vecV{j}{\sigma(x)}} - \dElmntsGen{i}{\sigma(i|x)}{x}{\sigma(x)} \right)
    \right)
    \nonumber\\
    &\phantom{\leq}
    + 
    \left(
    \sum_{x\in[c]}\sum_{y\in[c]\setminus \sigma(x)} \sum_{i\in[g]} \sum_{j\in[g]} \kUsrs{i}{j}{x}{y}
    \left(\hamDist{\vecV{\sigma(i|x)}{\sigma(x)}}{\vecV{j}{y}} - \dElmntsGen{i}{\sigma(i|x)}{x}{\sigma(x)} \right)
    \right)
    \label{ineq:Lambda_2} 
    \\
    &\geq
    (1-\relabelConst) \frac{n}{gc}
    \left(
	\sum\limits_{x \in [c]} \sum\limits_{i \in [g]} 
	\dElmntsGen{i}{\sigma(i|x)}{x}{\sigma(x)}
	\right)
	+
    \left(\intraTau m - \intraTau\tau m \right)
    \left(\sum_{x\in[c]}\sum_{i\in[g]}\sum_{j \in [g] \setminus \sigma(i|x)} \kUsrs{i}{j}{x}{\sigma(x)}\right)
    \nonumber\\
    &\phantom{\leq}
    + 
    \left( \interTau m -  \interTau\tau m \right)
    \left(\sum_{x\in[c]}\sum_{y\in[c]\setminus \sigma(x)} \sum_{i\in[g]} \sum_{j\in[g]} \kUsrs{i}{j}{x}{y}\right)
    \label{ineq:Lambda_2_1} 
    \\
    &=
    (1-\tau)\!
    \left(
    \frac{n}{gc}
	\left(
	\sum\limits_{x \in [c]} \sum\limits_{i \in [g]} 
	\dElmntsGen{i}{\sigma(i|x)}{x}{\sigma(x)}
	\right)
	+
    \intraTau m\left(\sum_{x\in[c]}\sum_{i\in[g]}\sum_{j \in [g] \setminus \sigma(i|x)} \kUsrs{i}{j}{x}{\sigma(x)}\right)
    + 
    \interTau m\left(\sum_{x\in[c]}\sum_{y\in[c]\setminus \sigma(x)} \sum_{i\in[g]} \sum_{j\in[g]} \kUsrs{i}{j}{x}{y}\right)\!\!\! 
	\right)\!\!,
    \label{ineq:Lambda_3} 
\end{align}
where \eqref{ineq:Lambda_1} follows from the definitions in \eqref{eq:N1_defn}, \eqref{eq:kUsrs_defn} and \eqref{eq:dElmntsGen_defn}; 
\eqref{ineq:Lambda_2} follows from 
\eqref{eq:Tlarge_complement_0} and the triangle inequality; 
and \eqref{ineq:Lambda_2_1} follows from \eqref{eq:Tlarge_complement_0} and the fact that the minimum hamming distance between any two different rating vectors in $\cV$ is $\min\{\intraTau,\interTau\}m$.

\noindent \underline{(2-2) Lower Bound on $\lvert \misClassfSet{\betaEdge}{\alphaEdge} \rvert$ and $\lvert \misClassfSet{\alphaEdge}{\betaEdge} \rvert$:} 
For $T\in \TsmallErr$, a lower bound on $\lvert \misClassfSet{\alphaEdge}{\betaEdge} \rvert$ is given by
\begin{align}
    \lvert\misClassfSet{\betaEdge}{\alphaEdge}\rvert
    &=
    \sum\limits_{x \in [c]}
    \sum\limits_{y \in [c]}
    \sum\limits_{i \in [g]}
    \sum\limits_{j \in [g] \setminus i}
    \sum\limits_{k \in [g]}
    \kUsrs{i}{k}{x}{y} \kUsrs{j}{k}{x}{y}
    \label{eq:Nbetaalpha}\\
    &\geq
    \sum\limits_{x \in [c]}
    \sum\limits_{i \in [g]}
    \sum\limits_{j \in [g] \setminus i}
    \sum\limits_{k \in [g]}
    \kUsrs{i}{k}{x}{\sigma(x)} \kUsrs{j}{k}{x}{\sigma(x)}
    \nonumber\\
    &\geq
    \sum\limits_{x \in [c]}
    \sum\limits_{i \in [g]}
    \sum\limits_{j \in [g] \setminus i}
    \sum\limits_{k \in [g] \setminus \sigma(i|x)}
    \kUsrs{i}{k}{x}{\sigma(x)} \kUsrs{j}{k}{x}{\sigma(x)}
    \nonumber\\
    &=
    \sum\limits_{x \in [c]}
    \sum\limits_{i \in [g]}
    \sum\limits_{k \in [g] \setminus \sigma(i|x)}
    \kUsrs{i}{k}{x}{\sigma(x)}
    \left(
    \sum\limits_{j \in [g] \setminus i}
    \kUsrs{j}{k}{x}{\sigma(x)}
    \right)
    \nonumber\\
    &\geq
    \sum\limits_{x \in [c]}
    \sum\limits_{i \in [g]}
    \sum\limits_{k \in [g] \setminus \sigma(i|x)}
    \kUsrs{i}{k}{x}{\sigma(x)}
    \left(
    \left(1 -\tau\right) \frac{n}{gc}
    \right)
    \label{ineq:Nbetaalpha_1}\\
    &=
    \left(1 -\tau\right) \frac{n}{gc}
    \sum\limits_{x \in [c]}
    \sum\limits_{i \in [g]}
    \sum\limits_{j \in [g] \setminus \sigma(i|x)}
    \kUsrs{i}{j}{x}{\sigma(x)},
    \label{ineq:Nbetaalpha_2}
\end{align}
where \eqref{eq:Nbetaalpha} follows from the definitions in  \eqref{eq:kUsrs_defn} and \eqref{eq:N2rev_defn}; 
and \eqref{ineq:Nbetaalpha_1} follows from \eqref{eq:Tlarge_complement_0}.
Similarly, for $T\in \TsmallErr$, a lower bound on $\lvert \misClassfSet{\alphaEdge}{\betaEdge} \rvert$ is given by
\begin{align}
    \lvert\misClassfSet{\alphaEdge}{\betaEdge}\rvert
    &=
    \sum_{x\in[c]}
    \sum_{y\in[c]}
    \sum_{i\in[g]} 
    \sum_{j\in[g]} 
    \sum_{k\in[g] \setminus j}
    \kUsrs{i}{j}{x}{y} \kUsrs{i}{k}{x}{y}
    \label{eq:Nalphabeta}\\
    &\geq
    \sum_{x\in[c]}
    \sum_{i\in[g]} 
    \sum_{j\in[g]} 
    \sum_{k\in[g] \setminus j}
    \kUsrs{i}{j}{x}{\sigma(x)} \kUsrs{i}{k}{x}{\sigma(x)}
    \nonumber\\
    &\geq
    \sum_{x\in[c]}
    \sum_{i\in[g]} 
    \sum_{k\in[g] \setminus \sigma(i|x)}
    \kUsrs{i}{\sigma(i|x)}{x}{\sigma(x)}
    \kUsrs{i}{k}{x}{\sigma(x)}
    \nonumber\\
    &\geq
    \sum_{x\in[c]}
    \sum_{i\in[g]} 
    \sum_{k\in[g] \setminus \sigma(i|x)}
    \left(1 -\tau\right) \frac{n}{gc}
    \kUsrs{i}{k}{x}{\sigma(x)}
    \nonumber\\
    &=
    \left(1 -\tau\right) \frac{n}{gc}
    \sum\limits_{x\in[c]} 
    \sum_{i\in[g]} 
    \sum_{j\in[g] \setminus \sigma(i|x)} 
    \kUsrs{i}{j}{x}{\sigma(x)},
    \label{ineq:Nalphabeta_3}
\end{align}
where \eqref{eq:Nalphabeta} follows from the definitions in \eqref{eq:kUsrs_defn} and \eqref{eq:N2_defn}.
Therefore, by \eqref{ineq:Nbetaalpha_2} and \eqref{ineq:Nalphabeta_3}, we obtain
\begin{align}
    \frac{
    \lvert\misClassfSet{\betaEdge}{\alphaEdge}\rvert
    + \lvert\misClassfSet{\alphaEdge}{\betaEdge}\rvert}
    {2}
    \geq 
    \left(1 -\tau\right) \frac{n}{gc}
    \sum\limits_{x\in[c]} 
    \sum_{i\in[g]} 
    \sum_{j\in[g] \setminus \sigma(i|x)} 
    \kUsrs{i}{j}{x}{\sigma(x)}.
    \label{ineq:Ng}
\end{align}

\noindent \underline{(2-3) Lower Bound on $\lvert \misClassfSet{\gammaEdge}{\alphaEdge} \rvert$ and $\lvert \misClassfSet{\alphaEdge}{\gammaEdge} \rvert$:}
For $T\in \TsmallErr$, a lower bound on $\lvert \misClassfSet{\gammaEdge}{\alphaEdge} \rvert$ is given by
\begin{align}
    \lvert\misClassfSet{\gammaEdge}{\alphaEdge}\rvert
    &=
    \sum_{x \in [c]}
    \sum_{y \in [c]}
    \sum_{z \in [c] \setminus x}
    \sum_{i \in [g]}
    \sum_{j \in [g]}
    \sum_{k \in [g]}
    \kUsrs{i}{j}{x}{y}  \kUsrs{k}{j}{z}{y} 
    \label{eq:Ngammaalpha}\\
    &\geq
    \sum_{x \in [c]}
    \sum_{y \in [c] \setminus \sigma(x)}
    \sum_{z \in [c] \setminus x}
    \sum_{i \in [g]}
    \sum_{j \in [g]}
    \sum_{k \in [g]}
    \kUsrs{i}{j}{x}{y}  \kUsrs{k}{j}{z}{y} 
    \nonumber\\
    &=
    \sum_{x \in [c]}
    \sum_{y \in [c] \setminus \sigma(x)}
    \sum_{i \in [g]}
    \sum_{j \in [g]}
    \kUsrs{i}{j}{x}{y}
    \left(
    \sum_{z \in [c] \setminus x}
    \sum_{k \in [g]}
    \kUsrs{k}{j}{z}{y} 
    \right)
    \nonumber\\
    &\geq
    \sum_{x \in [c]}
    \sum_{y \in [c] \setminus \sigma(x)}
    \sum_{i \in [g]}
    \sum_{j \in [g]}
    \kUsrs{i}{j}{x}{y}
    \left(
    \left(1 -\tau\right) \frac{n}{gc} 
    \right)
    \label{ineq:Ngammaalpha_2}\\
    &=
    \left(1 -\tau\right) \frac{n}{gc} 
    \sum_{x \in [c]}
    \sum_{y \in [c] \setminus \sigma(x)}
    \sum_{i \in [g]}
    \sum_{j \in [g]}
    \kUsrs{i}{j}{x}{y}, 
    \label{ineq:Ngammaalpha_4}
\end{align}
where \eqref{eq:Ngammaalpha} follows from the definitions in \eqref{eq:kUsrs_defn} and \eqref{eq:N3rev_defn}; 
and \eqref{ineq:Ngammaalpha_2} follows from \eqref{eq:Tlarge_complement_0}.
Similarly, for $T\in \TsmallErr$, a lower bound on $\lvert \misClassfSet{\alphaEdge}{\gammaEdge} \rvert$ is given by
\begin{align}
    \lvert\misClassfSet{\alphaEdge}{\gammaEdge}\rvert
    &=
    \sum_{x \in [c]}
    \sum_{y \in [c]}
    \sum_{z \in [c] \setminus y}
    \sum_{i \in [g]}
    \sum_{j \in [g]}
    \sum_{k \in [g]}
    \kUsrs{i}{j}{x}{y}  \kUsrs{i}{k}{x}{z} 
    \label{eq:Nalphagamma}\\
    &\geq
    \sum_{x \in [c]}
    \sum_{z \in [c] \setminus \sigma(x)}
    \sum_{i \in [g]}
    \sum_{k \in [g]}
    \sum_{j \in [g]}
    \kUsrs{i}{j}{x}{\sigma(x)}  \kUsrs{i}{k}{x}{z} 
    \nonumber\\
    &\geq
    \sum_{x \in [c]}
    \sum_{z \in [c] \setminus \sigma(x)}
    \sum_{i \in [g]}
    \sum_{k \in [g]}
    \kUsrs{i}{\sigma(i|x)}{x}{\sigma(x)}  \kUsrs{i}{k}{x}{z}  
    \nonumber\\
    &=
    \sum_{x \in [c]}
    \sum_{z \in [c] \setminus \sigma(x)}
    \sum_{i \in [g]}
    \sum_{k \in [g]}
    \left(1 -\tau\right) \frac{n}{gc} 
    \kUsrs{i}{k}{x}{z}  
    \nonumber\\
    &=
    \left(1 -\tau\right) \frac{n}{gc} 
    \sum_{x \in [c]}
    \sum_{y \in [c] \setminus \sigma(x)}
    \sum_{i \in [g]}
    \sum_{j \in [g]}
    \kUsrs{i}{j}{x}{y}, 
    \label{ineq:Nalphagamma_2}
\end{align}
where \eqref{eq:Nalphagamma} follows from the definitions in \eqref{eq:kUsrs_defn} and \eqref{eq:N3_defn}.
Hence, by \eqref{ineq:Ngammaalpha_4} and \eqref{ineq:Nalphagamma_2}, we obtain
\begin{align}
    \frac{
    \lvert\misClassfSet{\gammaEdge}{\alphaEdge}\rvert
    +
    \lvert\misClassfSet{\alphaEdge}{\gammaEdge}\rvert}{2}
    \geq 
    \left(1 -\tau\right) \frac{n}{gc}
    \sum_{x \in [c]}
    \sum_{y \in [c] \setminus x}
    \sum_{i \in [g]}
    \sum_{j \in [g]}
    \kUsrs{i}{j}{x}{y}.
    \label{ineq:Nc1}
\end{align}

\noindent \underline{(2-4) Lower Bound on $\lvert \misClassfSet{\gammaEdge}{\betaEdge} \rvert$ and $\lvert \misClassfSet{\betaEdge}{\gammaEdge} \rvert$:}
For $T\in \TsmallErr$, a lower bound on $\lvert \misClassfSet{\gammaEdge}{\betaEdge} \rvert$ is given by
\begin{align}
    \lvert\misClassfSet{\gammaEdge}{\betaEdge}\rvert
    &=
    \sum_{x \in [c]}
    \sum_{y \in [c]}
    \sum_{z \in [c] \setminus x}
    \sum_{i \in [g]}
    \sum_{k \in [g]}
    \sum_{j \in [g]}
    \sum_{\ell \in [g] \setminus j}
    \kUsrs{i}{j}{x}{y} 
    \kUsrs{k}{\ell}{z}{y} 
    \label{eq:Ngammabeta}\\
    &\geq
    \sum_{x \in [c]}
    \sum_{y \in [c] \setminus \sigma(x)}
    \sum_{z \in [c] \setminus x}
    \sum_{i \in [g]}
    \sum_{k \in [g]}
    \sum_{j \in [g]}
    \sum_{\ell \in [g] \setminus j}
    \kUsrs{i}{j}{x}{y} 
    \kUsrs{k}{\ell}{z}{y} 
    \nonumber\\
    &=
    \sum_{x \in [c]}
    \sum_{y \in [c] \setminus \sigma(x)}
    \sum_{i \in [g]}
    \sum_{j \in [g]}
    \kUsrs{i}{j}{x}{y} 
    \left(
    \sum_{\ell \in [g] \setminus j}
    \sum_{z \in [c] \setminus x}
    \sum_{k \in [g]}
    \kUsrs{k}{\ell}{z}{y}
    \right)
    \nonumber\\
    &\geq
    \sum_{x \in [c]}
    \sum_{y \in [c] \setminus \sigma(x)}
    \sum_{i \in [g]}
    \sum_{j \in [g]}
    \kUsrs{i}{j}{x}{y} 
    \left(
    \sum_{\ell \in [g] \setminus j}
    \left(1 -\tau\right) \frac{n}{gc}
    \right)
    \label{ineq:Ngammabeta_2}\\
    &=
    (g-1) \left(1 -\tau\right) \frac{n}{gc}
    \sum_{x \in [c]}
    \sum_{y \in [c] \setminus x}
    \sum_{i \in [g]}
    \sum_{j \in [g]}
    \kUsrs{i}{j}{x}{y}, 
    \label{ineq:Ngammabeta_3}
\end{align}
where \eqref{eq:Ngammabeta} follows from the definitions in \eqref{eq:kUsrs_defn} and \eqref{eq:N4rev_defn}; and \eqref{ineq:Ngammabeta_2} follows from \eqref{eq:Tlarge_complement_0}.
Similarly, for $T\in \TsmallErr$, a lower bound on $\lvert \misClassfSet{\betaEdge}{\gammaEdge} \rvert$ is given by
\begin{align}
    \lvert\misClassfSet{\betaEdge}{\gammaEdge}\rvert
    &=
    \sum_{x \in [c]}
    \sum_{y \in [c]}
    \sum_{z \in [c] \setminus y}
    \sum_{i \in [g]}
    \sum_{k \in [g] \setminus i}
    \sum_{j \in [g]}
    \sum_{\ell \in [g]}
    \kUsrs{i}{j}{x}{y} 
    \kUsrs{k}{\ell}{x}{z} 
    \label{eq:Nbetagamma}\\
    &\geq
    \sum_{x \in [c]}
    \sum_{z \in [c] \setminus \sigma(x)}
    \sum_{i \in [g]}
    \sum_{k \in [g] \setminus i}
    \sum_{j \in [g]}
    \sum_{\ell \in [g]}
    \kUsrs{i}{j}{x}{\sigma(x)} 
    \kUsrs{k}{\ell}{x}{z} 
    \nonumber\\
    &=
    \sum_{x \in [c]}
    \sum_{z \in [c] \setminus \sigma(x)}
    \sum_{k \in [g]}
    \sum_{i \in [g] \setminus k}
    \kUsrs{i}{\sigma(x|i)}{x}{\sigma(x)} 
    \sum_{\ell \in [g]}
    \kUsrs{k}{\ell}{x}{z} 
    \nonumber\\
    &\geq
    \sum_{x \in [c]}
    \sum_{z \in [c] \setminus \sigma(x)}
    \sum_{k \in [g]}
    \sum_{i \in [g] \setminus k}
    \left(1 -\tau\right) \frac{n}{gc}
    \sum_{\ell \in [g]}
    \kUsrs{k}{\ell}{x}{z} 
    \nonumber\\    
    &=
    \sum_{x \in [c]}
    \sum_{z \in [c] \setminus \sigma(x)}
    \sum_{k \in [g]}
    (g-1) \left(1 -\tau\right) \frac{n}{gc}
    \sum_{\ell \in [g]}
    \kUsrs{k}{\ell}{x}{z} 
    \nonumber\\
    &=
    (g-1) \left(1 -\tau\right) \frac{n}{gc}
    \sum_{x \in [c]}
    \sum_{y \in [c] \setminus \sigma(x)}
    \sum_{i \in [g]}
    \sum_{j \in [g]}
    \kUsrs{i}{j}{x}{y}, 
    \label{ineq:Nbetagamma_3}
\end{align}
where \eqref{eq:Nbetagamma} follows from the definitions in \eqref{eq:kUsrs_defn} and \eqref{eq:N4_defn}.
Therefore, by \eqref{ineq:Ngammabeta_3} and \eqref{ineq:Nbetagamma_3}, we obtain
\begin{align}
    \frac{
    \lvert\misClassfSet{\gammaEdge}{\betaEdge}\rvert
    +
    \lvert\misClassfSet{\betaEdge}{\gammaEdge}\rvert}{2}
    \geq 
    \left(1 -\tau\right) (g-1) \frac{n}{gc}
    \sum_{x \in [c]}
    \sum_{y \in [c] \setminus x}
    \sum_{i \in [g]}
    \sum_{j \in [g]}
    \kUsrs{i}{j}{x}{y}. 
    \label{ineq:Nc2}
\end{align}

By \eqref{ineq:Lambda_3}, \eqref{ineq:Ng}, \eqref{ineq:Nc1} and \eqref{ineq:Nc2}, an upper bound on $\mathsf{Term_2}$ is given by
\begin{align}
    \mathsf{Term_2}
    &=
    \exp\left(-\left(1+o(1)\right)  \left(\lvert\diffEntriesSet\rvert \Ir 
    + \Pleftrightarrow{\alphaEdge}{\betaEdge} \: \Ig \frac{\log n}{n}
    + \Pleftrightarrow{\alphaEdge}{\gammaEdge} \: \IcOne \frac{\log n}{n}
    + \Pleftrightarrow{\betaEdge}{\gammaEdge} \: \IcTwo \frac{\log n}{n} \right)\right)
    \nonumber\\
    &\leq
    \exp
    \left(
    - (1-\tau)
    \left(
    \frac{n \Ir}{gc}
	\left(
	\sum\limits_{x \in [c]} \sum\limits_{i \in [g]} 
	\dElmntsGen{i}{\sigma(i|x)}{x}{\sigma(x)}
	\right)
	+
	\left(
    \intraTau m \Ir + \frac{n \Ig}{gc} \frac{\log n}{n}
    \right)
    \left(\sum_{x\in[c]}\sum_{i\in[g]}\sum_{j \in [g] \setminus \sigma(i|x)} \kUsrs{i}{j}{x}{\sigma(x)}\right)
    \right.
    \right.
    \nonumber\\
    &\phantom{\leq\exp
    \left(- (1-\tau)\left(\right.\right.} \:\:\:
    \left.
    \left.
    + 
    \:
    \left(
    \interTau m \Ir + \frac{n \IcOne}{gc} \frac{\log n}{n} + \frac{ (g-1) n \IcTwo}{gc} \frac{\log n}{n}
    \right)
    \left(\sum_{x \in [c]}
    \sum_{y \in [c] \setminus \sigma(x)}
    \sum_{i \in [g]}
    \sum_{j \in [g]}
    \kUsrs{i}{j}{x}{y}\right) 
	\right)
    \right)
    \nonumber\\
    &\leq
    \exp
    \left(
    - (1-\tau) (1+\epsilon)
    \left(
    \frac{\log m}{g-r+1}
	\left(
	\sum\limits_{x \in [c]} \sum\limits_{i \in [g]} 
	\dElmntsGen{i}{\sigma(i|x)}{x}{\sigma(x)}
	\right)
	+
	\log n
    \left(\sum_{x\in[c]}\sum_{i\in[g]}\sum_{j \in [g] \setminus \sigma(i|x)} \kUsrs{i}{j}{x}{\sigma(x)}\right)
    \right.
    \right.
    \nonumber\\
    &\phantom{\leq\exp
    \left(- (1-\tau)(1+\epsilon)\left(\right.\right.} \:\:\:
    \left.
    \left.
    + 
    \:
    \log n
    \left(\sum_{x \in [c]}
    \sum_{y \in [c] \setminus \sigma(x)}
    \sum_{i \in [g]}
    \sum_{j \in [g]}
    \kUsrs{i}{j}{x}{y}\right) 
	\right)
    \right)
    \label{ineq:upp_exp_2}\\
    &\leq
    \exp
    \left(
    - \left( 1+\frac{\epsilon}{2} \right)
    \left(
    \frac{\log (c_1 m)}{g-r+1}
	\left(
	\sum\limits_{x \in [c]} \sum\limits_{i \in [g]} 
	\dElmntsGen{i}{\sigma(i|x)}{x}{\sigma(x)}
	\right)
	+
	\log n 
    \left(\sum_{x\in[c]}\sum_{i\in[g]}\sum_{j \in [g] \setminus \sigma(i|x)} \kUsrs{i}{j}{x}{\sigma(x)}\right)
    \right.
    \right.
    \nonumber\\
    &\phantom{\leq\exp
    \left(- (1-\tau)(1+\epsilon)\left(\right.\right.} \:\:\:
    \left.
    \left.
    + 
    \:
    \log n
    \left(\sum_{x \in [c]}
    \sum_{y \in [c] \setminus \sigma(x)}
    \sum_{i \in [g]}
    \sum_{j \in [g]}
    \kUsrs{i}{j}{x}{y}\right) 
	\right)
    \right)
    \label{ineq:upp_exp_3}\\
    &=
    \exp
    \left(
    - \left( 1+\frac{\epsilon}{2} \right)
    \left(
    \frac{\log (c_1 m)}{g-r+1}
	\left(
	\sum\limits_{x \in [c]} \sum\limits_{i \in [g]} 
	\dElmntsGen{i}{\sigma(i|x)}{x}{\sigma(x)}
	\right)
    + 
    \log n
    \left(
    \sum_{x \in [c]} 
    \sum_{i \in [g]} 
    \sum_{(y,j) \neq (\sigma(x),\sigma(i|x))} 
    \kUsrs{i}{j}{x}{y}\right) 
	\right)
    \right),
    \label{ineq:upp_exp_4}
\end{align}
where \eqref{ineq:upp_exp_2} follows from the sufficient conditions in \eqref{eq:suffCond_1}, \eqref{eq:suffCond_2} and \eqref{eq:suffCond_3}; 
and \eqref{ineq:upp_exp_3} holds since $\tau \leq  (\epsilon\log m - (2+\epsilon)\log(2q)) / (2 (1 + \epsilon)\log m))$ implies that $(1-\tau)(1+\epsilon)\log m \geq (1+\epsilon/2)\log(c_1m)$ and $(1-\tau)(1+\epsilon) \geq (1+(\epsilon/2))$.

Finally, by \eqref{eq:term1_lemma_large} and \eqref{ineq:upp_exp_4}, the function in the RHS of \eqref{eq:largeErr_LHS_app} is upper bounded by
\begin{align}
    &\sum\limits_{T \in \TsmallErr} 
	\left\vert \mathcal{X}(T) \right\vert
	\:\:
	\exp\left(-\left(1+o(1)\right) \left(\lvert\diffEntriesSet\rvert \Ir 
    + \Pleftrightarrow{\alphaEdge}{\betaEdge} \: \Ig \frac{\log n}{n}
    + \Pleftrightarrow{\alphaEdge}{\gammaEdge} \: \IcOne \frac{\log n}{n}
    + \Pleftrightarrow{\betaEdge}{\gammaEdge} \: \IcTwo \frac{\log n}{n} \right)\right).
    \nonumber \\
    &\quad\leq
    \sum\limits_{T \in \TsmallErr} 
    c_0
    \exp
    \left(
    - \frac{\epsilon}{2}
    \left(
    \frac{\log (c_1m)}{g-r+1}
	\left(
	\sum\limits_{x \in [c]} \sum\limits_{i \in [g]} 
	\dElmntsGen{i}{\sigma(i|x)}{x}{\sigma(x)}
	\right)
    + 
    \log n
    \left(
    \sum_{x \in [c]} 
    \sum_{i \in [g]} 
    \sum_{(y,j) \neq (\sigma(x),\sigma(i|x))} 
    \kUsrs{i}{j}{x}{y}\right) 
	\right)
    \right).
    \nonumber\\
    &\quad=
    \sum_{\ell_1 = 0}^{gc\tau  \min\{\intraTau, \interTau\} m}\:
    \sum_{\ell_2 = 0}^{\relabelConst n}
    \left| \left\{
    \sum\limits_{x \in [c]} \sum\limits_{i \in [g]} 
	\dElmntsGen{i}{\sigma(i|x)}{x}{\sigma(x)} = \ell_1, \:
	\sum_{x \in [c]} 
    \sum_{i \in [g]} 
    \sum_{(y,j) \neq (\sigma(x),\sigma(i|x))} 
    \kUsrs{i}{j}{x}{y} = \ell_2
    \right\} \right|
    \nonumber\\
    &\quad\phantom{=
    \sum_{\ell_1 = 1}^{gc\tau \min\{\intraTau, \interTau\} m}
    \sum_{\ell_2 = 1}^{\relabelConst n}}
    \times
    \exp
    \left(
    - \frac{\epsilon \log (c_1m)}{2(g-r+1)}
    \ell_1
    - \frac{\epsilon \log n}{2}
    \ell_2
    \right)
    \label{eq:largeErr_errComvg_0}\\
    &\quad=
    \sum_{\ell_1 = 1}^{gc\tau \min\{\intraTau, \interTau\} m}
    \left| \left\{
    \sum\limits_{x \in [c]} \sum\limits_{i \in [g]} 
	\dElmntsGen{i}{\sigma(i|x)}{x}{\sigma(x)} = \ell_1
    \right\} \right|
    \left| \left\{
	\sum_{x \in [c]} 
    \sum_{i \in [g]} 
    \sum_{(y,j) \neq (\sigma(x),\sigma(i|x))} 
    \kUsrs{i}{j}{x}{y} = 0
    \right\} \right|
    \exp
    \left(
    - \frac{\epsilon \log (c_1m)}{2(g-r+1)}
    \ell_1
    \right)
    \nonumber\\
    &\quad\phantom{=}\:
    + \sum_{\ell_2 = 1}^{\relabelConst n}
    \left| \left\{
    \sum\limits_{x \in [c]} \sum\limits_{i \in [g]} 
	\dElmntsGen{i}{\sigma(i|x)}{x}{\sigma(x)} = 0
    \right\} \right|
    \left| \left\{
	\sum_{x \in [c]} 
    \sum_{i \in [g]} 
    \sum_{(y,j) \neq (\sigma(x),\sigma(i|x))} 
    \kUsrs{i}{j}{x}{y} = \ell_2
    \right\} \right|
    \exp
    \left(
    - \frac{\epsilon \log n}{2}
    \ell_2
    \right)
    \nonumber\\
    &\quad\phantom{=}\:
    + \sum_{\ell_1 = 1}^{gc\tau \min\{\intraTau, \interTau\} m}\:
    \sum_{\ell_2 = 1}^{\relabelConst n}
    \left| \left\{
    \sum\limits_{x \in [c]} \sum\limits_{i \in [g]} 
	\dElmntsGen{i}{\sigma(i|x)}{x}{\sigma(x)} = \ell_1
    \right\} \right|
     \left| \left\{
	\sum_{x \in [c]} 
    \sum_{i \in [g]} 
    \sum_{(y,j) \neq (\sigma(x),\sigma(i|x))} 
    \kUsrs{i}{j}{x}{y} = \ell_2
    \right\} \right|
    \nonumber\\
    &\quad\phantom{\leq
    \sum_{\ell_1 = 1}^{gc\tau \min\{\intraTau, \interTau\} m}
    \sum_{\ell_2 = 1}^{\relabelConst n}}
    \times
    \exp
    \left(
    - \frac{\epsilon \log (c_1m)}{2(g-r+1)}
    \ell_1
    - \frac{\epsilon \log n}{2}
    \ell_2
    \right)
    \label{eq:largeErr_errComvg_1}\\
    &\quad=
    \sum_{\ell_1 = 1}^{gc\tau \min\{\intraTau, \interTau\} m}
    \binom{\ell_1 + gc}{gc}
    \exp
    \left(
    - \frac{\epsilon \log (c_1m)}{2(g-r+1)}
    \ell_1
    \right)
    +
    \sum_{\ell_2 = 1}^{\relabelConst n}
    \binom{\ell_2 + gc}{gc}
    \exp
    \left(
    - \frac{\epsilon \log n}{2}
    \ell_2
    \right)
    \nonumber\\
    &\quad\phantom{=}\:
    + \sum_{\ell_1 = 1}^{gc\tau \min\{\intraTau, \interTau\} m}\:
    \sum_{\ell_2 = 1}^{\relabelConst n}
    \binom{\ell_1 + gc -1}{gc -1}
    \binom{\ell_2 + gc -1}{gc -1}
    \exp
    \left(
    - \frac{\epsilon \log (c_1m)}{2(g-r+1)}
    \ell_1
    - \frac{\epsilon \log n}{2}
    \ell_2
    \right)
    \label{eq:largeErr_errComvg_2}\\
    &\quad\leq
    \sum_{\ell_1 = 1}^{gc\tau \min\{\intraTau, \interTau\} m}
    2^{\left(\displaystyle{\ell_1 + gc}\right)}\:\:
    (c_1m)^{\left(\displaystyle - \frac{\epsilon}{2(g-r+1)}
    \ell_1\right)}
    \:\:+\:\:
    \sum_{\ell_2 = 1}^{\relabelConst n}
    2^{\left(\displaystyle \ell_2 + gc\right)}\:\:
    n^{\left( \displaystyle -  \frac{\epsilon}{2} \ell_2\right)}
    \nonumber\\
    &\quad\phantom{\leq}\:
    + \sum_{\ell_1 = 1}^{gc\tau \min\{\intraTau, \interTau\} m}
    2^{\left(\displaystyle{\ell_1 + gc}\right)}\:\:
    (c_1m)^{\left(\displaystyle - \frac{\epsilon}{2(g-r+1)}
    \ell_1\right)}
    \left(
    \sum_{\ell_2 = 1}^{\relabelConst n}
    2^{\left(\displaystyle \ell_2 + gc\right)}\:\:
    n^{\left( \displaystyle -  \frac{\epsilon}{2} \ell_2\right)}
    \right)
    \label{eq:largeErr_errComvg_3}\\
    &\quad\leq
    2^{\displaystyle gc}
    \sum_{\ell_1 = 1}^{\infty}
    \left(2\:(c_1m)^{\left(\displaystyle - \frac{\epsilon}{2(g-r+1)}\right)}\right)^{\displaystyle \ell_1}
    \:\:+\:\:
    2^{\displaystyle gc}\:
    \sum_{\ell_2 = 1}^{\infty}
    \left(2 n^{\left( \displaystyle -  \frac{\epsilon}{2}\right)}\right)^{\displaystyle \ell_2}
    \nonumber\\
    &\quad\phantom{\leq}\:
    + 2^{\displaystyle 2gc}\:
    \sum_{\ell_1 = 1}^{\infty}
    \left(2 \:(c_1 m)^{\left(\displaystyle - \frac{\epsilon}{2(g-r+1)}\right)}\right)^{\displaystyle \ell_1}
    \left[
    \sum_{\ell_2 = 1}^{\infty}
    \left(2 n^{\left( \displaystyle -  \frac{\epsilon}{2}\right)}\right)^{\displaystyle \ell_2}
    \right]
    \nonumber\\
    &\quad=
    2^{\displaystyle gc}\:
    \frac{2 \:(c_1 m)^{\left(\displaystyle - \frac{\epsilon}{2(g-r+1)}\right)}}{1 - 2 \:(c_1m)^{\left(\displaystyle - \frac{\epsilon}{2(g-r+1)}\right)}}
    \:+\:
    2^{\displaystyle gc}\:
    \frac{2 n^{\left( \displaystyle -  \frac{\epsilon}{2}\right)}}{1 - 2 n^{\left( \displaystyle -  \frac{\epsilon}{2}\right)}}
    \:+\: 
    2^{\displaystyle 2gc}\:
    \frac{2\:(c_1m)^{\left(\displaystyle - \frac{\epsilon}{2(g-r+1)}\right)}}{1 - 2 \:(c_1m)^{\left(\displaystyle - \frac{\epsilon}{2(g-r+1)}\right)}}
    \frac{2 n^{\left( \displaystyle -  \frac{\epsilon}{2}\right)}}{1 - 2 n^{\left( \displaystyle - \frac{\epsilon}{2}\right)}},
    \label{eq:largeErr_errComvg_4}
\end{align}
where
\begin{itemize}
    \item \eqref{eq:largeErr_errComvg_0} readily follows from \eqref{eq:Tlarge_complement_0};
    \item in \eqref{eq:largeErr_errComvg_1}, we break the summation into three summations, and use the fact that the enumeration of the first element of the set is independent of the enumeration of the second element;
    \item in \eqref{eq:largeErr_errComvg_2}, we use the fact that the number of integer solutions of $\sum_{i=1}^{n} x_i = s$ is equal to $\binom{s+n-1}{n-1}$;
    \item in \eqref{eq:largeErr_errComvg_3}, we bound each binomial coefficient by
    $\binom{a}{b} \leq \sum_{i=0}^a \binom{a}{i} \leq 2^a$, for $a \geq b$;
    \item and finally in \eqref{eq:largeErr_errComvg_4}, we evaluate the infinite geometric series, where $\epsilon > \max\{(2 \log 2) / \log n, \: (2(g-r+1) \log 2) / \log m\}$.
\end{itemize}
Therefore, by \eqref{eq:largeErr_errComvg_4}, the RHS of \eqref{eq:largeErr_LHS_app} is given by
\begin{align}
    \lim_{n,m\rightarrow \infty}
    \sum\limits_{T \in \TsmallErr} 
	\sum\limits_{X \in \mathcal{X}(T)}
	\exp\left(-\left(1+o(1)\right) \left(\lvert\diffEntriesSet\rvert \Ir 
    + \Pleftrightarrow{\alphaEdge}{\betaEdge} \: \Ig \frac{\log n}{n}
    + \Pleftrightarrow{\alphaEdge}{\gammaEdge} \: \IcOne \frac{\log n}{n}
    + \Pleftrightarrow{\betaEdge}{\gammaEdge} \: \IcTwo \frac{\log n}{n} \right)\right) = 0.
\end{align}
Note that as $n$ tends to infinity, the condition on $\epsilon$ becomes 
\begin{align}
    \epsilon > \lim_{n,m\rightarrow \infty}
    \max\left\{
    \frac{2 \log 2}{\log n}, \: 
    \frac{2(g-r+1) \log 2}{\log(c_1m)},
    \frac{2\log c_1}{\log(m/c_1)}
    \right\} = 0.
\end{align}
This completes the proof of Lemma~\ref{lemma:T1_related}.
\hfill $\blacksquare$
\end{JA_AE}

\section{Proof of Lemma~\ref{lemma:eta_omega(nm)}}
\label{proof:eta_omega(nm)}
\begin{JA_AE}
The LHS of \eqref{ineq:eta_omega(nm)} is upper bounded by
\begin{align}
    & \lim_{n,m\rightarrow \infty}
    \sum\limits_{T \in \TlargeErr} 
	\sum\limits_{X \in \mathcal{X}(T)}
	\exp\left(-\left(1+o(1)\right) \left(\lvert\diffEntriesSet\rvert \Ir 
    + \Pleftrightarrow{\alphaEdge}{\betaEdge} \: \Ig \frac{\log n}{n}
    + \Pleftrightarrow{\alphaEdge}{\gammaEdge} \: \IcOne \frac{\log n}{n}
    + \Pleftrightarrow{\betaEdge}{\gammaEdge} \: \IcTwo \frac{\log n}{n} \right)\right)
    \nonumber \\
    &\qquad \leq
    \lim_{n,m\rightarrow \infty}
    \sum\limits_{T \in \TlargeErr} 
	\left\vert \mathcal{X}(T) \right\vert
	\exp\left(-\left(\lvert\diffEntriesSet\rvert \Ir 
    + \Pleftrightarrow{\alphaEdge}{\betaEdge} \: \Ig \frac{\log n}{n}
    + \Pleftrightarrow{\alphaEdge}{\gammaEdge} \: \IcOne \frac{\log n}{n}
    + \Pleftrightarrow{\betaEdge}{\gammaEdge} \: \IcTwo \frac{\log n}{n} \right)\right).
	\label{ineq:largeErr}
\end{align}
We first partition the set $\TlargeErr$ into two subsets (regimes), denoted by $\TlargeOne$ and $\TlargeTwo$. 
They are defined as follows:
\begin{align}
    \TlargeOne
    &=  
    \left\{ T \in \TlargeErr : 
    \exists (x,i) \in [c] \times [g],\:
    \text{ such that }
    \left(\left\vert \sigma(x,i) \right\vert = 0 \right)
    \right\}, \label{eq:Tlarge_1}
    \\
    \TlargeTwo
    &=  
    \left\{ T \in \TlargeErr : 
    \forall (x,i) \in [c] \!\times\! [g]
    \text{ such that }
    \left\vert \sigma(x,i) \right\vert = 1,
    \text{ and }
    \exists (x,i) \in [c] \times [g]
    \text{ such that }
    \dElmntsGen{i}{\sigma(i|x)}{x}{\sigma(x)} > \tau m \min\{\interTau, \intraTau\}
    \right\}.
    \label{eq:Tlarge_2}
\end{align}
Therefore, the RHS of \eqref{ineq:largeErr} is upper bounded by
\begin{align}
    &\lim_{n,m\rightarrow \infty}
    \sum\limits_{T \in \TlargeErr} 
	\sum\limits_{X \in \mathcal{X}(T)}
	\exp\left(-\left(1+o(1)\right) \left(\lvert\diffEntriesSet\rvert \Ir 
    + \Pleftrightarrow{\alphaEdge}{\betaEdge} \: \Ig \frac{\log n}{n}
    + \Pleftrightarrow{\alphaEdge}{\gammaEdge} \: \IcOne \frac{\log n}{n}
    + \Pleftrightarrow{\betaEdge}{\gammaEdge} \: \IcTwo \frac{\log n}{n} \right)\right)
    \nonumber \\
    &\qquad \leq
    \lim_{n,m\rightarrow \infty}
    \left[
    \sum\limits_{T \in \TlargeOne} 
	\left\vert \mathcal{X}(T) \right\vert
	\exp\left(-\left(\lvert\diffEntriesSet\rvert \Ir 
    + \Pleftrightarrow{\alphaEdge}{\betaEdge} \: \Ig \frac{\log n}{n}
    + \Pleftrightarrow{\alphaEdge}{\gammaEdge} \: \IcOne \frac{\log n}{n}
    + \Pleftrightarrow{\betaEdge}{\gammaEdge} \: \IcTwo \frac{\log n}{n} \right)\right)
    \right.
    \nonumber\\
    &\qquad \phantom{\leq}
    \left.
    + 
    \sum\limits_{T \in \TlargeTwo} 
	\left\vert \mathcal{X}(T) \right\vert
	\exp\left(-\left(\lvert\diffEntriesSet\rvert \Ir 
    + \Pleftrightarrow{\alphaEdge}{\betaEdge} \: \Ig \frac{\log n}{n}
    + \Pleftrightarrow{\alphaEdge}{\gammaEdge} \: \IcOne \frac{\log n}{n}
    + \Pleftrightarrow{\betaEdge}{\gammaEdge} \: \IcTwo \frac{\log n}{n} \right)\right)
    \right].
	\label{ineq:largeErr_1}
\end{align}
In what follows, we derive upper bounds on each summation term in \eqref{ineq:largeErr_1}.

\subsection{Large Grouping Error Regime:}
This regime corresponds to $\TlargeOne$ characterized by \eqref{eq:Tlarge_1}.
Suppose that there exist a cluster $x_0 \in [c]$ and a group $i_0 \in [g]$ such that $\left\vert \sigma(x_0,i_0) \right\vert = 0$. 
By \eqref{eq:sigma(x,i)}, this implies that
\begin{align}
    \kUsrs{i_0}{j}{x_0}{y}
    =
    \left\vert Z_0(x_0,i_0) \cap Z(y,j) \right\vert
    \leq 
    (1-\tau) \frac{n}{gc},
    \:\:\forall (y,j) \in [c] \times [g].
    \label{eq:regime2_1}
\end{align}
We further partition the set $\TlargeOne$ into three subregimes $\TlargeOneOne$, $\TlargeOneTwo$, and $\TlargeOneThree$ that are defined as follows:
\begin{align}
    \TlargeOneOne
    &=
    \left\{
    T \in \TlargeOne:
    \exists \mu > 0,\:
    \exists (y_1, j_1) \in [c] \times [g],\:
    \exists (y_2, j_2) \in [c] \times [g]
    \text{ such that }
    \kUsrs{i_0}{j_1}{x_0}{y_1} \geq \mu n,\:
    \kUsrs{i_0}{j_2}{x_0}{y_2} \geq \mu n
    \right\}, \label{eq:TlargeOneOne} 
    \\
    \TlargeOneTwo
    &=
    \left\{
    T \in \TlargeOne:
    \exists \mu > 0,\:
    \exists (y_1, j_1) \in [c] \times [g]
    \text{ such that }
    \kUsrs{i_0}{j_1}{x_0}{y_1} \geq \mu n
    \right\}, 
    \label{eq:TlargeOneTwo}\\
    \TlargeOneThree
    &=
    \left\{
    T \in \TlargeOne:
    \forall \mu > 0,\:
    \forall (y, j) \in [c] \times [g]
    \text{ such that }
    \kUsrs{i_0}{j}{x_0}{y} < \mu n
    \right\}
    . \label{eq:TlargeOneThree}
\end{align}

\noindent \underline{(1) Subregime 1-1:}
This subregime corresponds to $\TlargeOneOne$ characterized by \eqref{eq:TlargeOneOne}.
Suppose that there exist a constant $\mu > 0$, and two distinct pairs $(y_1,j_1), (y_2,j_2) \in [c]\times[g]$ such that 
\begin{align}
    \kUsrs{i_0}{j_1}{x_0}{y_1}
    \geq 
    \mu n,
    \text{ and }
    \kUsrs{i_0}{j_2}{x_0}{y_2}
    \geq
    \mu n.
    \label{prop:R2_largeErr}
\end{align}
There are $n$ users, each of which belong to one of the $gc$ groups. Moreover, each user rates $m$ items, where each item rating can be one of $q^{gc}$ possible ratings across all users.
Therefore, a loose upper bound on the number of matrices that belong to matrix class $\mathcal{X}(T)$ is given by
\begin{align}
    \left\vert \mathcal{X}(T) \right\vert 
    \leq 
    (gc)^n \left(q^{gc}\right)^{m},
    \quad \forall T \in \tupleSetDelta.
    \label{ineq:cardinality}
\end{align}
Next, a lower bound on $\lvert\diffEntriesSet\rvert$ is given by
\begin{align}
    \lvert\diffEntriesSet\rvert
	&=
	\sum\limits_{x \in [c]}
	\sum\limits_{i \in [g]}
	\sum\limits_{y \in [c]}
	\sum\limits_{j \in [g]}
	\kUsrs{i}{j}{x}{y} \dElmntsGen{i}{j}{x}{y}
	\label{eq:regime2_Pd_0}\\
	&\geq
	\kUsrs{i_0}{j_1}{x_0}{y_1} \dElmntsGen{i_0}{j_1}{x_0}{y_1}
	+
	\kUsrs{i_0}{j_2}{x_0}{y_2} \dElmntsGen{i_0}{j_2}{x_0}{y_2}
	\nonumber\\
	&> 
	\mu n
	\left(
	\dElmntsGen{i_0}{j_1}{x_0}{y_1}
	+ \dElmntsGen{i_0}{j_2}{x_0}{y_2}
	\right)
	\label{eq:regime2_Pd_1}\\
	&\geq
	\mu n\:
	\hamDist{\vecV{j_1}{y_1}}{\vecV{j_2}{y_2}} 
	\label{eq:regime2_Pd_2}\\
	&\geq
	\mu\min\left\{\intraTau,\interTau \right\}
	nm,
	\label{eq:regime2_Pd_3}
\end{align}
where \eqref{eq:regime2_Pd_0} follows from the definitions of $\diffEntriesSet$, $\kUsrs{i}{j}{x}{y}$ and $\dElmntsGen{i}{j}{x}{y}$ in \eqref{eq:N1_defn}, \eqref{eq:kUsrs_defn} and \eqref{eq:dElmntsGen_defn}, respectively; 
\eqref{eq:regime2_Pd_1} follows from \eqref{prop:R2_largeErr}; 
\eqref{eq:regime2_Pd_2} follows from the triangle inequality; and  
\eqref{eq:regime2_Pd_3} holds since the minimum hamming distance between any two different rating vectors in $\cV$ is $\min\{\intraTau,\interTau\}m$. 
Furthermore, if $y_1 = y_2$, then $\lvert\misClassfSet{\alphaEdge}{\betaEdge} \rvert$ is lower bounded by
\begin{align}
    \lvert\misClassfSet{\alphaEdge}{\betaEdge} \rvert
    &=
    \sum_{x\in[c]}
    \sum_{y\in[c]}
    \sum_{i\in[g]}
    \sum_{k\in[g]}
    \sum_{j\in[g]\setminus k}
    \kUsrs{i}{j}{x}{y}\kUsrs{i}{k}{x}{y} 
    \label{eq:regime2_Palphabeta_0}\\
    &\geq
    \kUsrs{i_0}{j_1}{x_0}{y_1}\kUsrs{i_0}{j_2}{x_0}{y_2}
    \nonumber \\
    &\geq 
    (\mu n)^2,
    \label{eq:regime2_Palphabeta}
\end{align}
where \eqref{eq:regime2_Palphabeta_0} follows from the definitions in \eqref{eq:kUsrs_defn} and \eqref{eq:N2_defn}.
On the other hand, if $y_1\neq y_2$, then $\lvert\misClassfSet{\alphaEdge}{\gammaEdge}\rvert$ is lower bounded by
\begin{align}
    \lvert\misClassfSet{\alphaEdge}{\gammaEdge}\rvert
    &=
    \sum_{x \in [c]}
    \sum_{z \in [c]}
    \sum_{y \in [c] \setminus z}
    \sum_{i \in [g]}
    \sum_{k \in [g]}
    \sum_{j \in [g]}
    \kUsrs{i}{j}{x}{y}  \kUsrs{i}{k}{x}{z} 
    \label{eq:regime2_Palphagamma_0}\\
    &\geq
    \kUsrs{i_0}{j_1}{x_0}{y_1}\kUsrs{i_0}{j_2}{x_0}{y_2}
    \nonumber \\
    &\geq 
    (\mu n)^2,
    \label{eq:regime2_Palphagamma}
\end{align}
where \eqref{eq:regime2_Palphagamma_0} follows from the definitions in \eqref{eq:kUsrs_defn} and \eqref{eq:N3_defn}.
Finally, the first summation term in the RHS of \eqref{ineq:largeErr} is upper bounded by
\begin{align}
    &\sum\limits_{T \in \TlargeOneOne} 
	\left\vert \mathcal{X}(T) \right\vert
	\exp\left(-\left(\lvert\diffEntriesSet\rvert \Ir 
    + \Pleftrightarrow{\alphaEdge}{\betaEdge} \: \Ig \frac{\log n}{n}
    + \Pleftrightarrow{\alphaEdge}{\gammaEdge} \: \IcOne \frac{\log n}{n}
    + \Pleftrightarrow{\betaEdge}{\gammaEdge} \: \IcTwo \frac{\log n}{n} \right)\right)
    \nonumber\\
    &\leq
    \sum\limits_{T \in \TlargeOneOne} 
	\left\vert \mathcal{X}(T) \right\vert
	\exp\left(
	-\left(\lvert\diffEntriesSet\rvert \Ir 
    +  \frac{\Ig}{2} \frac{\log n}{n} \lvert \misClassfSet{\alphaEdge}{\betaEdge} \rvert
    +  \frac{\IcOne}{2} \frac{\log n}{n} \lvert \misClassfSet{\alphaEdge}{\gammaEdge} \rvert
    \right)\right)
    \nonumber\\
    &=
    \exp\left(
	-\left(c_2 nm \frac{\log m}{n}
    +  c_3 \frac{\log n}{n} n^2
    \right)\right)
    \sum\limits_{T \in \TlargeOneOne} 
	\left\vert \mathcal{X}(T) \right\vert
	\label{eq:largeErrConvg_R2_0}\\
	&\leq 
    \exp\left(
	-\left(c_2 m \log m + c_3 n \log n
    \right)\right)
    \exp\left(n \log (gc)\right)
    \exp\left(m gc \log q\right)
	\label{eq:largeErrConvg_R2_1}\\
	&= 
	\exp\left(
	-\left(m \left(c_2 \log m - gc \log q\right) 
	+ n \left(c_3 \log n - \log (gc)\right)
    \right)\right),
	\label{eq:largeErrConvg_R2_2}
\end{align}
where $c_2$ and $c_3$ in \eqref{eq:largeErrConvg_R2_0} are some positive constants; 
\eqref{eq:largeErrConvg_R2_0} follows from \eqref{eq:suffCond_1}, \eqref{eq:regime2_Pd_3}, \eqref{eq:regime2_Palphabeta} and \eqref{eq:regime2_Palphagamma}; and 
\eqref{eq:largeErrConvg_R2_1} follows from \eqref{ineq:cardinality}. 

\noindent \underline{(2) Subregime 1-2:}
This subregime corresponds to $\TlargeOneTwo$ characterized by \eqref{eq:TlargeOneTwo}.
Suppose that there exist only one pair $(y_1,j_1) \in [c] \times [g]$, and a constant $\mu > 0$ such that 
\begin{align}
    \kUsrs{i_0}{j_1}{x_0}{y_1} \geq \mu n.
\end{align}
This implies that
\begin{align}
    \kUsrs{i_0}{j}{x_0}{y} < \frac{\tau}{(gc-1) gc} n, 
    \:\:\text{ for }
    (y,j) \neq (y_0, j_0). 
    \label{ineq:regime1_2_1}
\end{align}
Therefore, by \eqref{ineq:regime1_2_1}, we have
\begin{align}
    \kUsrs{i_0}{j_0}{x_0}{y_0} 
    &=
    \frac{n}{gc} - \sum_{(y,j) \neq (y_0, j_0)} \kUsrs{i_0}{j}{x_0}{y}
    \nonumber \\
    &> 
    \frac{n}{gc} - (gc-1) \frac{\tau}{(gc-1)} \frac{n}{gc}
    \nonumber\\
    &=
    (1-\tau)\frac{n}{gc}. \label{ineq:regime1_2_2}
\end{align}
However, this is in contradiction with \eqref{eq:regime2_1}.
Hence, we conclude that subregime $\TlargeOneTwo$ is impossible to exist.

\noindent \underline{(3) Subregime 1-3:}
This subregime corresponds to $\TlargeOneThree$ characterized by \eqref{eq:TlargeOneThree}.
Due to the fact that
\begin{align}
    \sum_{y\in[c]}\sum_{j\in[g]}\kUsrs{i_0}{j}{x_0}{y} = \frac{n}{gc}, 
\end{align}
there should be at least one pair $(y_1, j_1)$ such that 
\begin{align}
    \kUsrs{i_0}{j_1}{x_0}{y_1} \geq \mu n,
\end{align}
for some $\mu > 0$. However, this is in contradiction with \eqref{eq:TlargeOneThree}.
Thus, we conclude that subregime $\TlargeOneThree$ is impossible to exist.

As a result, we conclude that
\begin{align}
    &\sum\limits_{T \in \TlargeOne} 
	\left\vert \mathcal{X}(T) \right\vert
	\exp\left(-\left(\lvert\diffEntriesSet\rvert \Ir 
    + \Pleftrightarrow{\alphaEdge}{\betaEdge} \: \Ig \frac{\log n}{n}
    + \Pleftrightarrow{\alphaEdge}{\gammaEdge} \: \IcOne \frac{\log n}{n}
    + \Pleftrightarrow{\betaEdge}{\gammaEdge} \: \IcTwo \frac{\log n}{n} \right)\right)
    \nonumber\\
    &\qquad \leq
    \exp\left(
	-\left(m \left(c_2 \log m - gc \log q\right) 
	+ n \left(c_3 \log n - \log (gc)\right)
    \right)\right).
    \label{eq:R1_upperBound}
\end{align}

\subsection{Large Rating Estimation Error Regime:}
This regime corresponds to $\TlargeTwo$ characterized by \eqref{eq:Tlarge_2}.
Suppose the following conditions hold:
\begin{itemize}
    \item For every $(x,i) \in [c]\times [g]$, there exists a pair $(y,j) \in [c]\times[g]$ such that $\left\vert \sigma(x,i) \right\vert = 1$.
    That is, by \eqref{eq:sigma(x,i)}, we have
    \begin{align}
        \exists (y,j) = (\sigma(x), \sigma(i|x)) \in [c]\times [g] : 
        \kUsrs{i}{j}{x}{y} = 
         \left| Z_0(x,i) \cap  Z(y,j)\right| 
         &\geq (1-\relabelConst) \frac{n}{gc}, 
        \quad \forall (x,i) \in [c]\times [g].
        \label{eq:regime3_large_1}
    \end{align}
    \item There exists $(x_0, i_0) \in [c]\times[g]$ such that $\left\vert \sigma(x_0,i_0) \right\vert = 1$, and
    \begin{align}
    \dElmntsGen{i_0}{j_0}{x_0}{y_0}
    = \dElmntsGen{i_0}{\sigma(i_0 | x_0)}{x_0}{\sigma(x_0)} > \tau m \min\{\interTau, \intraTau\}.
        \label{eq:regime3_large_2}
    \end{align}
\end{itemize}

We first provide an upper bound on $\left\vert \mathcal{X}(T) \right\vert$. By \eqref{eq:term1_11_12}, \eqref{eq:term11_1} and \eqref{ineq:cardinality}, an upper bound on $\left\vert \mathcal{X}(T) \right\vert$ is given by
\begin{align}
    \left\vert \mathcal{X}(T) \right\vert
    &\leq
    \left(q^{gc}\right)^{m}
    \exp
    \left(
    \left(
    \sum_{x \in [c]} 
    \sum_{i \in [g]} 
    \sum_{(y,j) \neq (\sigma(x), \sigma(i|x))} 
    \kUsrs{i}{j}{x}{y}
    \right)
    \log n
    \right).
    \label{eq:XT_regime2}
\end{align}

Next, we provide a lower bound on $\lvert\diffEntriesSet\rvert$.
Based on \eqref{eq:regime3_large_1}, if there exists at least one other pair $(\widehat{y},\widehat{j}) \neq (\sigma(x), \sigma(i|x))$ for some $(x,i) \in [c] \times [g]$ such that
\begin{align}
    \kUsrs{i}{\widehat{j}}{x}{\widehat{y}} 
    = \left| Z_0(x,i) \cap  Z(\widehat{y},\widehat{j})\right| 
    \geq \mu n,
\end{align}
for some constant $\mu > 0$, then the analysis of this case boils down to Subregime 1-1. 
Therefore, we assume that for every $(x,i) \in [c]\times [g]$, we have
\begin{align}
    \kUsrs{i}{j}{x}{y} < \mu n,
    \quad \forall (y,j) \neq (\sigma(x), \sigma(i|x)),
    \: \forall \mu > 0.
    \label{eq:regime3_large_3}
\end{align} 
Consequently, a lower bound on $\lvert\diffEntriesSet\rvert$ is given by
\begin{align}
    \lvert\diffEntriesSet\rvert
	&=
	\sum\limits_{x \in [c]}
	\sum\limits_{i \in [g]}
	\sum\limits_{y \in [c]}
	\sum\limits_{j \in [g]}
	\kUsrs{i}{j}{x}{y} \dElmntsGen{i}{j}{x}{y}
	\nonumber\\
	&\geq
	\kUsrs{i_0}{j_0}{x_0}{y_0} \dElmntsGen{i_0}{j_0}{x_0}{y_0}
	\nonumber\\
	&>
	\left((1-\relabelConst) \frac{n}{gc}\right)
	\left(\tau m \min\{\interTau, \intraTau\} \right)
	\label{eq:regime3_Pd_0}\\
	&=
	\left( \frac{(1-\relabelConst) \tau \min\{\interTau, \intraTau\}}{2gc} \right)
	m n
	+
    \left( \frac{(1-\relabelConst) \tau \min\{\interTau, \intraTau\}}{2gc} \right)
	m n
	\nonumber \\
	&\geq
	\left( \frac{(1-\relabelConst) \tau \min\{\interTau, \intraTau\}}{2gc} \right)
	m n
	+
	((1-\relabelConst)m)
	\left(
	\sum_{x \in [c]} 
    \sum_{i \in [g]} 
    \sum_{(y,j) \neq \sigma(x,i)} 
    \kUsrs{i}{j}{x}{y}
    \right)
    \label{eq:regime3_Pd_1}\\
    &=
    \left( \frac{(1-\relabelConst) \tau \min\{\interTau, \intraTau\}}{2gc} \right)
	m n
	+
	((1-\relabelConst)m)
	\left(
	\sum\limits_{x \in [c]}
	\sum\limits_{i \in [g]}
	\sum\limits_{j \in [g]\setminus \sigma(i|x)}
	\kUsrs{i}{j}{x}{\sigma(x)} 
	+
	\sum\limits_{x \in [c]}
	\sum\limits_{y \in [c] \setminus \sigma(x)}
	\sum\limits_{i \in [g]}
	\sum\limits_{j \in [g]}
	\kUsrs{i}{j}{x}{y}
    \right)
    \nonumber\\
    &\geq
    c_4 m n
    +
    ((1-\relabelConst)m)
    \left[
    \intraTau
    \left(
	\sum\limits_{x \in [c]}
	\sum\limits_{i \in [g]}
	\sum\limits_{j \in [g]\setminus \sigma(i|x)}
	\kUsrs{i}{j}{x}{\sigma(x)} 
	\right)
	+
	\interTau
	\left(
	\sum\limits_{x \in [c]}
	\sum\limits_{y \in [c] \setminus \sigma(x)}
	\sum\limits_{i \in [g]}
	\sum\limits_{j \in [g]}
	\kUsrs{i}{j}{x}{y}
	\right)
	\right],
	\label{eq:regime3_Pd_2}
\end{align}
where \eqref{eq:regime3_Pd_0} follows from \eqref{eq:regime3_large_1} and \eqref{eq:regime3_large_2}; 
\eqref{eq:regime3_Pd_1} follows from \eqref{eq:regime3_large_3} for $\mu = (\tau \min\{\intraTau, \interTau\}) / (2(gc-1)(gc)^2)$; and \eqref{eq:regime3_Pd_2} follows by setting $c_4 = ((1-\relabelConst) \tau \min\{\interTau, \intraTau\})/(2gc)$ where $0 \leq c_4 < 1$.

On the other hand, recall from \eqref{ineq:Ng}, \eqref{ineq:Nc1} and \eqref{ineq:Nc2} the following  lower bounds: 
\begin{align}
\begin{split}
    \frac{
    \lvert\misClassfSet{\betaEdge}{\alphaEdge}\rvert
    + \lvert\misClassfSet{\alphaEdge}{\betaEdge}\rvert}
    {2}
    &\geq
    \left(1 -\tau\right) \frac{n}{gc}
    \sum\limits_{x\in[c]} 
    \sum_{i\in[g]} 
    \sum_{j\in[g] \setminus \sigma(i|x)} 
    \kUsrs{i}{j}{x}{\sigma(x)},
    \\
    \frac{
    \lvert\misClassfSet{\gammaEdge}{\alphaEdge}\rvert
    +
    \lvert\misClassfSet{\alphaEdge}{\gammaEdge}\rvert}{2}
    &\geq
    \left(1 -\tau\right) \frac{n}{gc}
    \sum_{x \in [c]}
    \sum_{y \in [c] \setminus \sigma(x)}
    \sum_{i \in [g]}
    \sum_{j \in [g]}
    \kUsrs{i}{j}{x}{y},
    \\
    \frac{
    \lvert\misClassfSet{\gammaEdge}{\betaEdge}\rvert
    +
    \lvert\misClassfSet{\betaEdge}{\gammaEdge}\rvert}{2}
    &\geq
    \left(1 -\tau\right) (g-1) \frac{n}{gc}
    \sum_{x \in [c]}
    \sum_{y \in [c] \setminus \sigma(x)}
    \sum_{i \in [g]}
    \sum_{j \in [g]}
    \kUsrs{i}{j}{x}{y}.
    \label{eq:regime3_abg_2}
\end{split}
\end{align}

Finally, the second summation term in the RHS of \eqref{ineq:largeErr} is upper bounded by
\begin{align}
    &\sum\limits_{T \in \TlargeTwo} 
	\left\vert \mathcal{X}(T) \right\vert
	\exp\left(-\left(\lvert\diffEntriesSet\rvert \Ir 
    + \Pleftrightarrow{\alphaEdge}{\betaEdge} \: \Ig \frac{\log n}{n}
    + \Pleftrightarrow{\alphaEdge}{\gammaEdge} \: \IcOne \frac{\log n}{n}
    + \Pleftrightarrow{\betaEdge}{\gammaEdge} \: \IcTwo \frac{\log n}{n} \right)\right)
    \nonumber\\
    &\leq
    \sum\limits_{T \in \TlargeTwo} 
    \left\vert \mathcal{X}(T) \right\vert
    \exp\left(
    - c_4 m n \frac{\log m}{n}
    \right)
    \exp \left(
    - (1-\relabelConst)
    \left(
    \left(
    \intraTau m \Ir + \frac{n}{gc} \Ig \frac{\log n}{n}
    \right)
    \sum\limits_{x \in [c]}
	\sum\limits_{i \in [g]}
	\sum\limits_{j \in [g]\setminus \sigma(i|x)}
	\kUsrs{i}{j}{x}{\sigma(x)} 
    \right.
    \right.
    \nonumber\\
    &\phantom{\leq}
    \left.
    \left.
    + \left(
    \interTau m \Ir + \frac{n \IcOne}{gc} \frac{\log n}{n} + \frac{ (g-1) n \IcTwo}{gc} \frac{\log n}{n}
    \right)
    \sum_{x\in[c]}\sum_{y\in[c]\setminus x} \sum_{i\in[g]} \sum_{j\in[g]} \kUsrs{i}{j}{x}{y} 
	\right)
    \right)
	\label{eq:largeErrConvg_R1_2}\\
	&\leq 
	\sum\limits_{T \in \TlargeTwo} 
    \exp
    \left(
    \log n
    \left(
    \sum_{x \in [c]} 
    \sum_{i \in [g]} 
    \sum_{(y,j) \neq \sigma(x,i)} 
    \kUsrs{i}{j}{x}{y}
    \right)
    \right)
    \times
    \left(q^{gc}\right)^{m}
    \exp\left(
    - c_4 m \log m
    \right)
    \nonumber\\
    &\phantom{\leq}
    \times 
	\exp \left(
    - (1-\relabelConst) (1+\epsilon) \log n
    \left(
    \sum\limits_{x \in [c]}
	\sum\limits_{i \in [g]}
	\sum\limits_{j \in [g]\setminus \sigma(i|x)}
	\kUsrs{i}{j}{x}{\sigma(x)} 
    +
    \sum_{x\in[c]}\sum_{y\in[c]\setminus x} \sum_{i\in[g]} \sum_{j\in[g]} \kUsrs{i}{j}{x}{y} 
	\right)
    \right)
    \label{eq:largeErrConvg_R1_3}\\
    &\leq 
	\sum\limits_{T \in \TlargeTwo} 
	\left(q^{gc}\right)^{m}
    \exp\left(
    - c_4 m \log m
    \right)
	\exp
    \left(
    \log n
    \left(
    \sum_{x \in [c]} 
    \sum_{i \in [g]} 
    \sum_{(y,j) \neq \sigma(x,i)} 
    \kUsrs{i}{j}{x}{y}
    \right)
    \right)
    \nonumber\\
    &\phantom{\leq}
    \times 
	\exp \left(
    - \left(1+\frac{\epsilon}{2}\right) \log n
    \left(
    \sum_{x \in [c]} 
    \sum_{i \in [g]} 
    \sum_{(y,j) \neq \sigma(x,i)} 
    \kUsrs{i}{j}{x}{y}
	\right)
    \right)
    \label{eq:largeErrConvg_R1_3_1}\\
    &=
    \exp\left(
	-m \left(c_4 \log m - gc \log q\right) 
    \right)
    \sum\limits_{T \in \TlargeTwo} 
	\exp \left(
    - \frac{\epsilon}{2} \log n
    \left(
    \sum_{x \in [c]} 
    \sum_{i \in [g]} 
    \sum_{(y,j) \neq \sigma(x,i)} 
    \kUsrs{i}{j}{x}{y}
    \right)
    \right)
    \nonumber\\
    &=
    \exp\left(
	-m \left(c_4 \log m - gc \log q\right) 
    \right)
    \sum_{\ell = 0}^{\relabelConst n}
    \left| \left\{
	\sum_{x \in [c]} 
    \sum_{i \in [g]} 
    \sum_{(y,j) \neq (\sigma(x),\sigma(i|x))} 
    \kUsrs{i}{j}{x}{y} = \ell
    \right\} \right|
    \exp
    \left(
    - \frac{\epsilon \log n}{2}
    \ell
    \right)
    \label{eq:largeErrConvg_R1_5}\\
    &=
    \exp\left(
	-m \left(c_4 \log m - gc \log q\right) 
    \right)
    \nonumber\\
    &\phantom{=}
    \times \left[
    \left| \left\{
	\sum_{x \in [c]} 
    \sum_{i \in [g]} 
    \sum_{(y,j) \neq (\sigma(x),\sigma(i|x))} 
    \kUsrs{i}{j}{x}{y} = 0
    \right\} \right|
    + 
    \sum_{\ell = 1}^{\relabelConst n}
    \left| \left\{
	\sum_{x \in [c]} 
    \sum_{i \in [g]} 
    \sum_{(y,j) \neq (\sigma(x),\sigma(i|x))} 
    \kUsrs{i}{j}{x}{y} = \ell
    \right\} \right|
    \exp
    \left(
    - \frac{\epsilon \log n}{2}
    \ell
    \right)
    \right]
    \label{eq:largeErrConvg_R1_5_1}\\
    &=
    \exp\left(
	-m \left(c_4 \log m - gc \log q\right) 
    \right)
    \left[
    1
    +
    \sum_{\ell = 1}^{\relabelConst n}
    \binom{\ell + gc}{gc}
    \exp
    \left(
    - \frac{\epsilon \log n}{2}
    \ell
    \right)
    \right]
    \label{eq:largeErrConvg_R1_6}\\
    &\leq
    \exp\left(
	-m \left(c_4 \log m - gc \log q\right) 
    \right)
    \left[
    1
    +
    \sum_{\ell_2 = 1}^{\relabelConst n}
    2^{\left(\displaystyle \ell_2 + gc\right)}\:\:
    n^{\left( \displaystyle -  \frac{\epsilon}{2} \ell_2\right)}
    \right]
    \label{eq:largeErrConvg_R1_7}\\
    &\leq
    \exp\left(
	-m \left(c_4 \log m - gc \log q\right) 
    \right)
    \left[
    1
    +
    2^{\displaystyle gc}
    \sum_{\ell_2 = 1}^{\infty}
    \left(2 n^{\left( \displaystyle -  \frac{\epsilon}{2}\right)}\right)^{\displaystyle \ell_2}
    \right]
    \nonumber\\
    &=
    \exp\left(
	-m \left(c_4 \log m - gc \log q\right) 
    \right)
    \left[
    1
    +
    2^{\displaystyle gc}
    \frac{2 n^{\left( \displaystyle -  \frac{\epsilon}{2}\right)}}{1 - 2 n^{\left( \displaystyle -  \frac{\epsilon}{2}\right)}}
    \right],
    \label{eq:largeErrConvg_R1_8}
\end{align}
where 
\begin{itemize}
    \item \eqref{eq:largeErrConvg_R1_2} follows from \eqref{eq:regime3_Pd_2} and \eqref{eq:regime3_abg_2};
    \item \eqref{eq:largeErrConvg_R1_3} follows from the sufficient conditions in \eqref{eq:suffCond_1}, \eqref{eq:suffCond_2} and \eqref{eq:suffCond_3};
    \item \eqref{eq:largeErrConvg_R1_3_1} follows from $\tau \leq  (\epsilon\log m - (2+\epsilon)\log(2q)) / (2 (1 + \epsilon)\log m))$ which implies that $(1-\tau)(1+\epsilon) \geq (1+(\epsilon/2))$;
    \item \eqref{eq:largeErrConvg_R1_5} readily follows from \eqref{eq:regime3_large_1};
    \item in \eqref{eq:largeErrConvg_R1_5_1}, we break the summation into two summations, and use the fact that the enumeration of the first element of the set is independent of the enumeration of the second element;
    \item in \eqref{eq:largeErrConvg_R1_6}, we use the fact that the number of integer solutions of $\sum_{i=1}^{n} x_i = s$ is equal to $\binom{s+n-1}{n-1}$;
    \item in \eqref{eq:largeErrConvg_R1_7}, we bound each binomial coefficient by
    $\binom{a}{b} \leq \sum_{i=0}^a \binom{a}{i} \leq 2^a$, for $a \geq b$;
    \item and finally in \eqref{eq:largeErrConvg_R1_8}, we evaluate the infinite geometric series, where $\epsilon > (2 \log 2) / \log n$.
\end{itemize}

By \eqref{eq:R1_upperBound} and \eqref{eq:largeErrConvg_R1_8}, the RHS of \eqref{ineq:largeErr_1} is upper bounded by
\begin{align}
    &
    \lim_{n,m\rightarrow \infty}
    \sum\limits_{T \in \TlargeErr} 
	\sum\limits_{X \in \mathcal{X}(T)}
	\exp\left(-\left(1+o(1)\right) \left(\lvert\diffEntriesSet\rvert \Ir 
    + \Pleftrightarrow{\alphaEdge}{\betaEdge} \: \Ig \frac{\log n}{n}
    + \Pleftrightarrow{\alphaEdge}{\gammaEdge} \: \IcOne \frac{\log n}{n}
    + \Pleftrightarrow{\betaEdge}{\gammaEdge} \: \IcTwo \frac{\log n}{n} \right)\right)
    \nonumber \\
    &\qquad \leq
    \lim_{n,m\rightarrow \infty}
    \exp\left(
	-\left(m \left(c_2 \log m - gc \log q\right) 
	+ n \left(c_3 \log n - \log (gc)\right)
    \right)\right)
    \nonumber\\
    &\qquad \phantom{\leq}
    + 
    \lim_{n,m\rightarrow \infty}
    \exp\left(
	-m \left(c_4 \log m - gc \log q\right) 
    \right)
    \left[
    1
    +
    2^{\displaystyle gc}
    \frac{2 n^{\left( \displaystyle -  \frac{\epsilon}{2}\right)}}{1 - 2 n^{\left( \displaystyle -  \frac{\epsilon}{2}\right)}}
    \right]
    \nonumber\\
    &\qquad =
    0.
\end{align}
Note that as $n$ tends to infinity, the condition on $\epsilon$ becomes 
\begin{align}
    \epsilon > \lim_{n,m\rightarrow \infty}
    \max\left\{
    \frac{2 \log 2}{\log n}, \: 
    \frac{2(g-r+1) \log 2}{\log(c_1m)},
    \frac{2\log c_1}{\log(m/c_1)}
    \right\} = 0.
\end{align}
This completes the proof of Lemma~\ref{lemma:eta_omega(nm)}.
\hfill $\blacksquare$
\end{JA_AE}

\section{Proof of Lemma~\ref{lm:inf_worstProbError}}
\label{app:inf_worstProbError}
We start with the proof with the minimax optimization approach in \eqref{eq:errorProb_wc} to minimize the maximum worst-case probability of error as follows: 
\begin{align}
	\inf_{\psi} P_e^{(\diff)} (\psi) 
	&=
	\inf_{\psi} 
	\max_{M \in \matSet}
	\mathbb{P}\left[\psi(Y^{\Omega}, G) \neq M\right]
	\nonumber\\
	&\geq
	\inf_{\psi} 
	\max_{M \in \matSet}
	\mathbb{P}\left[\psi(Y^{\Omega}, G) \neq M, \: \mathbf{M} = M\right]
	\nonumber\\
	&=
	\inf_{\psi} 
	\max_{M \in \matSet}
	\mathbb{P}\left[\psi(Y^{\Omega}, G) \neq M \:\vert\: \mathbf{M} = M\right]
	\label{eq:lowerB_Pe_M_cond}
	\\
	&=
	\inf_{\psi} 
	\max_{M \in \matSet}
	\sum_{X \neq M} \mathbb{P}\left[\psi(Y^{\Omega}, G) = X \:\vert\: \mathbf{M} = M\right]
	\nonumber\\
	&=
	\max_{M \in \matSet}
	\sum_{X \neq M} 
	\mathbb{P}\left[\estML(Y^{\Omega}, G) = X \:\vert\: \mathbf{M} = M\right]
	\label{eq:lowerB_Pe_noInf}
	\\
	&\geq
	\sum_{X \neq \gtMat} 
	\mathbb{P}\left[\estML(Y^{\Omega}, G) = X \:\vert\: \mathbf{M} = \gtMat\right]
	\label{eq:lowerB_Pe_M_noMax}
	\\
	&=
	\sum_{X \neq \gtMat}
	\mathbb{P}\left[\mathsf{L}(X) \leq \mathsf{L}(\gtMat)\right]
	\label{eq:lowerB_Pe_L}
	\\
	&\geq
	\mathbb{P}\left[
	\bigcup_{X \neq \gtMat}
	\left(\mathsf{L}(X) \leq \mathsf{L}(\gtMat)\right)\right]
	\label{eq:lowerB_Pe_union}
	\\
	&=
	\mathbb{P}\left[S^c\right],
	\label{eq:lowerB_Pe}
\end{align} 
where \eqref{eq:lowerB_Pe_M_cond} follows because $\mathbf{M}$ is uniformly distributed; \eqref{eq:lowerB_Pe_noInf} follows due to the fact that the maximum likelihood estimator is optimal under uniform prior; \eqref{eq:lowerB_Pe_M_noMax} follows since $\gtMat \in \matSet$ whose construction is given in Section~\ref{sec:achv}; \eqref{eq:lowerB_Pe_L} follows by the definition of maximum likelihood estimation; \eqref{eq:lowerB_Pe_union} follows from the union bound; and finally \eqref{eq:lowerB_Pe} follows from \eqref{eq:defS}.
This completes the proof of Lemma~\ref{lm:inf_worstProbError}.
\hfill $\blacksquare$

\section{Proof of Lemma~\ref{lm:lowerB_prob}}
\label{app:lowerB_prob_proof}
Recall from Appendix~\ref{proof:upper_bound_B} the definition of $\mathbf{U}_i = \mathbf{U}_i(p,\theta,q)$ in \eqref{eq:app_x(p,t,q)}, and the expression of $-\log M_{\mathbf{U}_i(p,\theta,q)}\left(\frac{1}{2}\right)$ in \eqref{eq:log-M-X}.
Define a related random variable $\widehat{\mathbf{U}}_i=\widehat{\mathbf{U}}_i(p,\theta,q)$ that has the same sample space as $\mathbf{U}_i(p,\theta,q)$, but its probability mass function is given by
\begin{align}
	f_{\widehat{\mathbf{U}}_i(p,\theta,q)} (u) = \frac{\exp\left(\frac{1}{2}u\right) f_{\mathbf{U}_i(p,\theta,q)}(u)}{M_{\mathbf{U}_i(p,\theta,q)}(\frac{1}{2})}.
	\label{eq:pdf_u}
\end{align}
More formally, $\widehat{\mathbf{U}}_i(p,\theta,q)$ is defined as
\begin{align}
	\widehat{\mathbf{U}}_i(p,\theta,q) 
	= \left\{
	\begin{array}{ll}
	-\log \left((q-1)\frac{1-\theta}{\theta}\right) 
	& \textrm{w.p. } 
	\frac{1}{M_{\mathbf{U}_i}\left(\frac{1}{2}\right)} \sqrt{\frac{\theta(1-\theta)}{q-1}} p,\\
	0 
	& \textrm{w.p. } 
	\frac{1}{M_{\mathbf{U}_i}\left(\frac{1}{2}\right)} \left((1-p) + p \theta \left(1-\frac{1}{q-1}\right)\right),\\
	\log \left((q-1)\frac{1-\theta}{\theta}\right)  
	& \textrm{w.p. }\frac{1}{M_{\mathbf{U}_i}\left(\frac{1}{2}\right)} \sqrt{\frac{\theta(1-\theta)}{q-1}} p,
	\end{array}
	\right.
\end{align}
from which one can readily show that
\begin{align}
	\mathbb{E}\left[\widehat{\mathbf{U}}_i(p,\theta,q)\right] 
	&=0,
	\label{eq:E-hat-X}\\
	\mathsf{Var}\left[\widehat{\mathbf{U}}_i(p,\theta,q)\right]
	&= 
	2 \left(\log \left((q-1)\frac{1-\theta}{\theta}\right)\right)^2
	\frac{\sqrt{\frac{\theta(1-\theta)}{q-1}} p}{1 - \left(\sqrt{1-\theta} - \sqrt{\frac{\theta}{q-1}}\right)^2 p}  
	= O(p).
	\label{eq:Var-hat-X}
\end{align}
Similarly, recall from Appendix~\ref{proof:upper_bound_B} the definition of $\mathbf{V}_j = \mathbf{V}_j(\mu,\nu)$ in \eqref{eq:app_z(m,n)}, and the expression of $-\log M_{\mathbf{V}_j(\mu,\nu)}\left(\frac{1}{2}\right)$ in \eqref{eq:log-M-Y}.
Define a related random variable $\widehat{\mathbf{V}}_j = \widehat{\mathbf{V}}_j(\mu,\nu)$ that has the same sample space as $\mathbf{V}_j(\mu,\nu)$, but its probability mass function is given by
\begin{align}
	f_{\widehat{\mathbf{V}}_j(\mu,\nu)}(v) 
	= \frac{\exp(\frac{1}{2}v) f_{\mathbf{V}_j(\mu,\nu)}(v)}{M_{\mathbf{V}_j(\mu,\nu)}(\frac{1}{2})}.
	\label{eq:pdf_v}
\end{align} 
More formally, $\widehat{\mathbf{V}}_j(\mu,\nu)$ is defined as
\begin{align}
	\widehat{\mathbf{V}}_j(\mu,\nu)
	= \left\{
	\begin{array}{ll}
	-\log\frac{(1-\mu)\nu}{(1-\nu)\mu} 
	& \textrm{w.p. } 
	\frac{1}{M_{\mathbf{V}_j}\left(\frac{1}{2}\right)} \sqrt{(1-\mu)(1-\nu)\mu\nu},
	\vspace{1mm}\\
	0 
	& \textrm{w.p. } 
	\frac{1}{M_{\mathbf{V}_j}\left(\frac{1}{2}\right)} \left((1-\mu)(1-\nu) + \mu\nu\right),
	\vspace{1mm}\\
	\log\frac{(1-\mu)\nu}{(1-\nu)\mu}  
	& \textrm{w.p. }
	\frac{1}{M_{\mathbf{V}_j}\left(\frac{1}{2}\right)} \sqrt{(1-\mu)(1-\nu)\mu\nu},
	\end{array}
	\right.
\end{align}
from which one can readily show that
\begin{align}
	\mathbb{E}\left[\widehat{\mathbf{V}}_j(\mu,\nu)\right] 
	&= 0,
	\label{eq:E-hat-Y}\\
	\mathsf{Var}\left[\widehat{\mathbf{V}}_j(\mu,\nu)\right] 
	&= 
	2 \left(\log\frac{(1-\mu)\nu}{(1-\nu)\mu}\right)^2
	\frac{\sqrt{(1-\mu)(1-\nu)\mu\nu}}{\left(\sqrt{\mu\nu}+\sqrt{(1-\mu)(1-\nu)}\right)^2}  
	= O\left(\sqrt{\mu\nu} \right).
	\label{eq:Var-hat-Y}
\end{align}

Let $\{\mathbf{U}_i (p,\theta,q): i\in \diffEntriesSet \}$, $\{\mathbf{V}_j(\betaEdge,\alphaEdge) : j\in\misClassfSet{\betaEdge}{\alphaEdge}\}$, $\{\mathbf{V}_k(\gammaEdge,\alphaEdge) : k\in\misClassfSet{\gammaEdge}{\alphaEdge}\}$ and $\{\mathbf{V}_\ell(\gammaEdge,\betaEdge) : \ell\in\misClassfSet{\gammaEdge}{\betaEdge}\}$ be sets of independent and identically distributed random variables defined as per \eqref{eq:app_x(p,t,q)} and \eqref{eq:app_z(m,n)}.
Note that the sets $\diffEntriesSet$, $\misClassfSet{\betaEdge}{\alphaEdge}$, $\misClassfSet{\gammaEdge}{\alphaEdge}$ and $\misClassfSet{\gammaEdge}{\betaEdge}$ (defined by \eqref{eq:N1_defn}, \eqref{eq:N2rev_defn}, \eqref{eq:N3rev_defn} and \eqref{eq:N4rev_defn}, respectively) are disjoint sets.
Similarly, let $\{\widehat{\mathbf{U}}_i (p,\theta,q) : i\in\diffEntriesSet \}$, $\{\widehat{\mathbf{V}}_j(\betaEdge,\alphaEdge) : j\in\misClassfSet{\betaEdge}{\alphaEdge}\}$, $\{\widehat{\mathbf{V}}_k(\gammaEdge,\alphaEdge) : k\in\misClassfSet{\gammaEdge}{\alphaEdge} \}$ and $\{\widehat{\mathbf{V}}_\ell(\gammaEdge,\betaEdge) : \ell\in\misClassfSet{\gammaEdge}{\betaEdge} \}$ be sets of independent and identically distributed random variables defined as per \eqref{eq:pdf_u} and \eqref{eq:pdf_v}.
For $\{\misClassfSet{\mu}{\nu} : \left\vert \misClassfSet{\mu}{\nu} \right\vert = \left\vert \misClassfSet{\nu}{\mu} \right\vert,  \:\mu, \nu \in \{\alphaEdge,\betaEdge,\gammaEdge \}, \: \mu \neq \nu\}$, we express the random variable of interest $\mathbf{B}$ from \eqref{eq:B_middle} as 
\begin{align}
	\mathbf{B} &=
	\sum_{i\in\diffEntriesSet} 
	\log \left((q-1)\frac{1-\theta}{\theta}\right) \mathsf{B}_i^{(p)} \left(\left(1+\mathsf{B}_i^{\left(\frac{1}{q-1}\right)}\right)\mathsf{B}_i^{(\theta)}-1\right)
	+ \sum_{j\in\misClassfSet{\betaEdge}{\alphaEdge}} 
	\log\left(\frac{(1-\betaEdge)\alphaEdge}{(1-\alphaEdge)\betaEdge}\right) \left(\mathsf{B}_j^{(\betaEdge)} - \mathsf{B}_j^{(\alphaEdge)} \right)
	\nonumber\\
	&\phantom{\coloneqq} 
	+ \sum_{k\in\misClassfSet{\gammaEdge}{\alphaEdge}} 
	\log\left(\frac{(1-\gammaEdge)\alphaEdge}{(1-\alphaEdge)\gammaEdge}\right)  \left(\mathsf{B}_k^{(\gammaEdge)} - \mathsf{B}_k^{(\alphaEdge)} \right)
	+ \sum_{\ell\in\misClassfSet{\gammaEdge}{\betaEdge}} 
	\log\left(\frac{(1-\gammaEdge)\betaEdge}{(1-\betaEdge)\gammaEdge}\right)  \left( \mathsf{B}_\ell^{(\gammaEdge)} - \mathsf{B}_\ell^{(\betaEdge)} \right),
	\nonumber\\
	&=
	\sum_{i\in\diffEntriesSet}  {\mathbf{U}}_i(p,\theta,q)
	+ \sum_{j\in\misClassfSet{\betaEdge}{\alphaEdge}}  {\mathbf{V}}_j(\betaEdge,\alphaEdge) 
	+ \sum_{k\in\misClassfSet{\gammaEdge}{\alphaEdge}} {\mathbf{V}}_k(\gammaEdge,\alphaEdge)
	+ \sum_{\ell\in\misClassfSet{\gammaEdge}{\betaEdge}} {\mathbf{V}}_\ell(\gammaEdge,\betaEdge).
	\label{eq:B-in-YX}
\end{align}
Following a similar proof technique to that of \cite[Lemma~5.2]{zhang2016minimax}, the probability that $\mathbf{B}$ is non-negative is lower bounded by
\begin{align}
	&\mathbb{P} \left[\mathbf{B} \geq 0 \right] 
	\nonumber\\
	&= \mathbb{P}\left[\sum_{i\in\diffEntriesSet}  {\mathbf{U}}_i(p,\theta,q)
	+ \sum_{j\in\misClassfSet{\betaEdge}{\alphaEdge}}  {\mathbf{V}}_j(\betaEdge,\alphaEdge) 
	+ \sum_{k\in\misClassfSet{\gammaEdge}{\alphaEdge}}  {\mathbf{V}}_k(\gammaEdge,\alphaEdge)
	+ \sum_{\ell\in\misClassfSet{\gammaEdge}{\betaEdge}}  {\mathbf{V}}_\ell(\gammaEdge,\betaEdge) \geq 0\right]
	\nonumber\\
	&\geq 
	\mathbb{P}\left[0\leq \sum_{i\in\diffEntriesSet}  {\mathbf{U}}_i(p,\theta,q)
	+ \sum_{j\in\misClassfSet{\betaEdge}{\alphaEdge}}  {\mathbf{V}}_j(\betaEdge,\alphaEdge) 
	+ \sum_{k\in\misClassfSet{\gammaEdge}{\alphaEdge}}  {\mathbf{V}}_k(\gammaEdge,\alphaEdge)
	+ \sum_{\ell\in\misClassfSet{\gammaEdge}{\betaEdge}}  {\mathbf{V}}_\ell(\gammaEdge,\betaEdge) < \xi\right]
	\nonumber\\
	&=
	\sum_{\mathcal{R}(\xi)}\left[
	\prod_{i\in\diffEntriesSet} f_{\mathbf{U}_i(p,\theta,q)} (u_i)
	\prod_{j\in\misClassfSet{\betaEdge}{\alphaEdge}} f_{\mathbf{V}_j(\betaEdge,\alphaEdge)} (v_j)
	\prod_{k\in\misClassfSet{\gammaEdge}{\alphaEdge}} f_{\mathbf{V}_k(\gammaEdge,\alphaEdge)} (v_k)
	\prod_{\ell\in\misClassfSet{\gammaEdge}{\betaEdge}} f_{\mathbf{V}_\ell(\gammaEdge,\betaEdge)} (v_\ell)
	\right]
	\label{eq:Blarge0_bound_1}\\ 
	&\geq
	\frac{
	\left(M_{\mathbf{U}_i(p,\theta,q)}\left(\frac{1}{2}\right)\right)^{\lvert\diffEntriesSet\rvert}
	\left(M_{\mathbf{V}_j(\betaEdge,\alphaEdge)}\left(\frac{1}{2}\right)\right)^{\lvert\misClassfSet{\betaEdge}{\alphaEdge}\rvert}
	\left(M_{\mathbf{V}_k(\gammaEdge,\alphaEdge)}\left(\frac{1}{2}\right)\right)^{\lvert\misClassfSet{\gammaEdge}{\alphaEdge}\rvert}
	\left(M_{\mathbf{V}_\ell(\gammaEdge,\betaEdge)}\left(\frac{1}{2}\right)\right)^{\lvert\misClassfSet{\gammaEdge}{\betaEdge}\rvert}
	}{\exp\left(\frac{1}{2} \xi\right)} 
	\nonumber\\
	&\phantom{\geq}
	\times
	\sum_{\mathcal{R}(\xi)}
	\left[\prod_{i\in\diffEntriesSet} \frac{\exp\left(\frac{1}{2} u_i \right) f_{\mathbf{U}_i(p,\theta,q)} (u_i)}{M_{\mathbf{U}_i(p,\theta)}\left(\frac{1}{2}\right)}
	\prod_{j\in\misClassfSet{\betaEdge}{\alphaEdge}} \frac{\exp\left(\frac{1}{2} v_j \right) f_{\mathbf{V}_j(\betaEdge,\alphaEdge)} (v_j)}{M_{\mathbf{V}_j(\betaEdge,\alphaEdge)}\left(\frac{1}{2}\right)}\right.
	\nonumber\\
	&\phantom{
	\geq 
	\times 
	\sum_{\mathcal{R}(\xi)}
	\left[
	\right.
	}
	\left. \cdot
	\prod_{k\in\misClassfSet{\gammaEdge}{\alphaEdge}} \frac{\exp\left(\frac{1}{2} v_k \right) f_{\mathbf{V}_k(\gammaEdge,\alphaEdge)} (v_k)}{M_{\mathbf{V}_k(\gammaEdge,\alphaEdge)}\left(\frac{1}{2}\right)}
	\prod_{\ell\in\misClassfSet{\gammaEdge}{\betaEdge}} \frac{\exp\left(\frac{1}{2} v_\ell \right) f_{\mathbf{V}_\ell(\gammaEdge,\betaEdge)} (v_\ell)}{M_{\mathbf{V}_\ell(\gammaEdge,\betaEdge)}\left(\frac{1}{2}\right)}\right]
	\label{eq:Blarge0_bound_2}\\
	& = 
	\exp\left(\!
	\lvert\diffEntriesSet\rvert \log M_{\mathbf{U}_i(p,\theta,q)} \!\left(\!\frac{1}{2}\!\right) 
	+ \lvert\misClassfSet{\betaEdge}{\alphaEdge}\rvert \log M_{\mathbf{V}_j(\betaEdge,\alphaEdge)} \!\left(\!\frac{1}{2}\!\right) 
	+ \lvert\misClassfSet{\gammaEdge}{\alphaEdge}\rvert \log M_{\mathbf{V}_k(\gammaEdge,\alphaEdge)} \!\left(\!\frac{1}{2}\!\right)
	+\lvert\misClassfSet{\gammaEdge}{\betaEdge}\rvert \log M_{\mathbf{V}_\ell(\gammaEdge,\betaEdge)} \!\left(\!\frac{1}{2}\!\right)-\frac{1}{2} \xi\right)
	\nonumber\\
	&\phantom{=} \times 
	\sum_{\mathcal{R}(\xi)} 
	\left[\prod_{i\in\diffEntriesSet} f_{\widehat{\mathbf{U}}_i(p,\theta,q)} (u_i) 
	\prod_{j\in\misClassfSet{\betaEdge}{\alphaEdge}} f_{\widehat{\mathbf{V}}_j(\betaEdge,\alphaEdge)} (v_j)
	\prod_{k\in\misClassfSet{\gammaEdge}{\alphaEdge}} f_{\widehat{\mathbf{V}}_k(\gammaEdge,\alphaEdge)} (v_k)
	\prod_{\ell\in\misClassfSet{\gammaEdge}{\betaEdge}}  f_{\widehat{\mathbf{V}}_\ell(\gammaEdge,\betaEdge)} (v_\ell)\right]
	\label{eq:Blarge0_bound_3}\\
	&=
	\exp\left( -(1+o(1))\left(
	\lvert\diffEntriesSet\rvert \Ir
	+ \lvert\misClassfSet{\betaEdge}{\alphaEdge}\rvert \Ig \frac{\log n}{n}
	+ \lvert\misClassfSet{\gammaEdge}{\alphaEdge}\rvert \IcOne \frac{\log n}{n}
	+ \lvert\misClassfSet{\gammaEdge}{\betaEdge}\rvert \IcTwo \frac{\log n}{n}
	\right)
	-\frac{1}{2} \xi\right)
	\nonumber\\ 
	&\phantom{=} \times \mathbb{P} \left[0\leq \sum_{i\in\diffEntriesSet}  \widehat{\mathbf{U}}_i(p,\theta,q)
	+  \sum_{j\in\misClassfSet{\betaEdge}{\alphaEdge}}  \widehat{\mathbf{V}}_j(\betaEdge,\alphaEdge) 
	+  \sum_{k\in\misClassfSet{\gammaEdge}{\alphaEdge}}  \widehat{\mathbf{V}}_k(\gammaEdge,\alphaEdge)
	+  \sum_{\ell\in\misClassfSet{\gammaEdge}{\betaEdge}}  \widehat{\mathbf{V}}_\ell(\gammaEdge,\betaEdge) < \xi\right],
	\label{eq:Xi_ineq_1}
\end{align}
where 
\begin{itemize}
	\item 
	the summation in \eqref{eq:Blarge0_bound_1} is over 
	\begin{align*}
		\mathcal{R}(\xi)=\left\{
		\left\{u_i\right\}_{i\in\diffEntriesSet},
		\left\{v_j\right\}_{j\in\misClassfSet{\betaEdge}{\alphaEdge}},
		\left\{v_k\right\}_{j\in\misClassfSet{\gammaEdge}{\alphaEdge}},
		\left\{v_\ell\right\}_{j\in\misClassfSet{\gammaEdge}{\betaEdge}}:   
		0 \leq \sum_{i\in\diffEntriesSet} u_i 
		+ \sum_{j\in\misClassfSet{\betaEdge}{\alphaEdge}} v_j
		+ \sum_{k\in\misClassfSet{\gammaEdge}{\alphaEdge}} v_k 
		+ \sum_{\ell\in\misClassfSet{\gammaEdge}{\betaEdge}} v_\ell  
		< \xi
		\right\}\!;
	\end{align*}
	moreover, \eqref{eq:Blarge0_bound_1} follows from the independence of the random variables $\{\mathbf{U}_i : i\in\diffEntriesSet\}$, $\{\mathbf{V}_j : j\in\misClassfSet{\betaEdge}{\alphaEdge}\}$, $\{\mathbf{V}_k : k\in\misClassfSet{\gammaEdge}{\alphaEdge}\}$ and $\{\mathbf{V}_\ell : \ell\in\misClassfSet{\gammaEdge}{\betaEdge}\}$.
	\item \eqref{eq:Blarge0_bound_2} holds since $\exp \left( \frac{1}{2} \left(\sum_{i\in\diffEntriesSet} u_i + \sum_{j\in\misClassfSet{\betaEdge}{\alphaEdge}} v_j
		+ \sum_{k\in\misClassfSet{\gammaEdge}{\alphaEdge}} v_k 
		+ \sum_{\ell\in\misClassfSet{\gammaEdge}{\betaEdge}} v_\ell \right) \right) < \exp\left(\frac{1}{2} \xi \right)$; 
	\item \eqref{eq:Blarge0_bound_3} follows from \eqref{eq:pdf_u} and \eqref{eq:pdf_v}; and
	\item \eqref{eq:Xi_ineq_1} follows from \eqref{eq:log-M-X} and \eqref{eq:log-M-Y}, and due to the independence of the random variables $\{\widehat{\mathbf{U}}_i: i\in\diffEntriesSet\}$, $\{\widehat{\mathbf{V}}_j: j\in\misClassfSet{\betaEdge}{\alphaEdge}\}$, $\{\widehat{\mathbf{V}}_k: k\in\misClassfSet{\gammaEdge}{\alphaEdge}\}$ and $\{\widehat{\mathbf{V}}_\ell: \ell\in\misClassfSet{\gammaEdge}{\betaEdge}\}$. 
\end{itemize}
It should be noted that \eqref{eq:Xi_ineq_1} holds for any value of $\xi$. In particular, we choose $\xi_n$ satisfying the following two conditions:
\begin{align}
&\lim_{n\rightarrow \infty} \frac{\xi_n}{\lvert\diffEntriesSet\rvert \Ir + \lvert\misClassfSet{\betaEdge}{\alphaEdge}\rvert\Ig \frac{\log n}{n} + \lvert\misClassfSet{\gammaEdge}{\alphaEdge}\rvert\IcOne \frac{\log n}{n} + \lvert\misClassfSet{\gammaEdge}{\betaEdge}\rvert\IcTwo \frac{\log n}{n}}=0,\label{eq:xi-small}\\
&\lim_{n\rightarrow \infty} \frac{\lvert\diffEntriesSet\rvert p + \lvert\misClassfSet{\betaEdge}{\alphaEdge}\rvert\sqrt{\alphaEdge \betaEdge} + \lvert\misClassfSet{\gammaEdge}{\alphaEdge}\rvert\sqrt{\alphaEdge \gammaEdge} + \lvert\misClassfSet{\gammaEdge}{\betaEdge}\rvert\sqrt{\betaEdge \gammaEdge}}{\xi_n^2} =0,
\label{eq:xi-large}
\end{align}
for any $\diffEntriesSet$, $\misClassfSet{\betaEdge}{\alphaEdge}$, $\misClassfSet{\gammaEdge}{\alphaEdge}$ and $\misClassfSet{\gammaEdge}{\betaEdge}$ such that at least one of these sets is non-empty\footnote{If the sets $\diffEntriesSet$, $\misClassfSet{\betaEdge}{\alphaEdge}$, $\misClassfSet{\gammaEdge}{\alphaEdge}$ and $\misClassfSet{\gammaEdge}{\betaEdge}$ are all empty, then \eqref{eq:lowerB_prob} is trivially true.}.
One instance of $\xi_n$ that satisfies both \eqref{eq:xi-small} and \eqref{eq:xi-large} is given by
\begin{align}
    \xi_n = \left(
    \max \left\{
    \left\vert\diffEntriesSet\right\vert, \left\vert\misClassfSet{\betaEdge}{\alphaEdge}\right\vert, \left\vert\misClassfSet{\gammaEdge}{\alphaEdge}\right\vert, \left\vert\misClassfSet{\gammaEdge}{\betaEdge}\right\vert
    \right\}
    \frac{\log n}{n}
    \right)^{\frac{2}{3}}.
\end{align}
Consequently, \eqref{eq:xi-small} implies that the exponent of the exponential term in \eqref{eq:Xi_ineq_1} can be asymptotically approximated as 
\begin{align}
    &-(1+o(1))\!\left(
    \lvert\diffEntriesSet\rvert \Ir
    + \lvert\misClassfSet{\betaEdge}{\alphaEdge}\rvert \Ig \frac{\log n}{n}
    + \lvert\misClassfSet{\gammaEdge}{\alphaEdge}\rvert \IcOne \frac{\log n}{n}
    + \lvert\misClassfSet{\gammaEdge}{\betaEdge}\rvert \IcTwo \frac{\log n}{n} \right)
    -\frac{1}{2} \xi_n 
    \nonumber\\
    &\qquad \qquad \simeq
    -(1+o(1))\!\left(
       \lvert\diffEntriesSet\rvert \Ir
    + \lvert\misClassfSet{\betaEdge}{\alphaEdge}\rvert \Ig \frac{\log n}{n}
    + \lvert\misClassfSet{\gammaEdge}{\alphaEdge}\rvert \IcOne \frac{\log n}{n}
    + \lvert\misClassfSet{\gammaEdge}{\betaEdge}\rvert \IcTwo \frac{\log n}{n} \right).
\label{eq:Xi_ineq_1:exp}
\end{align}
Furthermore, the probability term in \eqref{eq:Xi_ineq_1} can be lower bounded as
\begin{align}
	\mathbb{P} &\left[0\leq \sum_{i\in\diffEntriesSet} \widehat{\mathbf{U}}_i(p,\theta,q)
	+  \sum_{j\in\misClassfSet{\betaEdge}{\alphaEdge}}  \widehat{\mathbf{V}}_j(\betaEdge,\alphaEdge) 
	+  \sum_{k\in\misClassfSet{\gammaEdge}{\alphaEdge}}  \widehat{\mathbf{V}}_k(\gammaEdge,\alphaEdge)
	+  \sum_{\ell\in\misClassfSet{\gammaEdge}{\betaEdge}}  \widehat{\mathbf{V}}_\ell(\gammaEdge,\betaEdge) < \xi_n\right]
	\nonumber\\
	&\geq
	\frac{1}{2} - \mathbb{P} \left[ \sum_{i\in\diffEntriesSet}  \widehat{\mathbf{U}}_i(p,\theta,q)
	+  \sum_{j\in\misClassfSet{\betaEdge}{\alphaEdge}}  \widehat{\mathbf{V}}_j(\betaEdge,\alphaEdge) 
	+  \sum_{k\in\misClassfSet{\gammaEdge}{\alphaEdge}}  \widehat{\mathbf{V}}_k(\gammaEdge,\alphaEdge)
	+  \sum_{\ell\in\misClassfSet{\gammaEdge}{\betaEdge}}  \widehat{\mathbf{V}}_\ell(\gammaEdge,\betaEdge) \geq \xi_n\right]
	\label{eq:probB_bound_final_1}\\
	&\geq 
	\frac{1}{2} - \frac{\lvert\diffEntriesSet\rvert \mathsf{Var}\left[\widehat{\mathbf{U}}_i(p,\theta)\right] + 
	\lvert\misClassfSet{\betaEdge}{\alphaEdge}\rvert\mathsf{Var} \left[\widehat{\mathbf{V}}_j(\betaEdge,\alphaEdge)\right] + \lvert\misClassfSet{\gammaEdge}{\alphaEdge}\rvert\mathsf{Var}\left[\widehat{\mathbf{V}}_k(\gammaEdge,\alphaEdge)\right] + 
	\lvert\misClassfSet{\gammaEdge}{\betaEdge}\rvert\mathsf{Var}\left[\widehat{\mathbf{V}}_\ell(\gammaEdge,\betaEdge)\right]  
	}{\xi_n^2}
	\label{eq:probB_bound_final_2}\\
	&=
	\frac{1}{2} -\frac{\lvert\diffEntriesSet\rvert O(p) + \lvert\misClassfSet{\betaEdge}{\alphaEdge}\rvert O(\sqrt{\alphaEdge \betaEdge}) + \lvert\misClassfSet{\gammaEdge}{\alphaEdge}\rvert O(\sqrt{\alphaEdge \gammaEdge}) + \lvert\misClassfSet{\gammaEdge}{\betaEdge}\rvert O(\sqrt{\betaEdge \gammaEdge})}{\xi_n^2}
	\label{eq:probB_bound_final_3}\\
	&=
	\frac{1}{2} - o(1) \:>\: \frac{1}{4}, 
	\label{eq:Xi_ineq_1:prob}
\end{align}
where \eqref{eq:probB_bound_final_1} is due to the symmetry of the random variables $\widehat{\mathbf{U}}_i$, $\widehat{\mathbf{V}}_j$, $\widehat{\mathbf{V}}_k$ and $\widehat{\mathbf{V}}_\ell$; \eqref{eq:probB_bound_final_2} follows from Chebyshev's inequality; \eqref{eq:probB_bound_final_3} holds by substituting the variances with \eqref{eq:Var-hat-X} and \eqref{eq:Var-hat-Y}; and finally \eqref{eq:Xi_ineq_1:prob} is a consequence of \eqref{eq:xi-large}. 
Plugging \eqref{eq:Xi_ineq_1:exp} and \eqref{eq:Xi_ineq_1:prob} in \eqref{eq:Xi_ineq_1}, we obtain 
\begin{align*}
	&\mathbb{P}\left[ B\left(\diffEntriesSet,\misClassfSet{\betaEdge}{\alphaEdge},\misClassfSet{\gammaEdge}{\alphaEdge},\misClassfSet{\gammaEdge}{\betaEdge}\right) 
	\geq 0
	\right] 
	\nonumber\\
	& \qquad
	\geq 
	\frac{1}{4} \exp \left( -(1+o(1)) \left(\lvert\diffEntriesSet\rvert \Ir + \lvert\misClassfSet{\betaEdge}{\alphaEdge}\rvert \Ig \frac{\log n}{n} + \lvert\misClassfSet{\gammaEdge}{\alphaEdge}\rvert \IcOne \frac{\log n}{n}+ \lvert\misClassfSet{\gammaEdge}{\betaEdge}\rvert \IcTwo \frac{\log n}{n} \right) \right),
\end{align*}
which is the desired bound in \eqref{eq:lowerB_prob}.
This concludes the proof of Lemma~\ref{lm:lowerB_prob}.
\hfill $\blacksquare$

\section{Proof of Lemma~\ref{lm:randGraph}}
\label{app:randGraph}
The proof hinges on the alteration method \cite{alon2004probabilistic}. 
We present a constructive proof for the existence of subsets $\tG{i}{x}$ and $\tG{j}{y}$. Let $r= \frac{n}{\log^3 n}$. 
We start by sampling two random subsets $\uG{i}{x} \subset \grpG{i}{x}$ and $\uG{j}{y} \subset \grpG{j}{y}$ of cardinalities $\vert\uG{i}{x}\vert= 2r$ and $\vert\uG{j}{y}\vert= 2r$, respectively. 
Then, we prune these sets to obtain the desired edge free subsets. 
To this end, for any pair of nodes $a,b \in \uG{i}{x} \cup \uG{j}{y}$, we remove both $a$ and $b$ from $\uG{i}{x} \cup \uG{j}{y}$ if $(a,b)\in \mathcal{E}$. 
We continue this process until the remaining set of nodes is edge-free. 
Let $\cP$ be the set of nodes we remove from $\uG{i}{x} \cup \uG{j}{y}$ throughout the pruning process. The expected value of $|\cP|$ is upper bounded by 
\begin{align}
	\E[|\cP|] 
	&\leq 
	2\E\left[\sum_{a,b \:\in\: \uG{i}{x} \cup \uG{j}{y}} \indicatorFn{(a,b)\in \mathcal{E}}\right]
	\nonumber\\
	&= 
	2 \left(
	\sum_{a,b\in \uG{i}{x}} \E[\indicatorFn{(a,b)\in \mathcal{E}}] 
	+ \sum_{a,b\in \uG{j}{y}} \E[\indicatorFn{(a,b)\in \mathcal{E}}]
	+ \sum_{a\in \uG{i}{x}} \:\sum_{b\in \uG{j}{y}} \E[\indicatorFn{(a,b)\in \mathcal{E}}] 
	\right)
	\nonumber\\
	&= 
	2 \left(
	\sum_{a,b\in \uG{i}{x}} \alphaEdge 
	+ \sum_{a,b\in \uG{j}{y}} \alphaEdge 
	+ \sum_{a\in \uG{i}{x}} \: \sum_{b\in \uG{j}{y}} \betaEdge
	\right)
	\nonumber\\
	&= 
	2 \left(
	\binom{2r}{2} \alphaEdge 
	+ \binom{2r}{2} \alphaEdge 
	+ (2r)^2 \betaEdge
	\right)
	\:\leq\: 16 r^2 \alphaEdge,
\end{align}
where the last inequality holds since $\betaEdge \leq \alphaEdge$. Using Markov's inequality for the non-negative random variable $|\cP|$, we obtain
\begin{align}
  \mathbb{P} \left[|\cP| \geq r 
  \right]
  \:\leq\: \frac{\mathbb{E} \left[N\right]}{r}
  \:\leq\: \frac{16 n}{\log^3 n} \alphaEdge 
  \:=\:
  \Theta \left(\frac{n}{\log^3 n} \times \frac{\log n}{n}\right)
  \:=\: o(1).
  \label{eq:appC_markov}
\end{align}
Therefore, the number of remaining nodes (after pruning) satisfies
\begin{align*}
\mathbb{P} \left[|\uG{i}{x}\cup\uG{j}{y} \setminus \cP| > 3r\right] = \mathbb{P} \left[ |\cP| < r\right] = 1- \mathbb{P} \left[ |\cP| \geq r\right] = 1-o(1).
\end{align*}
Hence, $\uG{i}{x} \setminus \cP$ and $\uG{j}{y}\setminus \cP$ both have at least $3r$ elements. This, together with the fact that $|\uG{i}{x}| = |\uG{j}{y}| = 2r$, implies that each of $\uG{i}{x} \setminus \cP$ and $\uG{j}{y}\setminus \cP$ have at least $r$ elements. 
Therefore, we can choose $r$ elements from each of $\uG{i}{x} \setminus \cP$ and $\uG{j}{y} \setminus \cP$ to form the desired sets $\tG{i}{x}$ and $\tG{j}{y}$, respectively. This completes the proof of Lemma~\ref{lm:randGraph}.
\hfill $\blacksquare$

\end{document}

%% file: mySymb.tex
\usepackage{amsmath}
\usepackage{graphicx}
\usepackage{amssymb}
\usepackage[utf8]{inputenc}
\usepackage[english]{babel}

\newtheorem{lemma}{Lemma}

\newtheorem{prop}{Proposition}

\newtheorem{theorem}{Theorem}

\newtheorem{remark}{Remark}

\newcommand{\cE}{\mathcal{E}}

\newcommand{\cG}{\mathcal{G}}

\newcommand{\cI}{\mathcal{I}}
\newcommand{\cJ}{\mathcal{J}}

\newcommand{\cP}{\mathcal{P}}



%% file: matComp_HSBM_main.bbl
\begin{thebibliography}{10}
\providecommand{\url}[1]{#1}
\csname url@samestyle\endcsname
\providecommand{\newblock}{\relax}
\providecommand{\bibinfo}[2]{#2}
\providecommand{\BIBentrySTDinterwordspacing}{\spaceskip=0pt\relax}
\providecommand{\BIBentryALTinterwordstretchfactor}{4}
\providecommand{\BIBentryALTinterwordspacing}{\spaceskip=\fontdimen2\font plus
\BIBentryALTinterwordstretchfactor\fontdimen3\font minus
  \fontdimen4\font\relax}
\providecommand{\BIBforeignlanguage}[2]{{%
\expandafter\ifx\csname l@#1\endcsname\relax
\typeout{** WARNING: IEEEtran.bst: No hyphenation pattern has been}%
\typeout{** loaded for the language `#1'. Using the pattern for}%
\typeout{** the default language instead.}%
\else
\language=\csname l@#1\endcsname
\fi
#2}}
\providecommand{\BIBdecl}{\relax}
\BIBdecl

\bibitem{elmahdy2020matrix}
A.~Elmahdy, J.~Ahn, C.~Suh, and S.~Mohajer, ``Matrix completion with
  hierarchical graph side information,'' \emph{Advances in Neural Information
  Processing Systems (NeurIPS)}, vol.~33, 2020.

\bibitem{koren2009matrix}
Y.~Koren, R.~Bell, and C.~Volinsky, ``Matrix factorization techniques for
  recommender systems,'' \emph{Computer}, vol.~42, no.~8, pp. 30--37, 2009.

\bibitem{mcpherson2001birds}
M.~McPherson, L.~Smith-Lovin, and J.~M. Cook, ``Birds of a feather: Homophily
  in social networks,'' \emph{Annual review of sociology}, vol.~27, no.~1, pp.
  415--444, 2001.

\bibitem{tang2013social}
J.~Tang, X.~Hu, and H.~Liu, ``Social recommendation: a review,'' \emph{Social
  Network Analysis and Mining}, vol.~3, no.~4, pp. 1113--1133, 2013.

\bibitem{cai2010graph}
D.~Cai, X.~He, J.~Han, and T.~S. Huang, ``Graph regularized nonnegative matrix
  factorization for data representation,'' \emph{IEEE transactions on pattern
  analysis and machine intelligence}, vol.~33, no.~8, pp. 1548--1560, 2010.

\bibitem{jamali2010matrix}
M.~Jamali and M.~Ester, ``A matrix factorization technique with trust
  propagation for recommendation in social networks,'' \emph{Proceedings of the
  fourth ACM conference on Recommender systems}, pp. 135--142, 2010.

\bibitem{li2009relation}
W.-J. Li and D.-Y. Yeung, ``Relation regularized matrix factorization,''
  \emph{Twenty-First International Joint Conference on Artificial Intelligence
  (IJCAI)}, 2009.

\bibitem{ma2011recommender}
H.~Ma, D.~Zhou, C.~Liu, M.~R. Lyu, and I.~King, ``Recommender systems with
  social regularization,'' \emph{Proceedings of the fourth ACM international
  conference on Web search and data mining}, pp. 287--296, 2011.

\bibitem{kalofolias2014matrix}
V.~Kalofolias, X.~Bresson, M.~Bronstein, and P.~Vandergheynst, ``Matrix
  completion on graphs,'' \emph{arXiv preprint arXiv:1408.1717}, 2014.

\bibitem{ma2008sorec}
H.~Ma, H.~Yang, M.~R. Lyu, and I.~King, ``So{R}ec: social recommendation using
  probabilistic matrix factorization,'' \emph{Proceedings of the 17th ACM
  conference on Information and knowledge management}, pp. 931--940, 2008.

\bibitem{ma2009learning}
H.~Ma, I.~King, and M.~R. Lyu, ``Learning to recommend with social trust
  ensemble,'' \emph{Proceedings of the 32nd international ACM SIGIR conference
  on Research and development in information retrieval}, pp. 203--210, 2009.

\bibitem{guo2015trustsvd}
G.~Guo, J.~Zhang, and N.~Yorke-Smith, ``Trust{SVD}: Collaborative filtering
  with both the explicit and implicit influence of user trust and of item
  ratings,'' \emph{Twenty-Ninth AAAI Conference on Artificial Intelligence},
  2015.

\bibitem{zhao2017collaborative}
H.~Zhao, Q.~Yao, J.~T. Kwok, and D.~L. Lee, ``Collaborative filtering with
  social local models,'' \emph{IEEE International Conference on Data Mining
  (ICDM)}, pp. 645--654, 2017.

\bibitem{chouvardas2017robust}
S.~Chouvardas, M.~A. Abdullah, L.~Claude, and M.~Draief, ``Robust online matrix
  completion on graphs,'' \emph{IEEE International Conference on Acoustics,
  Speech and Signal Processing (ICASSP)}, pp. 4019--4023, 2017.

\bibitem{massa2005controversial}
P.~Massa and P.~Avesani, ``Controversial users demand local trust metrics: An
  experimental study on epinions. com community,'' \emph{AAAI}, pp. 121--126,
  2005.

\bibitem{golbeck2006filmtrust}
J.~Golbeck, J.~Hendler \emph{et~al.}, ``Filmtrust: Movie recommendations using
  trust in web-based social networks,'' \emph{Proceedings of the IEEE Consumer
  communications and networking conference}, vol.~96, no.~1, pp. 282--286,
  2006.

\bibitem{jamali2009trustwalker}
M.~Jamali and M.~Ester, ``Trustwalker: a random walk model for combining
  trust-based and item-based recommendation,'' \emph{Proceedings of the 15th
  ACM SIGKDD international conference on Knowledge discovery and data mining},
  pp. 397--406, 2009.

\bibitem{jamali2009using}
------, ``Using a trust network to improve top-n recommendation,''
  \emph{Proceedings of the third ACM conference on Recommender systems}, pp.
  181--188, 2009.

\bibitem{yang2012bayesian}
X.~Yang, Y.~Guo, and Y.~Liu, ``Bayesian-inference-based recommendation in
  online social networks,'' \emph{IEEE Transactions on Parallel and Distributed
  Systems}, vol.~24, no.~4, pp. 642--651, 2012.

\bibitem{yang2012top}
X.~Yang, H.~Steck, Y.~Guo, and Y.~Liu, ``On top-k recommendation using social
  networks,'' \emph{Proceedings of the sixth ACM conference on Recommender
  systems}, pp. 67--74, 2012.

\bibitem{monti2017geometric}
F.~Monti, M.~Bronstein, and X.~Bresson, ``Geometric matrix completion with
  recurrent multi-graph neural networks,'' \emph{Advances in Neural Information
  Processing Systems (NIPS)}, pp. 3697--3707, 2017.

\bibitem{berg2017graph}
R.~v.~d. Berg, T.~N. Kipf, and M.~Welling, ``Graph convolutional matrix
  completion,'' \emph{arXiv preprint arXiv:1706.02263}, 2017.

\bibitem{chiang2015matrix}
K.-Y. Chiang, C.-J. Hsieh, and I.~S. Dhillon, ``Matrix completion with noisy
  side information,'' \emph{Advances in Neural Information Processing Systems
  (NIPS)}, pp. 3447--3455, 2015.

\bibitem{rao2015collaborative}
N.~Rao, H.-F. Yu, P.~K. Ravikumar, and I.~S. Dhillon, ``Collaborative filtering
  with graph information: Consistency and scalable methods,'' \emph{Advances in
  neural information processing systems}, pp. 2107--2115, 2015.

\bibitem{ahn2018binary}
K.~Ahn, K.~Lee, H.~Cha, and C.~Suh, ``Binary rating estimation with graph side
  information,'' \emph{Advances in Neural Information Processing Systems
  (NeurIPS)}, pp. 4272--4283, 2018.

\bibitem{yoon2018joint}
J.~Yoon, K.~Lee, and C.~Suh, ``On the joint recovery of community structure and
  community features,'' \emph{56th Annual Allerton Conference on Communication,
  Control, and Computing}, pp. 686--694, 2018.

\bibitem{tan2019community}
Q.~Zhang, V.~Y.~F. Tan, and C.~Suh, ``Community detection and matrix completion
  with social and item similarity graphs,'' \emph{IEEE Transactions on Signal
  Processing}, vol.~69, pp. 917--931, 2021.

\bibitem{jo2020discrete}
C.~Jo and K.~Lee, ``Discrete-valued preference estimation with graph side
  information,'' \emph{arXiv preprint arXiv:2003.07040}, 2020.

\bibitem{nguyen2019low}
L.~T. Nguyen, J.~Kim, and B.~Shim, ``Low-rank matrix completion: A contemporary
  survey,'' \emph{IEEE Access}, vol.~7, pp. 94\,215--94\,237, 2019.

\bibitem{candes2009exact}
E.~J. Cand{\`e}s and B.~Recht, ``Exact matrix completion via convex
  optimization,'' \emph{Foundations of Computational Mathematics}, vol.~9,
  no.~6, pp. 717--772, 2009.

\bibitem{keshavan2010matrix}
R.~H. Keshavan, A.~Montanari, and S.~Oh, ``Matrix completion from a few
  entries,'' \emph{IEEE Transactions on Information Theory}, vol.~56, no.~6,
  pp. 2980--2998, 2010.

\bibitem{candes2010power}
E.~J. Cand{\`e}s and T.~Tao, ``The power of convex relaxation: Near-optimal
  matrix completion,'' \emph{IEEE Transactions on Information Theory}, vol.~56,
  no.~5, pp. 2053--2080, 2010.

\bibitem{fazel2002matrix}
M.~Fazel, ``Matrix rank minimization with applications,'' \emph{PhD thesis,
  Stanford University}, 2002.

\bibitem{cai2010singular}
J.-F. Cai, E.~J. Cand{\`e}s, and Z.~Shen, ``A singular value thresholding
  algorithm for matrix completion,'' \emph{SIAM Journal on Optimization},
  vol.~20, no.~4, pp. 1956--1982, 2010.

\bibitem{fornasier2011low}
M.~Fornasier, H.~Rauhut, and R.~Ward, ``Low-rank matrix recovery via
  iteratively reweighted least squares minimization,'' \emph{SIAM Journal on
  Optimization}, vol.~21, no.~4, pp. 1614--1640, 2011.

\bibitem{mohan2012iterative}
K.~Mohan and M.~Fazel, ``Iterative reweighted algorithms for matrix rank
  minimization,'' \emph{The Journal of Machine Learning Research (JMLR)},
  vol.~13, no. Nov, pp. 3441--3473, 2012.

\bibitem{lee2010admira}
K.~Lee and Y.~Bresler, ``{ADMiRA}: Atomic decomposition for minimum rank
  approximation,'' \emph{IEEE Transactions on Information Theory}, vol.~56,
  no.~9, pp. 4402--4416, 2010.

\bibitem{wang2014rank}
Z.~Wang, M.-J. Lai, Z.~Lu, W.~Fan, H.~Davulcu, and J.~Ye, ``Rank-one matrix
  pursuit for matrix completion,'' \emph{International Conference on Machine
  Learning (ICML)}, pp. 91--99, 2014.

\bibitem{tanner2016low}
J.~Tanner and K.~Wei, ``Low rank matrix completion by alternating steepest
  descent methods,'' \emph{Applied and Computational Harmonic Analysis},
  vol.~40, no.~2, pp. 417--429, 2016.

\bibitem{wen2012solving}
Z.~Wen, W.~Yin, and Y.~Zhang, ``Solving a low-rank factorization model for
  matrix completion by a nonlinear successive over-relaxation algorithm,''
  \emph{Mathematical Programming Computation}, vol.~4, no.~4, pp. 333--361,
  2012.

\bibitem{dai2705set}
W.~Dai and O.~Milenkovic, ``{SET}: an algorithm for consistent matrix
  completion,'' \emph{IEEE International Conference on Acoustics, Speech and
  Signal Processing}, pp. 3646--3649, 2010.

\bibitem{vandereycken2013low}
B.~Vandereycken, ``Low-rank matrix completion by riemannian optimization,''
  \emph{SIAM Journal on Optimization}, vol.~23, no.~2, pp. 1214--1236, 2013.

\bibitem{hu2012fast}
Y.~Hu, D.~Zhang, J.~Ye, X.~Li, and X.~He, ``Fast and accurate matrix completion
  via truncated nuclear norm regularization,'' \emph{IEEE transactions on
  pattern analysis and machine intelligence}, vol.~35, no.~9, pp. 2117--2130,
  2012.

\bibitem{gotoh2018dc}
J.-y. Gotoh, A.~Takeda, and K.~Tono, ``Dc formulations and algorithms for
  sparse optimization problems,'' \emph{Mathematical Programming}, vol. 169,
  no.~1, pp. 141--176, 2018.

\bibitem{candes2010matrix}
E.~J. Candes and Y.~Plan, ``Matrix completion with noise,'' \emph{Proceedings
  of the IEEE}, vol.~98, no.~6, pp. 925--936, 2010.

\bibitem{holland1983stochastic}
P.~W. Holland, K.~B. Laskey, and S.~Leinhardt, ``Stochastic blockmodels: First
  steps,'' \emph{Social networks}, vol.~5, no.~2, pp. 109--137, 1983.

\bibitem{abbe2017community}
E.~Abbe, ``Community detection and stochastic block models: recent
  developments,'' \emph{The Journal of Machine Learning Research (JMLR)},
  vol.~18, no.~1, pp. 6446--6531, 2017.

\bibitem{abbe2015exact}
E.~Abbe, A.~S. Bandeira, and G.~Hall, ``Exact recovery in the stochastic block
  model,'' \emph{IEEE Transactions on Information Theory}, vol.~62, no.~1, pp.
  471--487, 2015.

\bibitem{abbe2015community}
E.~Abbe and C.~Sandon, ``Community detection in general stochastic block
  models: Fundamental limits and efficient algorithms for recovery,''
  \emph{IEEE 56th Annual Symposium on Foundations of Computer Science}, pp.
  670--688, 2015.

\bibitem{jog2015information}
V.~Jog and P.-L. Loh, ``Information-theoretic bounds for exact recovery in
  weighted stochastic block models using the renyi divergence,'' \emph{arXiv
  preprint arXiv:1509.06418}, 2015.

\bibitem{mossel2015reconstruction}
E.~Mossel, J.~Neeman, and A.~Sly, ``Reconstruction and estimation in the
  planted partition model,'' \emph{Probability Theory and Related Fields}, vol.
  162, no. 3-4, pp. 431--461, 2015.

\bibitem{hajek2017information}
B.~Hajek, Y.~Wu, and J.~Xu, ``Information limits for recovering a hidden
  community,'' \emph{IEEE Transactions on Information Theory}, vol.~63, no.~8,
  pp. 4729--4745, 2017.

\bibitem{saad2018community}
H.~Saad and A.~Nosratinia, ``Community detection with side information: Exact
  recovery under the stochastic block model,'' \emph{IEEE Journal of Selected
  Topics in Signal Processing}, vol.~12, no.~5, pp. 944--958, 2018.

\bibitem{saad2018exact}
------, ``Exact recovery in community detection with continuous-valued side
  information,'' \emph{IEEE Signal Processing Letters}, vol.~26, no.~2, pp.
  332--336, 2018.

\bibitem{saad2018recovering}
------, ``Recovering a single community with side information,'' \emph{arXiv
  preprint arXiv:1809.01738}, 2018.

\bibitem{kemp2006learning}
C.~Kemp, J.~B. Tenenbaum, T.~L. Griffiths, T.~Yamada, and N.~Ueda, ``Learning
  systems of concepts with an infinite relational model,'' \emph{AAAI}, vol.~3,
  p.~5, 2006.

\bibitem{airoldi2008mixed}
E.~M. Airoldi, D.~M. Blei, S.~E. Fienberg, and E.~P. Xing, ``Mixed membership
  stochastic blockmodels,'' \emph{The Journal of machine learning research
  (JMLR)}, vol.~9, no. Sep, pp. 1981--2014, 2008.

\bibitem{ball2011efficient}
B.~Ball, B.~Karrer, and M.~E. Newman, ``Efficient and principled method for
  detecting communities in networks,'' \emph{Physical Review E}, vol.~84,
  no.~3, p. 036103, 2011.

\bibitem{karrer2011stochastic}
B.~Karrer and M.~E. Newman, ``Stochastic blockmodels and community structure in
  networks,'' \emph{Physical review E}, vol.~83, no.~1, p. 016107, 2011.

\bibitem{aicher2013adapting}
C.~Aicher, A.~Z. Jacobs, and A.~Clauset, ``Adapting the stochastic block model
  to edge-weighted networks,'' \emph{arXiv preprint arXiv:1305.5782}, 2013.

\bibitem{yang2013community}
J.~Yang, J.~McAuley, and J.~Leskovec, ``Community detection in networks with
  node attributes,'' \emph{2013 IEEE 13th International Conference on Data
  Mining}, pp. 1151--1156, 2013.

\bibitem{clauset2008hierarchical}
A.~Clauset, C.~Moore, and M.~E. Newman, ``Hierarchical structure and the
  prediction of missing links in networks,'' \emph{Nature}, vol. 453, no. 7191,
  pp. 98--101, 2008.

\bibitem{peixoto2014hierarchical}
T.~P. Peixoto, ``Hierarchical block structures and high-resolution model
  selection in large networks,'' \emph{Physical Review X}, vol.~4, no.~1, p.
  011047, 2014.

\bibitem{lyzinski2016community}
V.~Lyzinski, M.~Tang, A.~Athreya, Y.~Park, and C.~E. Priebe, ``Community
  detection and classification in hierarchical stochastic blockmodels,''
  \emph{IEEE Transactions on Network Science and Engineering}, vol.~4, no.~1,
  pp. 13--26, 2016.

\bibitem{cohen2019hierarchical}
V.~Cohen-Addad, V.~Kanade, F.~Mallmann-Trenn, and C.~Mathieu, ``Hierarchical
  clustering: Objective functions and algorithms,'' \emph{Journal of the ACM
  (JACM)}, vol.~66, no.~4, pp. 1--42, 2019.

\bibitem{ashtiani2016clustering}
H.~Ashtiani, S.~Kushagra, and S.~Ben-David, ``Clustering with same-cluster
  queries,'' \emph{Advances in neural information processing systems (NIPS)},
  pp. 3216--3224, 2016.

\bibitem{mazumdar2017query}
A.~Mazumdar and B.~Saha, ``Query complexity of clustering with side
  information,'' \emph{Advances in Neural Information Processing Systems
  (NIPS)}, pp. 4682--4693, 2017.

\bibitem{macwilliams1977theory}
F.~J. MacWilliams and N.~J.~A. Sloane, \emph{The theory of error correcting
  codes}.\hskip 1em plus 0.5em minus 0.4em\relax Elsevier, 1977, vol.~16.

\bibitem{zhang2016minimax}
A.~Y. Zhang and H.~H. Zhou, ``Minimax rates of community detection in
  stochastic block models,'' \emph{The Annals of Statistics}, vol.~44, no.~5,
  pp. 2252--2280, 2016.

\bibitem{alon2004probabilistic}
N.~Alon and J.~H. Spencer, \emph{The probabilistic method}.\hskip 1em plus
  0.5em minus 0.4em\relax John Wiley \& Sons, 2004.

\end{thebibliography}
